\documentclass[fleqn,usenatbib]{mnras}

\usepackage{newtxtext,newtxmath}
\usepackage[normalem]{ulem}  
\usepackage{booktabs}
\usepackage{multirow}
\usepackage{siunitx}   
\usepackage{placeins}

\usepackage[T1]{fontenc}

\DeclareRobustCommand{\VAN}[3]{#2}
\let\VANthebibliography\thebibliography
\def\thebibliography{\DeclareRobustCommand{\VAN}[3]{##3}\VANthebibliography}

\usepackage{graphicx}	
\usepackage{amsmath}	

\makeatletter
\newcommand{\floatpageon}{\def\fps@figure{p}}
\newcommand{\floatpageoff}{\def\fps@figure{tbp}}
\makeatother

\title[UnEZ]{UnEZ: A Fully Unsupervised Self-Templating Neural Network for Redshift and Line Strength Estimation}

\author[Abd et al.]{
Asha Abd,$^{1}$\thanks{E-mail: aabd@swin.edu.au}
Karl Glazebrook,$^{1}$
Colin Jacobs$^{1}$
\\
$^{1}$Centre for Astrophysics and Supercomputing, Swinburne University of Technology, PO Box 218, Hawthorn, VIC 3122, Australia\\
}

\date{Accepted XXX. Received YYY; in original form ZZZ}

\pubyear{2026}

\begin{document}
\label{firstpage}
\pagerange{\pageref{firstpage}--\pageref{lastpage}}
\maketitle

\begin{abstract}
As spectroscopic surveys grow, estimating galaxy properties such as metallicity, star formation rate, and redshift becomes computationally cumbersome. Methods of estimating such properties rely on template spectra or curated redshift catalogs, which fail to represent the diversity of objects seen in large-scale surveys. We present UnEZ, a fully unsupervised, physics-informed machine learning model that can predict spectroscopic redshifts and line strengths with no prior redshift or line strength labels. An encoder predicts emission and absorption line amplitudes/widths, and a fixed deterministic decoder constructs a unique template, estimated redshift, redshift confidence curve, and spectral model. UnEZ was tested on the Sloan Digital Sky Survey (SDSS) Data Release 18 due to its large dataset and thoroughly vetted redshift catalog which serves as an independent benchmark. Despite the lack of templates, line strength labels, or redshift labels, UnEZ's performance competes with modern redshift estimation algorithms. At $\text{SNR}>10$, UnEZ achieves an outlier fraction of 0.459\% and a normalized median absolute deviation (NMAD) of $1.237 \times 10^{-4}$. At $\text{SNR}>3$, UnEZ achieves an outlier fraction of 3.89\% and an NMAD of $1.471 \times 10^{-4}$. Furthermore, we demonstrate how UnEZ's line strength and redshift predictions can be utilized for various purposes such as constructing template spectra required in traditional redshift estimation algorithms. UnEZ illustrates how a fully unsupervised neural network can allow redshift, among other galaxy properties, to emerge naturally from the spectra themselves, providing a template-free, label-free approach that scales to the demands of future surveys.
\end{abstract}



\section{Introduction}
\label{sec:introduction} 
Recent efforts in astronomical surveying have yielded vast datasets ranging across many redshift regimes and areas of the sky. Such large datasets enable high-precision statistical modeling of the universe.

Redshift carries information that is vital to such cosmological models. Measures such as scale factor or lookback time \citep{hogg2000distancemeasurescosmology} can be calculated from redshift for example. Other important galaxy characteristics such as rotation curves \citep{2001ARA&A..39..137S} can also be isolated given the redshift of an object. Large-scale surveys allow statistical analyses across millions of spectra, such as measurements of the galaxy clustering power spectrum \citep{Tegmark_2004}, making them valuable for verifying cosmological models \citep{Abdul_Karim_2025}.

Photometry has also been used to estimate redshifts. Traditional algorithms fit template spectral energy distributions (SEDs) to the broad features traced out between photometric filters \citep{2000ApJ...536..571B}. Photometric redshift can suffer from degeneracies resulting in catastrophic redshift prediction failures \citep{Massarotti_2001} and have larger uncertainties due to the lack of highly resolved emission/absorption lines. While photometric redshifts are cheap in both observational resources and compute needs, they lack the precision of spectroscopic redshifts \citep{Newman_2022}.

Spectroscopy, while being more resource-expensive than photometry, is particularly important for high precision redshift measurements. Modern fiber-fed spectrographs can collect thousands of spectra in a single pointing allowing for large surveys \citep{Smee_2013}. Redshift surveys were not always conducted on the scale they are today. The first surveys were composed of only a few dozen redshifts due to the hours or sometimes nights required to collect a spectrum on photographic plates.

Endeavors such as mapping out galaxy clusters require a large number of precise redshift measurements to understand how distant they are. Later surveys such as the Center for Astrophysics (CfA) survey collected several thousand spectra automatically. Such surveys led to the discovery of large-scale structures in the universe \citep{1988IAUS..130..105H}. The 1990s and early 2000s brought even larger surveys such as the 2dF Galaxy Redshift Survey which produced almost 250,000 spectra \citep{2003astro.ph..6581C}, with later surveys such as the Sloan Digital Sky Survey (SDSS) producing millions of spectra \citep{2023ApJS..267...44A}. Future and current endeavors such as the Dark Energy Spectroscopic Instrument (DESI) survey will bring tens of millions more spectra \citep{desicollaboration2026datarelease1dark}, creating a unprecedented amount of data available for analysis.

Many methods of estimating the redshift of spectra have been adopted over the years. The first method required manually identifying spectral lines and comparing their distances to one another with a reference line list \citep{1956AJ.....61...97H}. This method was very labor-intensive although it did come with the benefit of being human-verified for every data point. Modern methods of estimating redshift generally utilize a set of templates and a $\chi^{2}$ or cross-correlation algorithm. The $\chi^{2}$ algorithm fits templates to a spectrum at trial redshifts to minimize $\chi^{2}$ \citep{Bolton_2012}. The cross-correlation algorithm, on the other hand, computes the correlation between templates and spectra in log-wavelength space to find the redshift corresponding with the highest correlation \citep{1979AJ.....84.1511T}. AUTOZ, an algorithm for estimating redshift, removes the continuum from spectra, removes potential skyline or cosmic ray contamination, and cross-correlates the processed spectrum with a set of templates to estimate redshift \citep{Baldry_2014}. Further endeavors such as MARZ have refined that process with a larger template library and a user interface \citep{Hinton_2016}. 

As astronomical datasets grow in area and depth, estimating redshifts quickly becomes cumbersome. The SDSS dataset alone contains millions of spectra, and future surveys promise tens of millions more. Machine learning models are a good candidate for analyzing such datasets due to their capacity to learn complex non-linear relationships in the data \citep{Fluke_2019}. The use of neural networks in astronomy dates as far back as the 1990s for tasks such as correcting atmospheric aberrations in adaptive optics and object classification \citep{1993VA.....36..141M}. As neural networks became larger and more sophisticated, their accuracy and precision made them a much more enticing candidate for tasks such as morphological classification \citep{Dieleman_2015}, photometric redshift estimation \citep{Collister_2004}, and stellar classification \citep{10.1093/mnras/sty3020}. Neural networks can be classified into two generic categories, supervised and unsupervised, which come with different benefits and shortcomings.

Supervised machine learning models are very well studied. Models have been trained to predict redshift using redshift labels from traditional fitting algorithms \citep{10.1093/mnras/stae1461}. SpecPT, for example, uses a unique transformer architecture to predict redshifts directly by training on spectra with their corresponding redshift label \citep{Pattnaik_2025}. Such supervised neural networks learn by comparing predictions with ground-truth labels \citep{1986Natur.323..533R}. The ground-truth redshift labels are often sourced from template-fitting algorithms that have been heavily vetted for bad fits \citep{2023ApJ...943...68L, 2023AJ....165..124A}.

Unsupervised learning, however, operates under fundamentally different constraints. Unsupervised models have no access to a label to directly compare their predictions to. These neural networks typically compress high-dimensional data (e.g. spectra) to a lower-dimensional space while preserving the important features of the data. For example, autoencoders compress their input data into a low-dimensional latent space before decompressing to recreate the input data \citep{autoencoder}. Another powerful unsupervised algorithm is the self-organizing map (SOM). SOMs map each input to a discrete and finite two-dimensional map \citep{Kohonen2004SelforganizedFO}. Self-organizing maps are ideal for clustering because they preserve the topology of the input space \citep{Carrasco_Kind_2014}. Further research has been conducted combining such methods in a joint learning effort. The deeply-embedded self-organizing map (DESOM) embeds a SOM in the latent-space of an autoencoder. Such models use a unique method of backpropagation that updates the encoder based on the typical autoencoder loss and the SOM loss by modifying the SOM to be differentiable \citep{2023MNRAS.518..752M}. DESOMs have been used to classify galaxy morphology, inclination angle, and star formation history \citep{Teimoorinia_2022}. All of the mentioned unsupervised models attempt to recreate the data from a compressed representation, therefore using the input data itself as the ground truth. Without prior labels for measures such as redshift, machine learning models cannot output directly interpretable values.

Numerous examples of using machine learning for redshift prediction have been introduced; however, they are all fully or semi-supervised models. Such supervised methods require spectra \citep{10.1093/mnras/stae1461} or photometry \citep{Collister_2004} with associated predetermined redshifts for training. These redshift labels are often created using templates. Examples of semi-supervised models for photometric redshift have been introduced such as SOMz \citep{Carrasco_Kind_2014}, which uses an underlying unsupervised machine learning mechanism with post-inference supervised steps that require redshift labels. However, all of the methods described thus far require prior redshift labels either in training, template construction, or post-inference calibration. Without prior labels, conventional unsupervised models compress and reconstruct data, but do not output directly interpretable physical measures such as redshift.

Modern conventional algorithms remain constrained by their reliance on human-made redshift catalogs and template spectra. SDSS templates are created by grouping objects with similar redshifts and spectral shapes together, stacking them, and using PCA to identify the eigenspectra or templates \citep{Bolton_2012}. However, such templates are prone to their own biases, which translate to estimated redshift measures \citep{Hewett_2010}. As surveys grow and the diversity of objects observed increases, a small set of templates is not sufficient to model spectra that exhibit extreme features \citep{2024AJ....168..124A}. Limitations in template-based redshift algorithms carry over to the redshift catalogs used to train neural networks, creating the need for a label-free method of estimating redshift.

We introduce UnEZ, a novel fully unsupervised redshift and line strength prediction model. UnEZ addresses the weaknesses of previous redshift estimation algorithms while maintaining high precision and accuracy. The model requires initialization with physical parameters such as a line list, redshift sampling resolution, and maximum expected redshift. UnEZ accurately predicts redshifts and line strength ratios while requiring no labels for training. These measures can then be used for further analysis such as creating templates for traditional redshift fitting algorithms, performing AGN diagnostics \citep{veilleux2002spectroscopicdiagnosticsagns}, and identifying outliers. Importantly, this model learns how to predict redshift and line strengths from first-principles by constraining the model to only output scientifically significant measures.

This paper will discuss the training, architecture, performance, and shortcomings of UnEZ. Beginning in Section \ref{sec:data} the data used for training/validation is reviewed alongside the various data cuts applied before training and testing. Section \ref{sec:architecture} describes the UnEZ pipeline from the model priors to the final output. Furthermore, Section \ref{sec:performance} analyzes UnEZ's precision and accuracy in both redshift and line strength prediction. Lastly, Section \ref{sec:discussion} concludes with discussion of the quality of UnEZ's outputs, shortcomings/benefits of the pipeline, and future endeavors.

\section{Data}
\label{sec:data} 

The spectra used for training and inference are sourced from the SDSS DR18 dataset \citep{2023ApJS..267...44A} due to its large spectral dataset, wide diversity of spectra, and robust redshift catalog. SDSS was designed to map a large portion of the sky in detail to understand the large-scale structure of the universe and to study various properties of galaxies and quasars. Spectra are taken using the Baryonic Oscillation Spectroscopic Survey (BOSS) \citep{Smee_2013} and original SDSS spectrographs \citep{2000AJ....120.1579Y}. These sophisticated systems separate the incoming light into a red and blue side that are split at around 6000$\text{\AA}$, with each arm having a slight overlap. The full BOSS spectrograph spans from 3,600$\text{\AA}$-10,400$\text{\AA}$ while the SDSS I/II spectrograph spans from 3,800$\text{\AA}$ to 9,200$\text{\AA}$. Both spectrographs have a resolution of R$\sim$1500 at 3800$\text{\AA}$ and R$\sim$2500 at 9000$\text{\AA}$ with a pixel spacing of $\Delta\log_{10}(\lambda) = 10^{-4}$. Flux calibration and sky subtraction were performed by the SDSS pipeline.

The SDSS team has implemented a redshift estimation algorithm that utilizes a $\chi^{2}$ fit between eigenspectra and observed spectra \citep{Bolton_2012}. Although they mention a catastrophic redshift error rate of well below 1\%, others have found that the error rate could be 1-2 times higher than cited by SDSS \citep{Wu_2022}. This is important to note because UnEZ is being directly compared to the SDSS pipeline redshifts. Some objects may be labeled as a catastrophic redshift error due to the reliance on SDSS redshifts as the ``ground truth.'' 

Although both the original SDSS and BOSS spectrographs contributed to the DR18 dataset, they use different target selection methods to yield a diverse set of spectra. The original SDSS spectrograph primarily contributed to the Legacy Survey's Main Galaxy Sample (MGS) \citep{2002AJ....124.1810S}, Luminous Red Galaxy (LRG) sample \citep{2001AJ....122.2267E}, and a quasar sample selected from objects with colors significantly different from the stellar locus \citep{2002AJ....123.2945R}. These samples cover galaxies and quasars up to redshift 5.46. The BOSS spectrograph further extended the legacy dataset to include dimmer and bluer targets alongside high redshift (z>2.1) quasars \citep{2012ApJS..199....3R}.

While UnEZ is designed to be implemented across different instruments and datasets, its performance on other instruments is not tested in this paper. The diverse dataset containing over 5 million objects provided by SDSS makes it ideal for training UnEZ and testing its reliability. The massive dataset provided by SDSS also provides robust performance metrics when testing UnEZ across various SNR and redshift regimes. The high count and diversity of objects available from SDSS is ideal for statistical analysis and is often required for training large neural networks.

\section{Architecture}
\label{sec:architecture} 

\begin{figure}
	\includegraphics[width=\columnwidth]{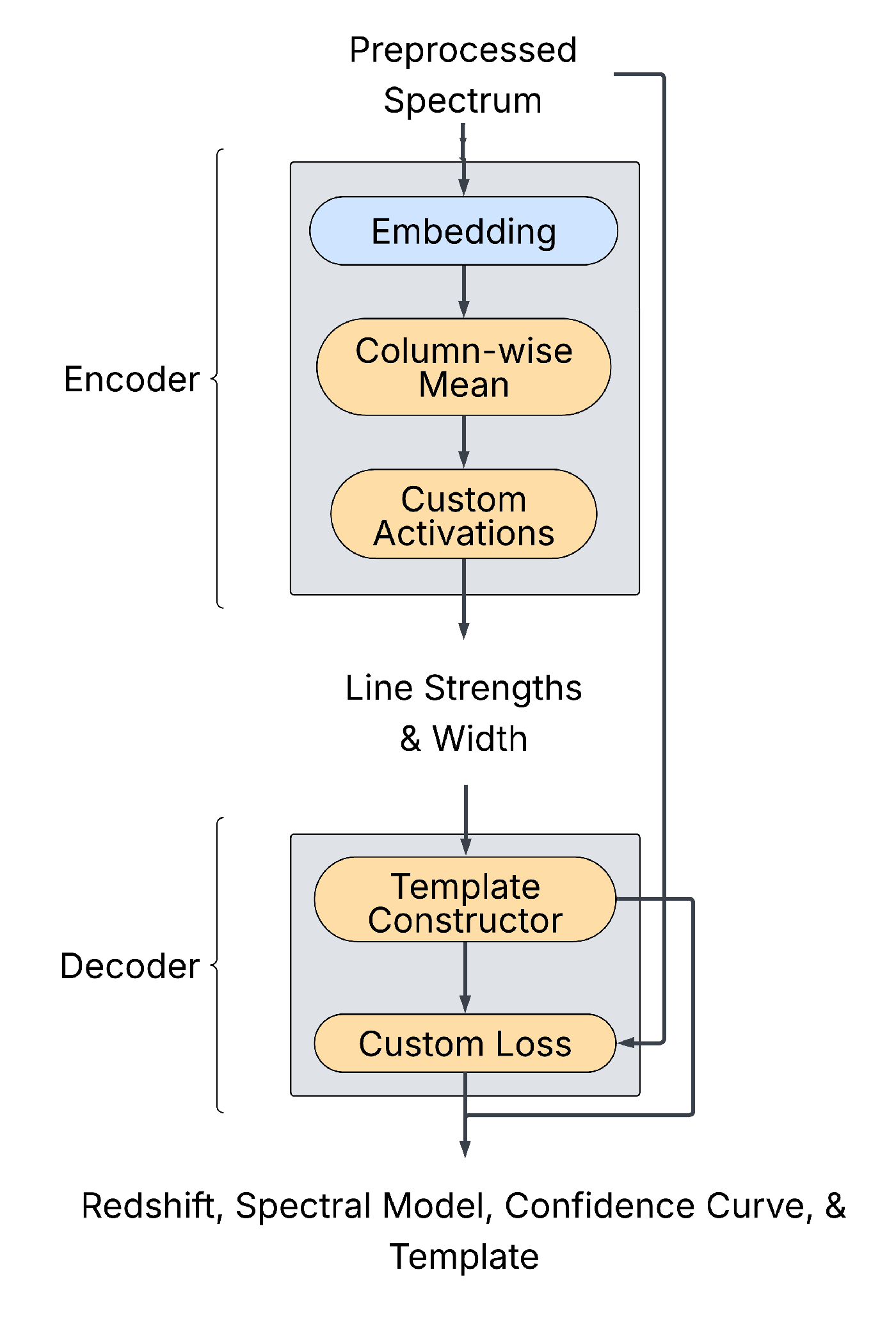}
    \caption{Diagram of the UnEZ neural network. The blue embedding block has learnable parameters while orange blocks are fixed deterministic functions with no learnable parameters.}
    \label{fig:architecture}
\end{figure}

\begin{figure}
	\includegraphics[width=\columnwidth]{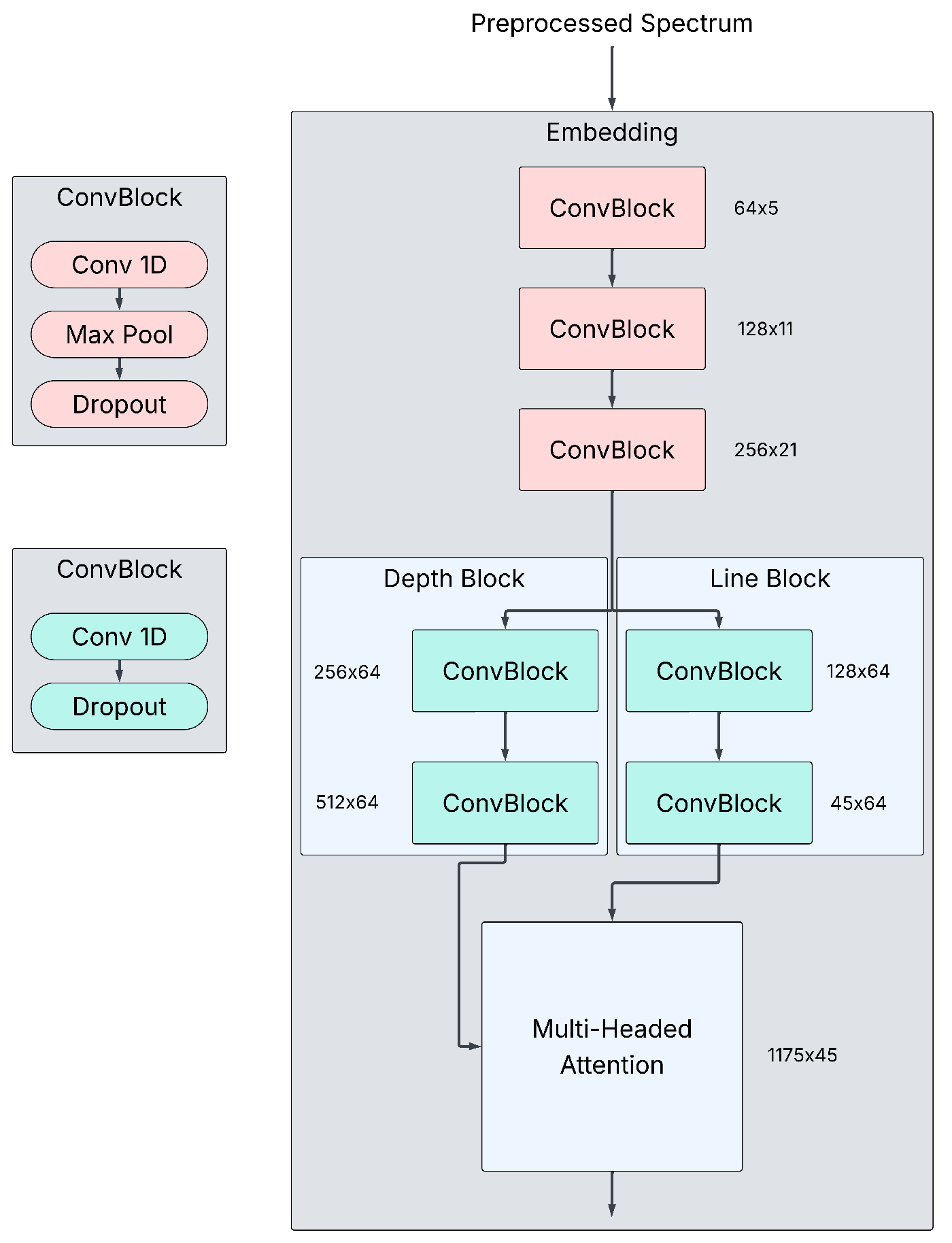}
    \caption{Diagram of the embedding layer in Figure \ref{fig:architecture}. The red blocks contain a max-pooling layer while the cyan blocks do not. The last cyan blocks feed into the multi-headed attention mechanism, with the right-side block being the keys and the left-side block being the queries. Detailed specifications and hyperparameters are listed in Table \ref{tab:unez architecture}.}
    \label{fig:embeddings}
\end{figure}

UnEZ consists of an encoder and decoder component as shown in Figure \ref{fig:architecture}. The decoder is composed of fixed algorithmic functions with no learnable parameters, ensuring that there is a systematic, unchanging reconstruction method for every spectrum. The encoder features several modern techniques such as a multi-headed attention mechanism and a series of convolutional embedding blocks to properly learn line strengths in each spectrum. Taking advantage of a sophisticated encoder and a fixed decoder, we show that UnEZ can accurately and precisely predict redshifts for millions of spectra. The configuration shown in Figures \ref{fig:architecture} and \ref{fig:embeddings} has a total of 15,899,724 parameters.

UnEZ was designed to function similar to an autoencoder while meeting an additional requirement not native to autoencoder architectures. While both autoencoders and UnEZ transform data into a highly compact latent space, UnEZ is tasked with ensuring that latent space is directly interpretable as well. While an autoencoder can compress a spectrum into a few variables that contain the spectrum's line strength features, those features cannot be directly read from the latent space. UnEZ fills that gap by ensuring that every value of the latent space represents a unique spectral line that can be directly read and interpreted without further processing. The architecture and loss function are carefully chosen to encourage learning behavior that results in a directly interpretable latent space.

The overall architecture follows a similar encoder-decoder architecture seen in autoencoders. The encoder receives a preprocessed spectrum as the input, and outputs a set of line strengths and a global line width. The decoder receives these line strengths as input, and deterministically outputs an unique template, a redshift, and a redshift confidence distribution. The deterministic nature of this decoder is not seen in traditional autoencoders which have a learned decoder. However, they both share the same goal of attempting to recreate the input spectrum from a latent space, hence we still refer to this portion of the pipeline as a decoder. The entire model is constrained by the physics information provided including spectral resolution, a maximum expected redshift, and line classifications among other variables discussed in Section \ref{sec:physics priors}.

\begin{table*}
\centering
\caption{%
  Architecture of the \textsc{UnEZ} model ($N$ = batch size).
  All Conv1D layers use same padding and ReLU activation.
  A uniform dropout rate $p = 0.2$ is applied after every Conv1D.
  The line branch provides the attention query; The depth branch provides the
  key and value.
}
\label{tab:unez architecture}
\setlength{\tabcolsep}{6pt}
\begin{tabular}{l l l l r}
\toprule
\textbf{Stage} & \textbf{Layer} & \textbf{Configuration}
  & \textbf{Output shape} & \textbf{Parameters} \\
\midrule
Input
  & InputLayer    & ---                         & $(N,\ 9400,\ 1)$   & 0 \\
\midrule
\multirow{3}{*}{Conv Block 1}
  & Conv1D        & 64 filters, $k=5$           & $(N,\ 9400,\ 64)$  & \num{384} \\
  & MaxPooling1D  & pool size $= 2$             & $(N,\ 4700,\ 64)$  & 0 \\
  & Dropout       & rate $= 0.2$                  & $(N,\ 4700,\ 64)$  & 0 \\
\midrule
\multirow{3}{*}{Conv Block 2}
  & Conv1D        & 128 filters, $k=11$         & $(N,\ 4700,\ 128)$ & \num{90240} \\
  & MaxPooling1D  & pool size $= 2$             & $(N,\ 2350,\ 128)$ & 0 \\
  & Dropout       & rate $= 0.2$                  & $(N,\ 2350,\ 128)$ & 0 \\
\midrule
\multirow{3}{*}{Conv Block 3}
  & Conv1D        & 256 filters, $k=21$         & $(N,\ 2350,\ 256)$ & \num{688384} \\
  & MaxPooling1D  & pool size $= 2$             & $(N,\ 1175,\ 256)$ & 0 \\
  & Dropout       & rate $= 0.2$                  & $(N,\ 1175,\ 256)$ & 0 \\
\midrule
\multirow{4}{*}{\shortstack[l]{Line Branch\\(Query)}}
  & Conv1D        & 128 filters, $k=64$         & $(N,\ 1175,\ 128)$ & \num{2097280} \\
  & Dropout       & rate $= 0.2$                  & $(N,\ 1175,\ 128)$ & 0 \\
  & Conv1D        & 45 filters, $k=64$          & $(N,\ 1175,\ 45)$  & \num{401457} \\
  & Dropout       & rate $= 0.2$                  & $(N,\ 1175,\ 45)$  & 0 \\
\midrule
\multirow{4}{*}{\shortstack[l]{Depth Branch\\(Key / Value)}}
  & Conv1D        & 256 filters, $k=64$         & $(N,\ 1175,\ 256)$ & \num{4194560} \\
  & Dropout       & rate $= 0.2$                  & $(N,\ 1175,\ 256)$ & 0 \\
  & Conv1D        & 512 filters, $k=64$         & $(N,\ 1175,\ 512)$ & \num{8389120} \\
  & Dropout       & rate $= 0.2$                  & $(N,\ 1175,\ 512)$ & 0 \\
\midrule
\multirow{1}{*}{\shortstack[l]{Attention}}
  & MultiHeadAttention & 4 heads, key dim of 32
                       & $(N,\ 1175,\ 45)$ & \num{38245} \\[2pt]
\midrule
\multirow{1}{*}{\shortstack[l]{Column-wise Mean}}
  & Custom  & ---                    & $(N,\ 45)$         & 0 \\
\midrule
\multirow{1}{*}{Output Heads}
  & Custom Activations + ScaledSigmoid & $44 \text{ line strengths} + 1 \text { velocity dispersion } (\sigma)$ & $(N,\ 45)$  & 45 \\
\midrule
\multicolumn{4}{r}{\textbf{Total trainable parameters}}
  & \textbf{\num{15899724}} \\
\bottomrule
\end{tabular}
\end{table*}

\subsection{Physics Priors}
\label{sec:physics priors} 

\label{sec:physicsinfo} 

Before training, UnEZ must be initialized with physics priors. These priors influence many aspects of the model, including the size of the model and the spectral lines the model can detect.

The initial use of the physics information is for preprocessing. Given a maximum expected redshift, maximum observed wavelength of the spectrograph, redshift sampling resolution, and number of pixels per spectrum, the pipeline defines a range of rest frame wavelengths the model can probe. The wavelength bins also serve as a basis for interpolating each spectrum onto a uniform wavelength axis, ensuring a constant length for all spectra. Details of the native and resampled parameters used for SDSS are highlighted in Table \ref{tab:preprocessing}.

The relationship between redshift and wavelength is a quotient in linear space as shown in Equation \ref{eq:redshift}, while being a difference in log-space as shown in Equation \ref{eq:redshift_log}.
\begin{equation}
z = \frac{\lambda_{\text{obs}}-\lambda_{\text{em}}}{\lambda_{\text{em}}}
    \label{eq:redshift}
\end{equation}
\begin{equation}
z=10^{\log_{10}(\lambda_{\text{obs}}) - \log_{10}(\lambda_{\text{em}})} - 1
    \label{eq:redshift_log}
\end{equation}
 We use new wavelength bins with equal distance in log-space rather than linear-space. This allows redshift to be computed directly as a function of shift in log-space rather than a quotient in linear-space.

The variability in the wavelength bins also allows the model to adjust its redshift probing resolution to various instruments and redshift regimes. Increasing probing resolution (super-resolution) beyond the telescope’s native capabilities can reduce scatter in redshift predictions \citep{bezanson2016}. Conversely, decreasing probing resolution allows the user to optimize for computational efficiency at the cost of redshift prediction precision.

In addition to preprocessing priors, UnEZ requires a comprehensive list of spectral lines. Each line entry must include:
\begin{enumerate}
\item \textbf{Line Center:} The rest-frame wavelength of the spectral line.
\item \textbf{Line Name:} A label used for visualization and interpretation.
\item \textbf{Line Classification:} An assignment of each line as either absorption-only, emission-only, or both.
\end{enumerate}

The line list used for this experiment, sourced from the SDSS DR6 pipeline line list \citep{sdss_dr6_linelist}, is further discussed in Appendix \ref{app:linelist} and listed in Table \ref{tab:linelist}. This line list with its corresponding classifications directly influences UnEZ’s architecture at almost every layer. Most notably, it affects the encoder's output dimension, the shape of the attention mechanism, and the Gaussian template construction layer found in Figures \ref{fig:architecture} and \ref{fig:embeddings}. 

This prior physics information is critical to UnEZ’s architecture. By constraining UnEZ's outputs to only what is physically possible, the probability of the model being trapped in a local minimum or fitting to degeneracies is greatly reduced. Constraining the parameter-space made available to UnEZ makes the problem of learning line strengths easier for the model by forcing its predictions into a smaller and more physical latent space. 

\begin{table}
	\centering
	\caption{Native and resampled spectral properties including resolution and wavelength range, and pixel spacing.}
	\label{tab:preprocessing}
	\begin{tabular}{lcc}
		\hline
		 & SDSS & BOSS \\
		\hline
		Native Min/Max Wavelength & 3800\AA--9200\AA & 3600\AA--10400\AA \\
		Native Resolution & \multicolumn{2}{c}{1500 at 3800\AA, 2500 at 9000\AA} \\
		Native Pixel Spacing & \multicolumn{2}{c}{$10^{-4}$ dex} \\
		Resampled Pixel Spacing & \multicolumn{2}{c}{4.84 $\times 10^{-5}$ dex} \\
		Resampled Min/Max Wavelength & \multicolumn{2}{c}{3600\AA -- 9400\AA} \\
		\hline
	\end{tabular}
\end{table}

\subsection{Preprocessing}
\label{sec:preprocessing}
\begin{figure*}
    \centering
    \includegraphics[width=\textwidth]{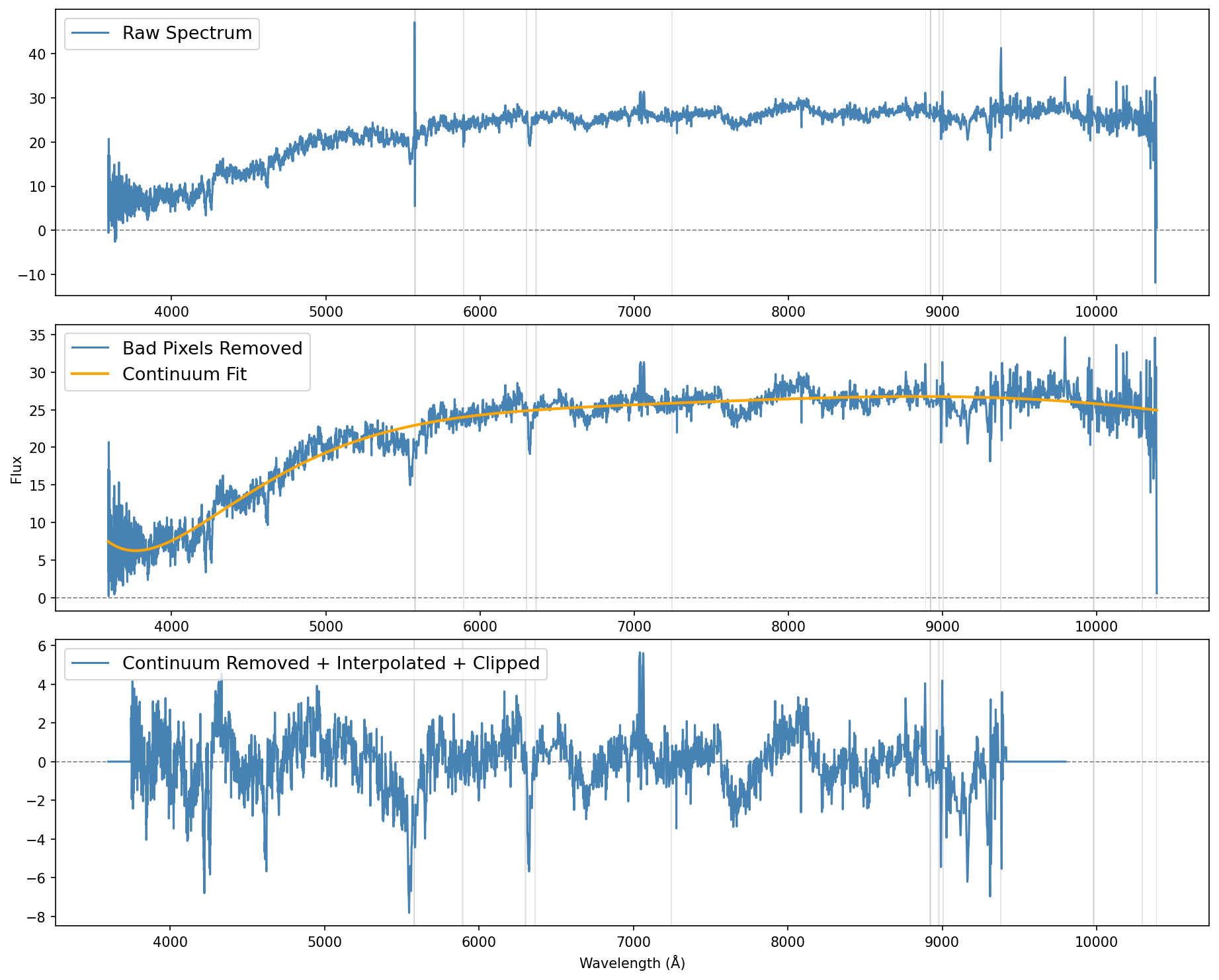}
    \caption{(top) Raw SDSS spectrum. (middle) Spectrum with bad pixels such as skylines, zero variance pixels, and cosmic rays removed. (bottom) Final preprocessed spectrum with the continuum removed that has been interpolated into uniform wavelength bins and clipped at the ends to remove edge effects. Bad pixels are denoted by the gray shaded region.}
    \label{fig:preprocessing}
\end{figure*}

The preprocessing pipeline is designed to emphasize the emission and absorption lines while removing any information that is not important for line strength and redshift prediction. This process involves three key steps: masking bad pixels, removing the continuum, and mitigating edge effects, as shown in Figure \ref{fig:preprocessing}. The spectral preprocessing steps largely follow those used in MARZ \citep{Hinton_2016}.

Prior to preprocessing, the SDSS catalog is utilized to apply a $z_{\text{SDSS}} > 0.003$ cut. This is one of the most important limitations of the UnEZ pipeline. Star spectra display a wide diversity of continua that cannot always be approximated with a polynomial as we do below. This major limitation will be addressed in future work, however, as a proof of concept we remove these low redshift spectra from our training and testing dataset.

In addition, two signal to noise ratio (SNR) cuts are applied to the data. SNR is defined in Equation \ref{eq:SNR} where pixel flux is defined as $f_{n}$, standard deviation of $f_{n}$ is defined as $\sigma_{n}$, and the number of pixels is defined as p.
\begin{equation}
\label{eq:SNR}
    \text{SNR} = \frac{1}{p}\sum^{p}_{n = 1} \frac{f_{n}}{\sigma_{n}} 
\end{equation}
For this experiment, a z < 1.0 cut was applied as it constrains the distribution of spectral lines in the dataset. Training and validation spectra are held to a high quality standard with a SNR cut of SNR > 10. Performance metrics are conducted on a wide range of SNR cuts as discussed later in Section \ref{sec:performance}. 

After all cuts have been applied, 771,086 spectra remain for testing. Of the over 700 thousand spectra left, 83,333 spectra are randomly selected for training and 16,667 random spectra are used for validation.

After all data cuts have been applied, we remove bad pixels from the spectra. Pixels that meet this criteria include cosmic rays, human-specified sky lines \citep{sdss_dr6_linelist}, NaN pixels, zero inverse-variance pixels, and pixels exceeding a predetermined power threshold. Cosmic rays are flagged as any pixel that exceeds 30 standard deviations above the mean of the four surrounding pixels. These pixels are replaced by the mean intensity of the nine neighboring pixels on both sides of the cosmic ray. If neighboring pixels are flagged as bad pixels, the pixel intensity defaults to zero. For all other bad pixels (sky line residuals, NaNs, zero inverse-variance, power threshold exceedances), each pixel is replaced by the mean intensity of the four neighboring pixels on either side. This attempts to remove any noise or anomalous data that may not be native to the object being observed. 

Once bad pixels have been masked, the pipeline moves on to the continuum removal step. A sixth-order polynomial is recursively fit to the spectrum and subtracted from the flux. Pixels with an intensity exceeding 3.5 standard deviations from the mean are excluded from the fitting process. Each time the polynomial is fit, it is then subtracted from the spectrum and the process is repeated again. This is done recursively until the spectrum no longer changes, or 15 iterations of this fitting method have been completed. This step aims to keep the sharp emission and absorption features in the spectrum while eliminating the broad and smooth continuum. Lastly, edge effects must be addressed, as they have been shown to critically degrade UnEZ’s ability to predict line strengths and redshift during testing. To reduce edge effects, the last 75 pixels on either side of the spectrum are set to zero intensity, effectively removing them completely from the spectrum. This last step is critical as testing UnEZ without the removal of edge effects has led to catastrophic failure in redshift prediction across the entire dataset due to the excess noise near the edge of the detector. Stellar spectra were omitted from this pipeline due to this continuum subtraction method failing for stars. Star continua cannot be systematically modeled using a low-order polynomial due to their wide diversity in continuum shapes.

In summary, the preprocessing pipeline removes bad pixels (e.g. cosmic rays, skylines, etc.), removes the continuum with a recursive polynomial fit, interpolates the spectrum into uniform wavelength bins, and clips the noisy edges. These preprocessing steps produce a spectrum that has been optimized for accurate line strength and redshift measurements.

\subsection{Encoder}
\label{sec:encoder} 

The encoder mechanism utilizes several modern and custom techniques to perform its sole task of predicting line strengths. It takes a preprocessed spectrum in the observed frame as an input and outputs a series of line strengths for the lines listed in the physics priors. The encoder's role is to map spectra to directly interpretable and redshift-invariant spectral line parameters. The specific parameters of the model are highlighted in Table \ref{tab:unez architecture}. 

The embedding layer, detailed in Figure \ref{fig:embeddings}, consists of several convolutional blocks and a multi-headed attention mechanism. The model begins with three convolutional blocks with each block including dropout and max pooling. Each block also has an increasing number of filters and kernel size. The last of these blocks are then fed into two separate and disconnected lines of convolutional blocks without max-pooling. While one branch is left unconstrained, the other is constrained by the number of lines, allowing the number of learnable parameters to scale with the complexity of the physics priors.  Testing has shown the lack of max-pooling yields more accurate results. One branch, referred to as the depth branch, ends with 512 filters and a kernel width of 64, as testing has shown little to no benefit from increasing the dimensionality of this branch. The other branch, referred to as the line branch, has $n_{e}+n_{w}$ filters, where $n_{e}$ is the number of lines in the provided line list and $n_{w}$ is the number of line widths predicted for each spectrum. To constrain the latent-space as much as possible we use $n_{w}=1$. The line branch output is constrained by the physics information, serving as a mechanism for the number of predicted lines, $n_{e}$, to affect the embedding architecture. The depth layer serves as a larger abstract interpretation of the line strengths. As the input size of a model increases, more parameters are often needed to prevent undertraining. The depth branch allows for this modification in future endeavors while the line branch maintains a physics-constraint on the architecture as well. The last layer in the line branch, with $n_{e}+n_{w}$ filters, is meant to represent the initial line strength predictions for each line. However, it has been shown that convolutional layers are inefficient in cross-referencing data that are far apart in high-dimensional inputs \citep{vaswani2023attentionneed}. Meaning, if there is a line on one end of the spectrum and another line on the other end of the spectrum, it cannot effectively compare the two to make an accurate prediction without increasing the depth of the model. With high-dimensional spectra this can become a critical problem. However, this can be mitigated with the multi-headed attention mechanism. 

Attention mechanisms are a modern technique that Google researchers developed to address the shortcomings of recurrent neural networks (RNNs) and convolutional neural networks (CNNs) in processing high-dimensional data \citep{vaswani2023attentionneed}. The creation of the attention mechanism led to a massive growth in neural network applications, particularly large language models such as GPT-3 \citep{brown2020languagemodelsfewshotlearners}. Attention mechanisms have been shown to be exceptional in processing high-dimensional inputs and identifying correlated features. In the context of large language models, some words have many different meanings depending on the context. For example, the word ``star'' might refer to an exceptional pupil such as a star athlete, or it could refer to an astrophysical object. The attention mechanism will correlate the word ``star'' with other words in the sentence, perhaps the word ``space'' or ``plasma'', to understand which meaning of the word is intended. In the context of UnEZ, the attention mechanism is meant to find correlations between different spectral lines. For example, a strong H$\alpha$ emission might correlate with a strong [O III] emission line, since they are both associated with star formation. The attention mechanism will learn to recognize these patterns between lines, even when they are far apart, to better predict which lines are present and at what strength. 

The multi-headed attention layer in UnEZ has 4 heads with a key dimension of 32. Although utilizing more heads or a higher dimensionality for the keys could be beneficial, these values were chosen due to hardware limitations. The purpose of the multi-headed attention mechanism is to use the initial line strength predictions and the depth layer in order to update the initial line strength predictions. By comparing the depth layer with the initial line strength predictions it is able to accurately compare two lines despite their distance from one another in the spectrum to make a more accurate prediction. In essence, it acts as an updating layer that updates the initial line strength predictions to a final line strength prediction. This multi-headed attention mechanism outputs an updated set of vectors, each representing an individual line's strength. The mean of each vector is then taken as the final line strength prediction in the column-wise mean layer. 

Lastly, these final line strength predictions are passed through the classification activation functions labeled as custom activations in Figure \ref{fig:architecture}. As discussed in Section \ref{sec:physicsinfo}, each line must be classified prior to training. If a line is classified as emission/absorption, then the line strength prediction is passed through a linear activation function. If the line is classified as emission-only, the line strength prediction is passed through a ReLU activation function. Lastly, if the line is absorption-only, it is passed through a negative ReLU activation function. This ensures that the line strength predictions are constrained to only what is physically possible, mitigating the chances of fitting to degeneracies and maintaining a directly interpretable latent space. The final vector corresponding to the line width prediction is passed through a scaled sigmoid activation function. While a typical sigmoid function is constrained to values from 0 to 1, the scaled sigmoid function is normalized to output values from a defined minimum and maximum value. This prevents the model from predicting line widths that are too narrow or broad causing a disappearing/exploding gradient problem. This final output of the encoder is a set of line strengths and widths that can be used for various tasks such as creating BPT diagrams as in Section \ref{sec:line strength predictions}, creating templates as in Section \ref{sec:self organizing map}, or estimating redshift as discussed in Section \ref{sec:decoder}.

\subsection{Decoder}
\label{sec:decoder}
The decoder reconstructs the input spectrum using a Gaussian curve to model each spectral line. The decoder's role is to transform the predicted line strengths into a unique template, spectral model, redshift confidence curve, and estimated redshift. Although we refer to this portion of the architecture as the decoder, it does not have any learned parameters that a traditional autoencoder decoder has. It is a fixed deterministic decoder (e.g. video decoder, text decoder, audio decoder, etc.), that recreates the input spectrum from its predicted line strengths.

The first layer of the decoder creates the unique Gaussian template. This template spans the length of the wavelength bins and includes all possible rest-frame wavelengths within the constraints of the provided physics information. Each line is modeled by a Gaussian shape with two primary parameters: amplitude and width. The Gaussians are centered on the wavelength of each spectral line $\mu_{n}$. The amplitude $S_{n}$ is determined by the expected line strength provided by the encoder. However, the width parameter $\sigma$ is uniform across all lines of a spectrum in this model. The final Gaussian template is modeled by the sum of all Gaussians $G$ shown in equation \ref{eq:gaussian_template}. 
\begin{equation}
G = \sum_{n=0}^{n_{e}-1} S_{n}e^{-\frac{(\lambda-\mu_{n})^{2}}{2\sigma^{2}}}
    \label{eq:gaussian_template}
\end{equation}
It is important to note that not all lines in a spectrum will have the same width. Variance in line widths can be caused by several different mechanisms. For example, lines can have different widths depending on the line of sight and disc inclination \citep{kollatschny2006}. However, this intentional constraint is enforced to stabilize unsupervised learning and constrain the parameter-space UnEZ learns from.

The subsequent decoder function predicts the redshift and calculates the training loss. This algorithm aligns the input spectrum $I$ with the Gaussian template $G$ and shifts the spectrum one pixel at a time, covering the entire range of possible redshifts. This is similar to how template fitting functions operate, however there are some critical differences. The Gaussian template used by UnEZ is unique to each spectrum, while template fitting methods utilize a few fixed templates for all spectra. Additionally, a unique loss function is utilized to compute a similarity while template fitting methods generally use either a $\chi^{2}$ or cross-correlation algorithm for computing similarity. At each incremental shift $z$, a composite loss function that utilizes both a cosine similarity and a mean squared error is calculated. The cosine similarity shown in equation \ref{eq:cosine_sim} 
\begin{equation}
\cos(\theta_{z}) = \frac{\sum_{j=0}^{m-1}G_{z+j} \cdot I_{j}}{\lVert \mathbf{G_{z}} \rVert  \lVert \mathbf{I} \rVert}
\label{eq:cosine_sim}
\end{equation}
is useful for measuring alignment between two spectra but is insensitive to a difference in amplitude. This is due to the L2 normalization factor required in calculating a cosine similarity exhibited by equations \ref{eq:G_norm} and \ref{eq:I_norm}. 
\begin{equation}
\lVert \mathbf{G_{z}} \rVert = \sqrt{\sum_{j=0}^{m-1}(G_{z+j})^{2}}
\label{eq:G_norm}
\end{equation}
\begin{equation}
\lVert \mathbf{I} \rVert = \sqrt{\sum_{j=0}^{m-1}(I_{j})^{2}}
\label{eq:I_norm}
\end{equation}
The mean squared error shown in equation \ref{eq:MSE} on the other hand, carries information about differences in magnitude and alignment between two spectra.
\begin{equation}
\text{MSE}_{z} = \frac{1}{m}\sum_{j=0}^{m-1}(G_{z+j}-I_{j})^{2}
\label{eq:MSE}
\end{equation}
By using a weighted sum of these two functions shown in equation \ref{eq:L_z} 
\begin{equation}
    L_{z}=\alpha MSE_{z} - (1-\alpha) \cos(\theta_{z})
    \label{eq:L_z}
\end{equation}
the model is able to robustly measure both alignment and magnitude measures, improving the accuracy and reliability of line strength and redshift predictions. With each shift, the loss is recalculated and assigned to the shift’s corresponding redshift. The collection of these loss values is referred to as the inverse confidence curve, and the lowest point on the inverse confidence curve corresponds to the most probable redshift of the spectrum. 

Traditional redshift estimation methods typically utilize template eigenspectra and either a cross-correlation or $\chi^{2}$ algorithm to estimate redshift \citep{Glazebrook_1998, Baldry_2014, Bolton_2012}. These methods are very similar to the custom loss layer in UnEZ's architecture in that they measure similarity across many trial redshifts. However, UnEZ predicts a unique set of line strengths for every object, rather than projecting a spectrum onto a set of templates. Given the deterministic nature of the decoder, one might ask whether the line strengths can also be estimated deterministically by projecting the spectrum onto Equation \ref{eq:gaussian_template} at every redshift and line width. This avenue is further explored in Appendix \ref{app:brute force}. We find that UnEZ performs better in every metric.

The final step of the decoder reconstructs the spectrum using the redshift prediction and Gaussian template. The redshift is obtained from the inverse confidence curve and the Gaussian template is precisely spliced at the corresponding redshift. The model effectively extracts the segment from the Gaussian template that best matches the input spectrum. Both the spectrum and Gaussian template segment are normalized via L2 normalization and multiplied by the spectrum normalization factor to ensure similar scaling. This segment is the decoder’s reconstruction of the input spectrum. Both UnEZ's final layer and autoencoders reconstruct the spectrum, however the UnEZ reconstruction is constrained to only what is physically possible.

Using such a decoder brings various benefits to redshift prediction. While modern automatic redshift algorithms use a series of fixed templates for their estimates, UnEZ's decoder creates a unique template for every spectrum. This diversity in templates will ideally fit a more diverse distribution of spectra.

\subsection{Training Loss}
\label{subsec:training loss}
The losses, $L_{z}$, computed at trial redshifts form the inverse confidence curve where low values correspond to the best fit. However, neural networks require a single scalar loss value to train \citep{damadi2023backpropagationalgorithmmathstudent}, not a collection of loss values. The mean or sum of the inverse confidence curve can be used as a training loss, but that comes with downsides. While UnEZ probes thousands of trial redshifts, only one is assumed to be correct. A vast majority of trial redshifts are bad fits, therefore a training method that focuses learning on only the most probable redshifts would reduce catastrophic redshift estimation failures.

The inverse confidence curve is further processed to encourage this ``focused'' learning. We negate this curve so we have the confidence curve, $-L_{z}$, and sort it in descending order such that $l=0$ corresponds to the most probable redshift. Only the top k values are retained (k=120 for this experiment) and weighted by a decaying exponential curve and normalized to sum one as in Equation \ref{eq:weights}.
\begin{equation}
    w_l = \frac{e^{-l/r}}{\sum_{j=0}^{k-1} e^{-j/r}}, \qquad l = 0, \dots, k-1,
    \label{eq:weights}
\end{equation}
The radius, $r$, exponentially decays with training epochs from $r=60$ to $r=0.1$ over the training epochs. This ensures the model initially learns from a broad range of candidate redshifts before narrowing its focus on only the most probable few.

The training loss is the re-inverted weighted sum of the retained values as in Equation \ref{eq:training_loss}.
\begin{equation}
    \mathcal{L} = -\sum_{l=0}^{k-1} w_l \, \mathrm{Sort}\{-L_z\}_l .
    \label{eq:training_loss}
\end{equation}

The gradient through the training loss, $\partial \mathcal{L} / \partial\, \mathrm{Sort}\{-L_z\}_l = -w_l$, scales with the exponential weights. This ensures that improbable redshifts contribute very little to the learned parameter-space during training. Restricting the loss to the top-k fits further prevents the model from learning from improbable redshifts and reduces its sensitivity to degeneracies between lines. This training loss updates only the embeddings highlighted in Figures \ref{fig:architecture} and \ref{fig:embeddings}.

\section{Performance}
\label{sec:performance} 
UnEZ was trained on a NVIDIA 80GB A100 GPU for approximately 6 hours while testing was conducted on up to five A100s simultaneously in approximately 2 hours. Varying the training set size and width of the wavelength bins have a linear impact on training/inference time and performance. Doubling the training set size will double training time, while doubling the wavelength bin width will halve the training time.

Various metrics can be used to understand the redshift and line strength performance of UnEZ. By comparing with commonly used software alongside the SDSS catalog measures, both the strengths and weaknesses of UnEZ emerge.

\subsection{Redshift}
\label{subsec: redshift}
\begin{figure*}
    \centering
    \includegraphics[width=\textwidth]{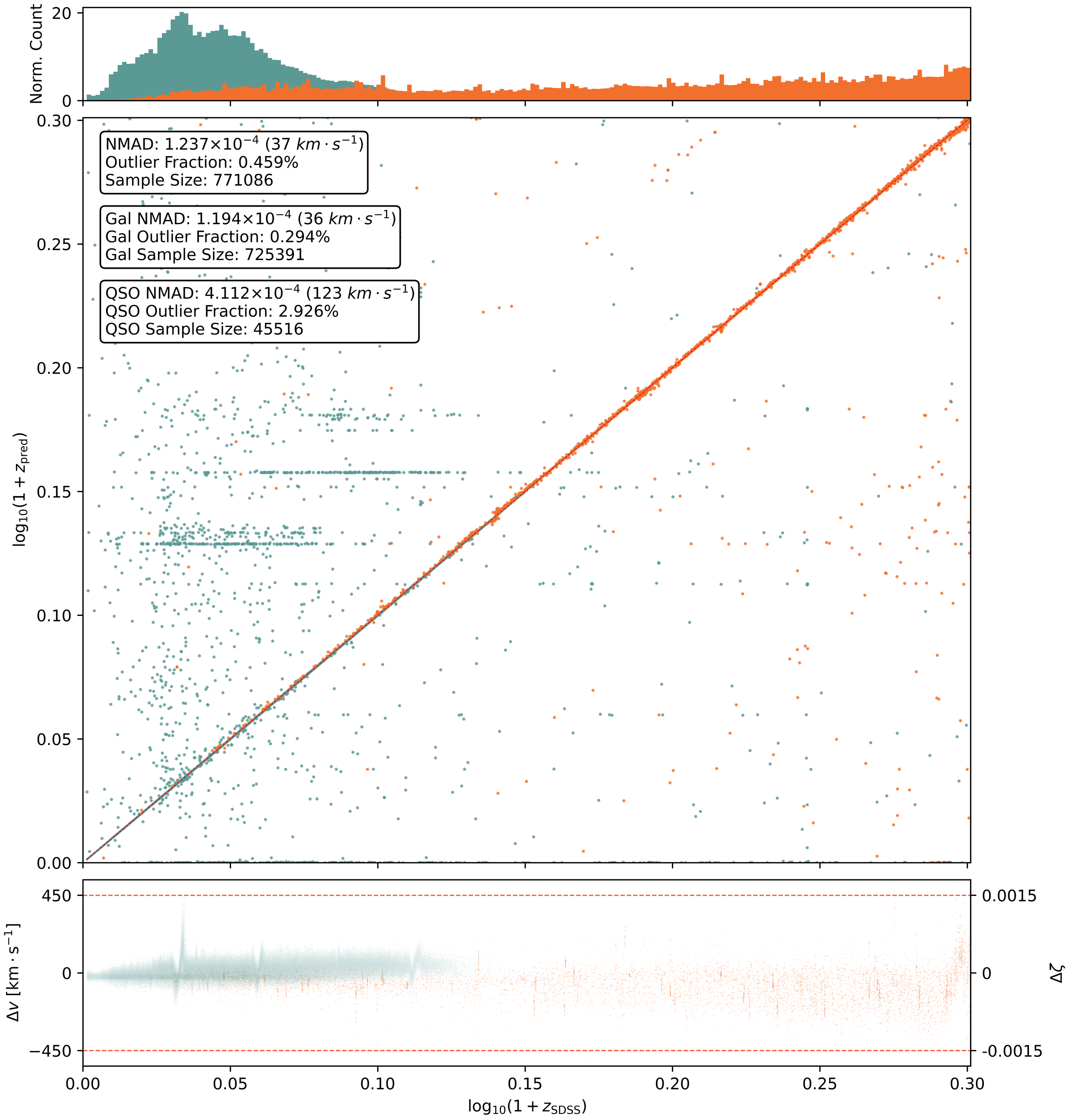}
    \caption{
    Comparison of SDSS catalog redshifts with UnEZ predictions for objects with $0.003 < z < 1.0$ and SNR $>10$. Top: Distribution of catalog redshifts, highlighting the dominance of galaxies (blue) and the increasing prevalence of quasars (orange) at higher redshift. Middle: Predicted versus catalog redshift, with the red dashed line marking the 1:1 relation. The concentration along this line indicates accurate recovery, while horizontal and parallel features mark common misidentifications. Bottom: $\Delta\zeta$ and $\Delta \text{v}$ (in [$\text{km}\cdot\text{s}^{-1}$]) versus catalog redshift showing how the spread trends with increasing redshift. UnEZ achieves NMAD = $1.24\times10^{-4}$ with an outlier fraction of 0.46\% across 771,086 spectra.}
    \label{fig:z performance}
\end{figure*}

\begin{figure*}
    \centering
    \includegraphics[width=\textwidth]{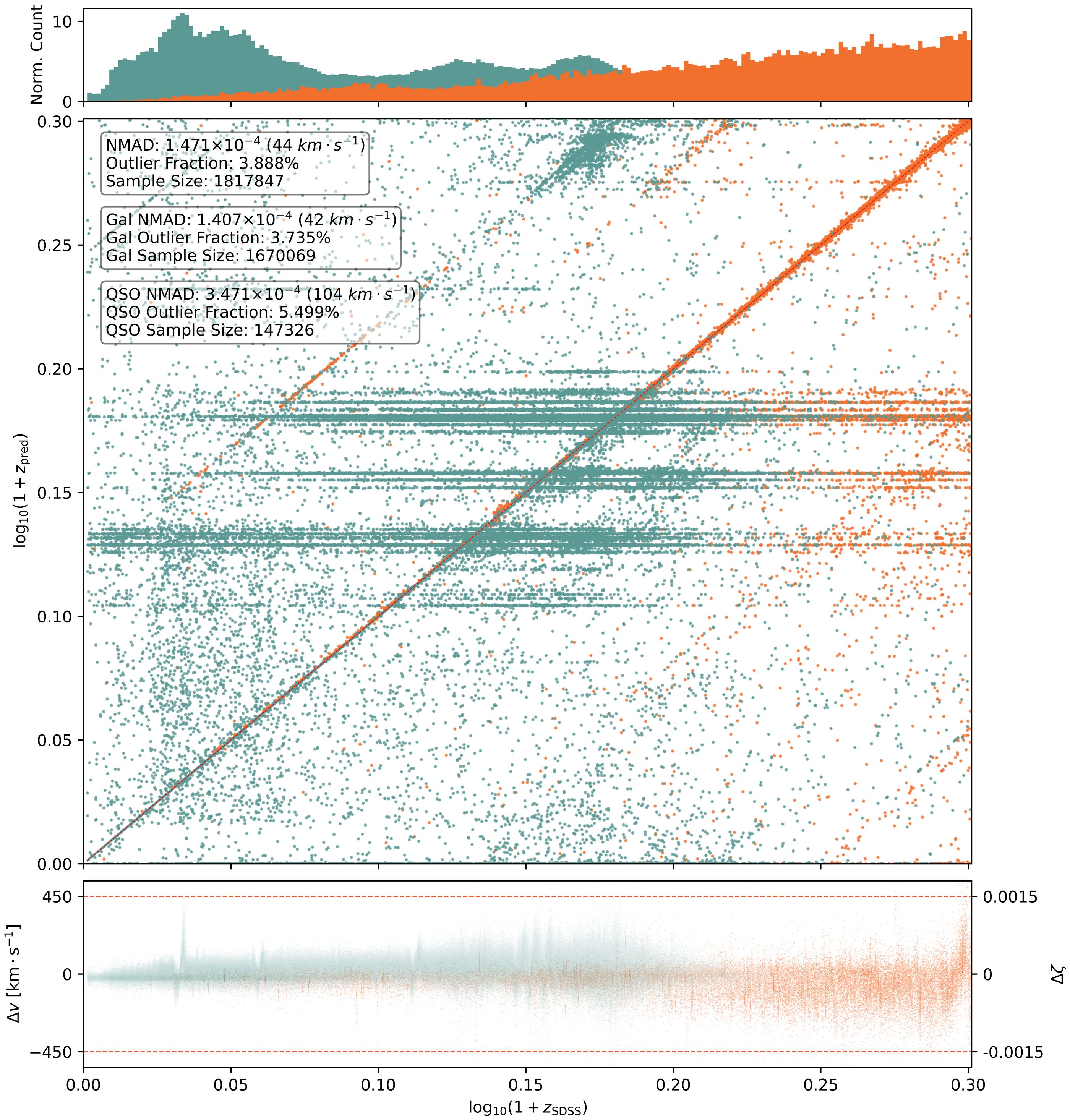}
    \caption{As Figure \ref{fig:z performance}, but for the lower SNR $>3$  sample. All panels, colors, and symbols are identical to Figure \ref{fig:z performance}. At this cut, UnEZ achieves NMAD = $1.47\times10^{-4}$ with an outlier fraction of 3.89\% across 1,817,847 spectra. Skyline contamination becomes markedly more prevalent as noise increases, and the strongest line-confusion degeneracies appear as ridges parallel to the 1:1 relation.}
    \label{fig:z performance low snr}
\end{figure*}

\begin{figure}
	\includegraphics[width=\columnwidth]{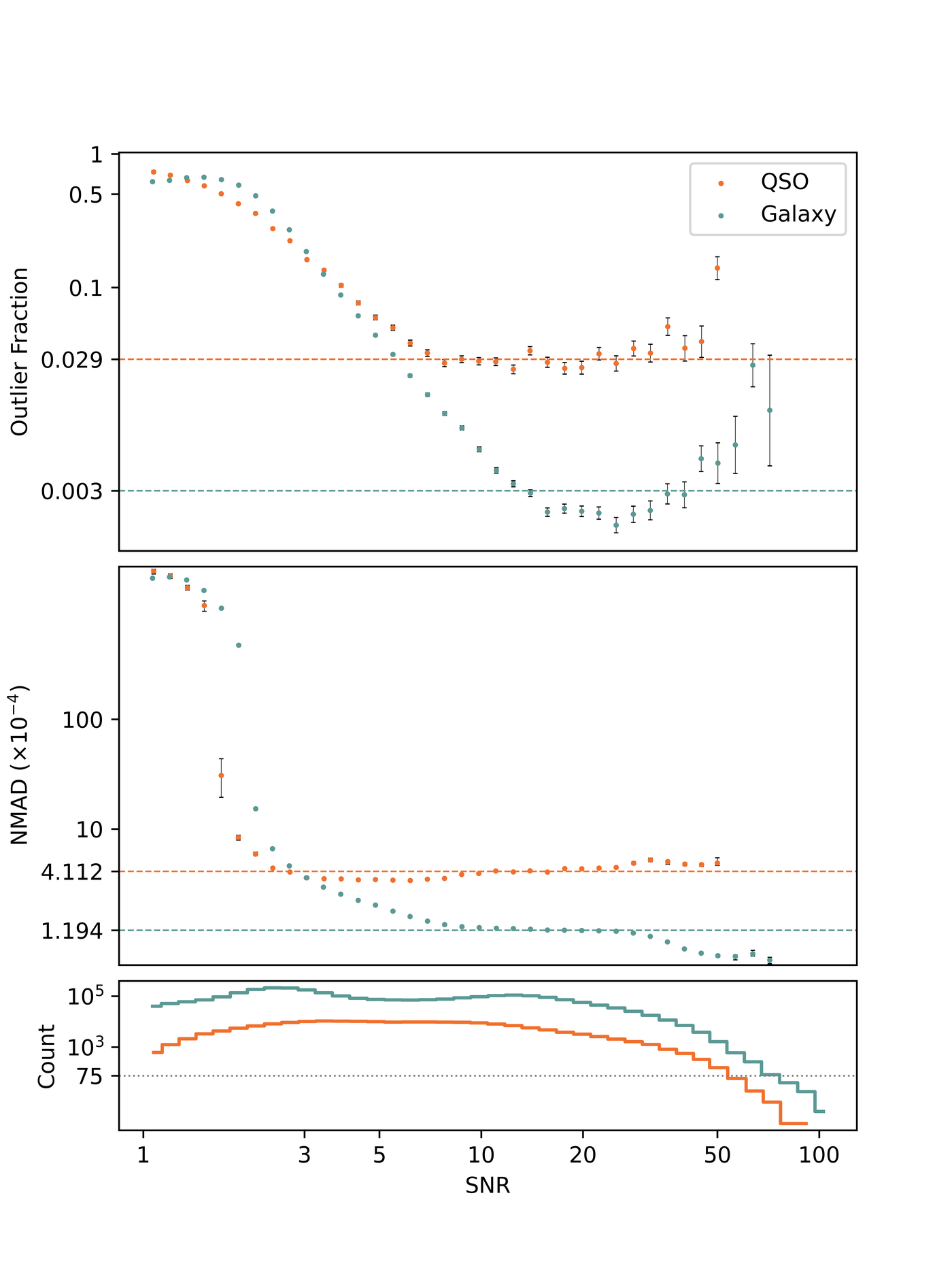}
    \caption{(top) Outlier fraction as a function of signal-to-noise ratio. (middle) NMAD as a function of SNR. The dashed lines in these plots  correspond to NMAD and outlier fraction values shown in the main plot of Figure \ref{fig:z performance}. (bottom) Distribution of SNR across the entire dataset, with the dashed line representing the cutoff for how many points are needed for a meaningful measure of outlier fraction and NMAD and their errors. Performance quickly degrades with lower SNR regardless of quasar (orange) or galaxy (blue) classification.}
    \label{fig:snr vs nmad and outlier fraction}
\end{figure}

UnEZ achieves performance comparable to state-of-the-art methods across a wide range of signal-to-noise ratios. Figure \ref{fig:z performance} shows an outlier fraction of 0.459\% at SNR > 10, and Figure \ref{fig:z performance low snr} shows an outlier fraction of 3.888\% at SNR > 3 when comparing to the SDSS redshifts. Figure \ref{fig:snr vs nmad and outlier fraction} further highlights UnEZ's abilities as the model maintains stable and predictable performance metrics across a wide range of noise levels. Despite the lack of templates or redshift labels UnEZ demonstrates the high accuracy and precision seen in traditional redshift estimation methods.  

An outlier cutoff of $\Delta\zeta > 0.0015$ (approximately 450 $\text{km}\cdot \text{s}^{-1}$) is chosen in accordance with traditional automatic redshift software such as MARZ \citep{Hinton_2016} and AUTOZ \citep{Baldry_2014}. Several valuable metrics, including $\Delta\zeta$, are used to evaluate UnEZ's performance to varying degrees.

\begin{figure}
	\includegraphics[width=\columnwidth]{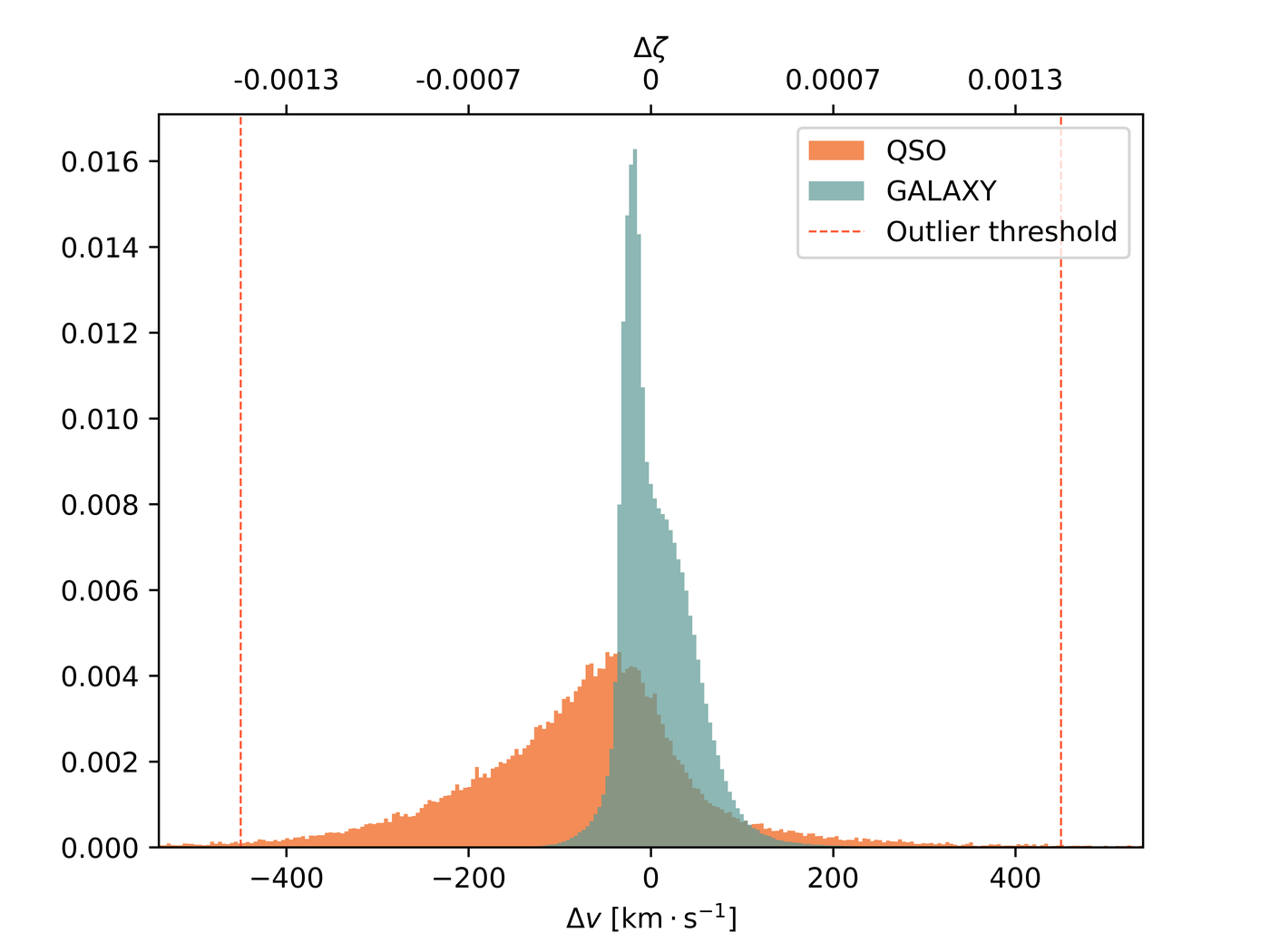}
    \caption{Distribution of $\Delta\zeta$ for galaxies (blue) and quasars (orange).  The outlier threshold is highlighted by the red dashed line at $\Delta\zeta = 0.0015$ or $\Delta$v = 450 $\text{km} \cdot \text{s}^{-1}$.}
    \label{fig:redshift scatter}
\end{figure}

$\Delta\zeta$, defined here in equation \ref{eq: delta zeta},
\begin{equation}
\label{eq: delta zeta}
    \Delta\zeta=\frac{\Delta z}{1+z} = \frac{\Delta v}{c}
\end{equation} approximates a difference in redshifts as a velocity \citep{baldry2018reinventingslideruleredshifts} making it widely used as a measure of precision in automatic redshift algorithms. The bottom panel of Figures \ref{fig:z performance} and \ref{fig:z performance low snr} show how $\Delta\zeta$ varies for inliers across various redshifts. Figure \ref{fig:redshift scatter} is a similar plot showing $\Delta$v in units of $\text{km}\cdot \text{s}^{-1}$. A clear trend between these plots is the wider scatter quasars exhibit when compared to galaxies. This is expected as quasars often have very broad lines creating higher statistical uncertainties in redshift estimation.

Another commonly used measure to gauge the precision of such a model is the normalized median absolute deviation, or NMAD shown in Equation \ref{eq:nmad} \citep{Pattnaik_2025}. 
\begin{equation}
\label{eq:nmad}
    \text{NMAD} = 1.48 \times \text{median}(\left|\frac{z_{\text{SDSS}} - z_{\text{pred}}}{1+z_{\text{SDSS}}}\right|)
\end{equation}
The NMAD is also a robust performance metric due to its lack of sensitivity to catastrophic redshift prediction failures \citep{Molino_2014}. The NMAD for SNR > 10 is 1.23 $\times 10^{-4}$ (37 $\text{km}\cdot \text{s}^{-1}$) while for SNR > 3 is 1.47 $\times 10^{-4}$ (44 $\text{km}\cdot \text{s}^{-1}$), exhibiting a high level of precision which is comparable to or better than existing automatic redshift algorithms such as MARZ \citep{Hinton_2016}. This demonstrates UnEZ's ability to predict redshifts with high precision and accuracy. It is important to note that the NMAD is heavily dependent on the wavelength bin width chosen in the model's physics priors. Decreasing the wavelength bin width leads to a lower NMAD, but only to an extent. As the bin width approaches the resolution of the telescope the NMAD no longer decreases.  Additionally, the computational resources required to train such a large model makes decreasing the bin width redundant. All of the mentioned performance metrics are highlighted in Table \ref{tab:performance} for SNR > 10 and Table \ref{tab:performance low snr} for SNR > 3.

Further analysis of the catastrophic redshift estimation failures revealed 3 systematic outlier classifications. At SNR > 10, stellar contamination from nearby stars make up approximately 500 of the outliers (0.06\% of the entire dataset), skyline contamination make up approximately 600 of the outliers (0.08\% of the dataset), and line confusion degeneracies are responsible for another 31 outliers (0.004\% of the dataset). These outliers are further discussed in Section \ref{subsec: output quality}.

\begin{table}
	\centering
	\caption{Sample size, inlier fraction, and NMAD of each spectral class at SNR > 10.}
	\label{tab:performance}
	\begin{tabular}{lccr} 
		\hline
		Class & Sample Size & Inlier \% & NMAD\\
		\hline
		QSO & 45,516 & 97.1\% & $4.112 \times 10^{-4}\ (123\ \text{km}\cdot \text{s}^{-1})$\\
		Galaxy & 725,391 & 99.7\% & $1.194 \times 10^{-4}\ (36\ \text{km}\cdot \text{s}^{-1})$\\
		Total & 771,086 & 99.5\% & $1.237 \times 10^{-4}\ (37\ \text{km}\cdot \text{s}^{-1})$\\
		\hline
	\end{tabular}
\end{table}

\begin{table}
	\centering
	\caption{As Table \ref{tab:performance}, but for the SNR > 3 sample.}
	\label{tab:performance low snr}
	\begin{tabular}{lccr} 
		\hline
		Class & Sample Size & Inlier \% & NMAD\\
		\hline
		QSO & 147,326 & 94.5\% & $3.471 \times 10^{-4}\ (104\ \text{km}\cdot \text{s}^{-1})$\\
		Galaxy & 1,670,069 & 96.3\% & $1.407 \times 10^{-4}\ (42\ \text{km}\cdot \text{s}^{-1}) $\\
		Total & 1,817,847 & 96.1\% & $1.471 \times 10^{-4}\ (44\ \text{km}\cdot \text{s}^{-1})$\\
		\hline
	\end{tabular}
\end{table}

\subsubsection{Outliers and Inliers}

\label{sec:outliers and inliers} 

\begin{figure*}
    \centering
    \includegraphics[width=\textwidth]{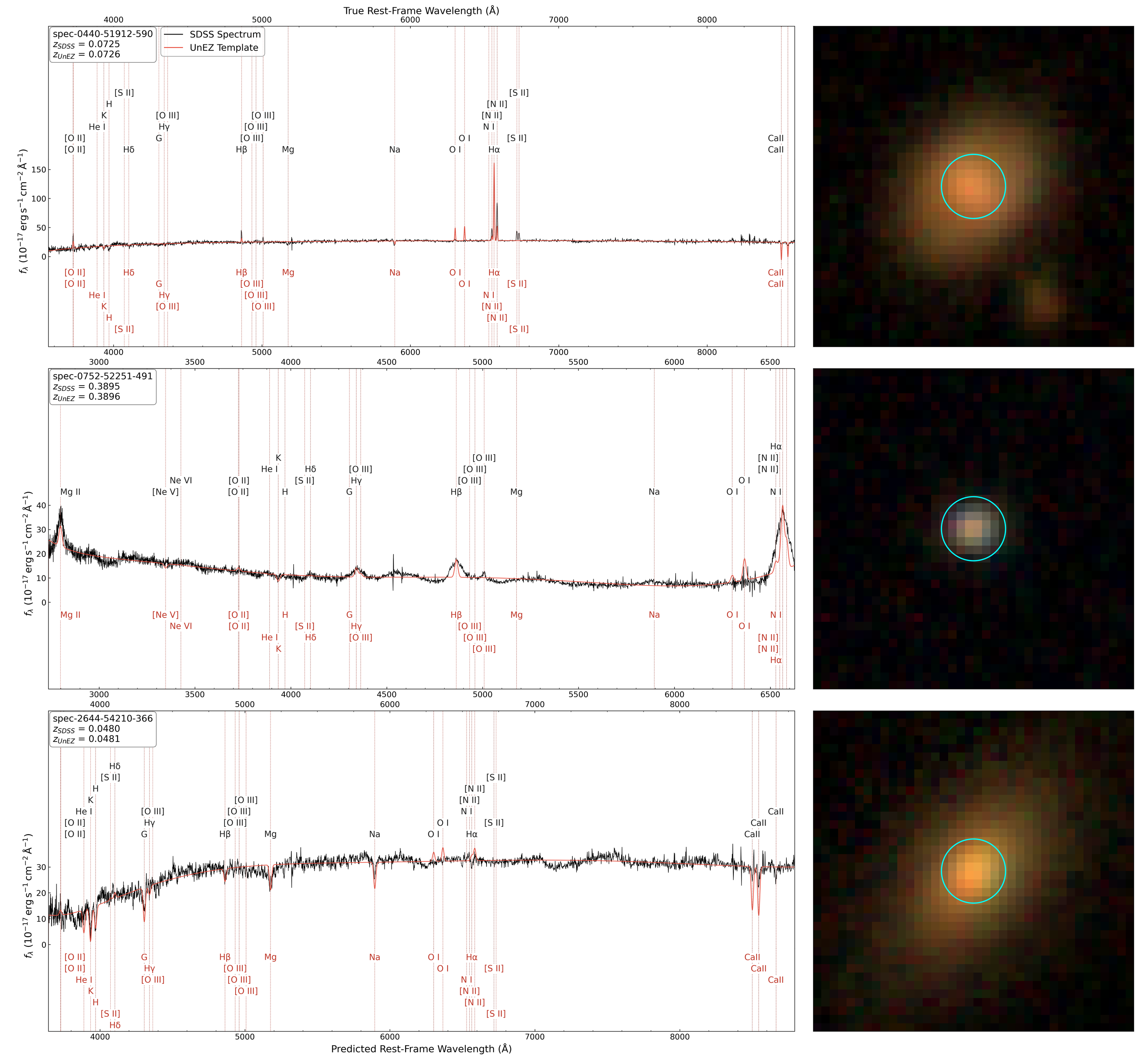}
    \caption{Inlier examples. (top) Starburst Galaxy. (middle) Quasar. (bottom) Quiescent Galaxy. SDSS spectra are shown in black while UnEZ's spliced template is shown in red. The right panels show the SDSS photometry for each object with the fiber placement overlaid in cyan.}
    \label{fig:inlier examples}
\end{figure*}

Both inliers and outliers in UnEZ's predicted redshift catalog reveal the strengths and shortcomings of the model. Figure \ref{fig:inlier examples} showcases representative examples of spectra where UnEZ predicts the redshift accurately. A starburst galaxy, quasar, and quiescent galaxy are all shown with their UnEZ models and rest frame lines. Here, a weakness of UnEZ is highlighted. Although the model is very robust when tasked with predicting the redshift, its line strength predictions are not always accurate. This is shown for example in the quiescent galaxy spectrum's Mg b 5176$\text{\AA}$ absorption where UnEZ overestimates the absorption line.

\begin{figure*}
    \centering
    \includegraphics[width=\textwidth]{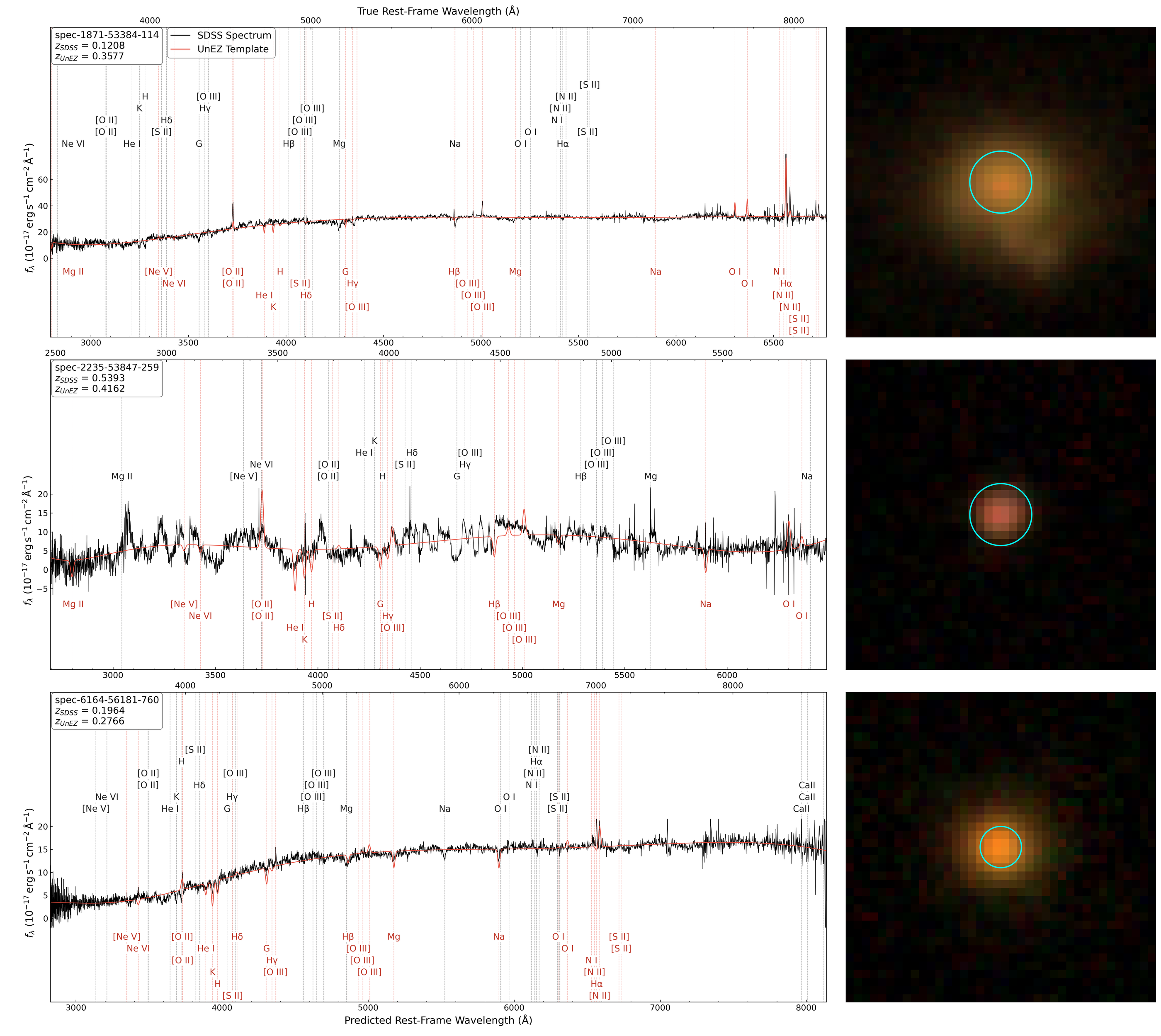}
    \caption{As Figure \ref{fig:inlier examples}, but for outliers. (top) Superimposed objects at different redshifts with multiple distinct sources in the photometry. (middle) Iron low-ionization broad absorption line (FeLoBAL) quasar spectrum. (bottom) Superimposed objects at different redshifts with the second source not visible in the SDSS photometry.}
    \label{fig:outlier examples}
\end{figure*}

\begin{figure*}
    \centering
    \includegraphics[width=\textwidth]{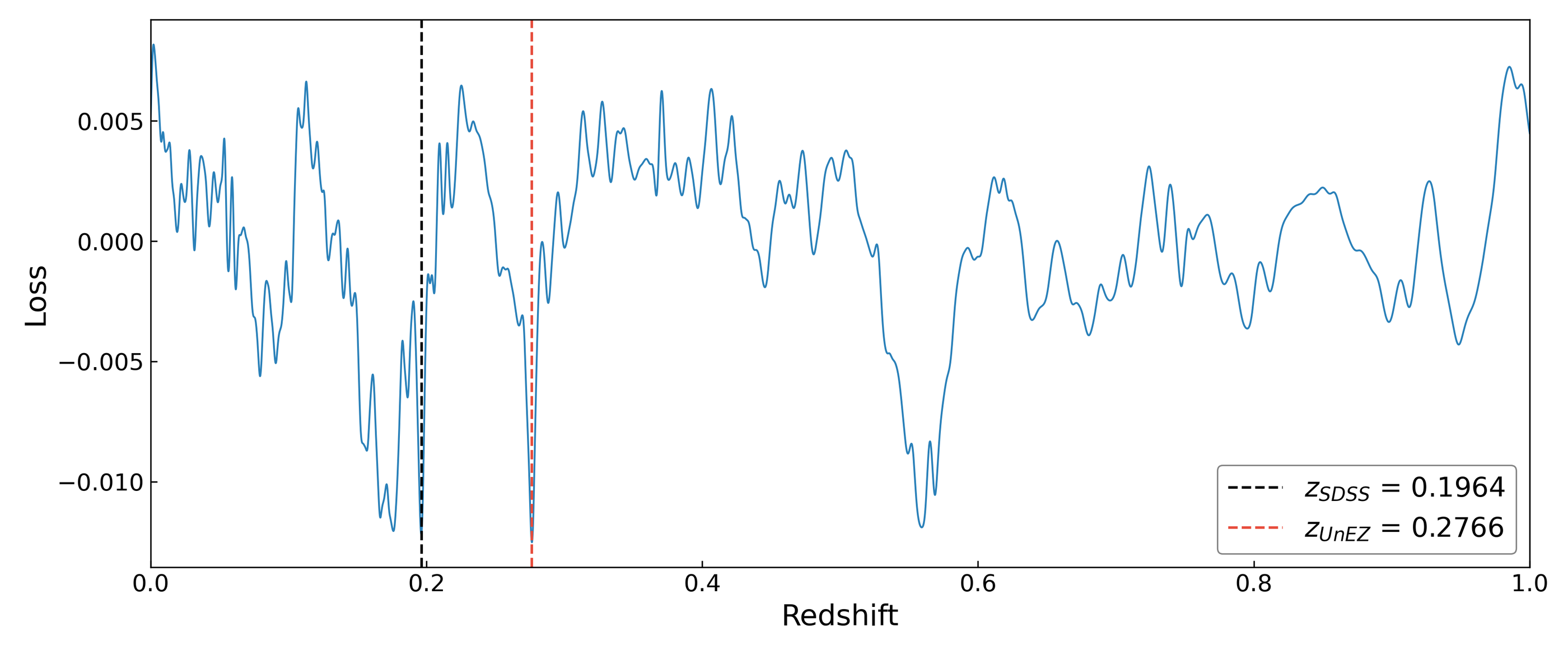}
    \caption{UnEZ's predicted loss landscape for superimposed galaxies shown in the bottom panel of Figure \ref{fig:outlier examples}. }
    \label{fig:superimposed confidence}
\end{figure*}

A majority of UnEZ's catastrophic failures in redshift prediction are caused by either sky line contamination or by line degeneracies. However, hidden among these outliers are some uniquely interesting spectra. Three examples are shown in Figure \ref{fig:outlier examples}. The middle panel displays an iron low-ionization broad absorption line (FeLoBAL) quasar candidate, while the top and bottom panels show spectra of superimposed galaxies. FeLoBAL quasars exhibit extreme properties that are not yet fully understood. This particular spectrum is a FeLoBAL candidate at a redshift of 2.58 \citep{leighly2025balmerabsorptionironlowionization} with thousands of iron absorption lines superimposed and velocity dispersions exceeding 3000 $\text{km} \cdot \text{s}^{-1}$. Models such as SimBAL have been utilized to model properties such as the outflow ionization factor, however the mechanism driving such extreme objects has not yet been confirmed \citep{Choi__2022}. The top and bottom panel showing superimposed galaxy spectra are also interesting because the SDSS automatic redshift prediction algorithm found one redshift, while UnEZ found the other. The photometry in the top right panel of Figure \ref{fig:outlier examples} reveals multiple sources overlapping. The spectrum shows a clear H and K absorption at the SDSS redshift ($z_{SDSS} = 0.1208$), with H$\alpha$ and several oxygen emission lines at the UnEZ redshift ($z_{UnEZ} = 0.3577$). The bottom panel's photometry does not show multiple sources in the photometry, but two clear sets of H and K lines are visible at both the SDSS and UnEZ redshifts. The superimposed nature of the spectrum is further confirmed by the two minima  highlighted on the inverse confidence plot in Figure \ref{fig:superimposed confidence}.

To quantify how often redshift outliers are real UnEZ failures, 980 randomly selected outlier spectra with SNR > 10 were visually inspected to further verify the redshift. These objects are categorized into the following six groups:

\begin{enumerate}
\item SDSS Correct
\item UnEZ Correct
\item Both Correct: Superimposed spectra at different redshifts
\item Close: Spectra were too noisy to distinguish between the two candidate redshifts
\item Neither: Both the UnEZ and SDSS redshift were incorrect
\item Unknown: Could not visually confirm a redshift
\end{enumerate}

The results in Table \ref{tab:unez truth outliers} show that approximately 63.8\% of the time SDSS recovers the correct redshift and UnEZ does not. On the other hand, in 12.8\% of cases UnEZ recovers the correct redshift where SDSS does not. 
\begin{table}
\centering
\caption{Visual classification of 980 inspected UnEZ redshift outliers.}
\label{tab:unez truth outliers}
\begin{tabular}{lrr}
\hline
Category & $N$ & Percentage \\
\hline
SDSS correct        & 625 & 63.8\% \\
UnEZ correct        & 125 & 12.8\% \\
Both (superimposed) & 88  & 9.0\%  \\
Close               & 58  & 5.9\%  \\
Neither             & 51  & 5.2\%  \\
Unknown             & 33  & 3.4\%  \\
\hline
\end{tabular}
\end{table}
A further 9\% are superimposed spectra with different redshifts where UnEZ recovers one redshift and SDSS the other. Another 5.9\% are too noisy to confirm one redshift or the other. Together, the UnEZ-correct and superimposed cases show that at least 21.8\% of redshift outliers are not failures of UnEZ, rising to 27.7\% if the ambiguous close cases are included. This further supports the interpretation of the quoted outlier fraction as a lower limit on the model’s performance in Section \ref{subsec: output quality}. 

\subsection{Line Strength Predictions}
\label{sec:line strength predictions}

\begin{figure*}
    \centering
    \includegraphics[width=\textwidth]{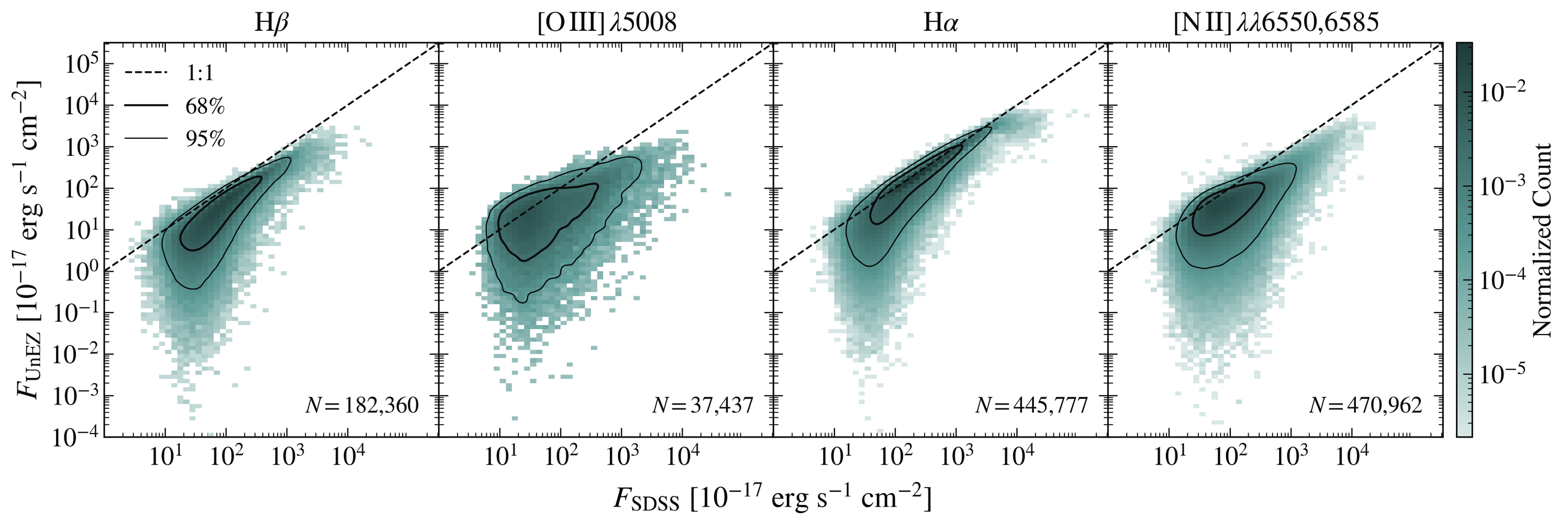}
    \caption{Comparison of emission line fluxes predicted by UnEZ ($F_{\rm UnEZ}$) with those from the SDSS catalog ($F_{\rm SDSS}$) for H$\beta$, [O III] $\lambda$5008, H$\alpha$, and the [N II] $\lambda\lambda$6550,6585 doublet. Only redshift inliers with SDSS line SNR > 5 and a non-zero UnEZ amplitude are shown. $N$ gives the number of objects in each panel. UnEZ fluxes are computed using Equation \ref{eq:line flux}. The [N II] doublet is the sum of both line fluxes in each catalog. The color bar shows the normalized count per bin. The thick and thin contours enclose 68\% and 95\% of the objects, while the dashed line marks the 1:1 relation.}
    \label{fig:flux comparison}
\end{figure*}

UnEZ has shown promising results with respect to redshift prediction, however Figure \ref{fig:architecture} highlights a potential weak point for the model. Line strengths are predicted first, then redshift is estimated as a proxy. Therefore, having inaccurate line strength predictions will lead to many catastrophic redshift estimation failures.

We compare UnEZ line fluxes with those in the SDSS catalog in Figure \ref{fig:flux comparison}. UnEZ fluxes are computed from the predicted Gaussian amplitude $S$ and shared line width $\sigma$ as shown in Equation \ref{eq:line flux} after rescaling $S$ by the spectrum’s normalization factor.
\begin{equation}
    \label{eq:line flux}
    F_{\text{UnEZ}} = S\sigma\sqrt{2\pi}
\end{equation}
The comparison is restricted to redshift inliers ($\Delta\zeta < 0.0015$), lines with SDSS SNR > 5, and lines with non-zero UnEZ amplitude, causing the sample size differences between lines.

UnEZ systematically under-predicts line fluxes, most strongly for [O III] $\lambda$5008, and the discrepancy grows as the line fluxes weaken. On the other hand, H$\alpha$ line fluxes are closest in agreement. Individual UnEZ fluxes are therefore not yet reliable for scientific use. Because the under-prediction is consistent in sign across lines, it may partly cancel in line ratios.

\begin{figure}
	\includegraphics[width=\columnwidth]{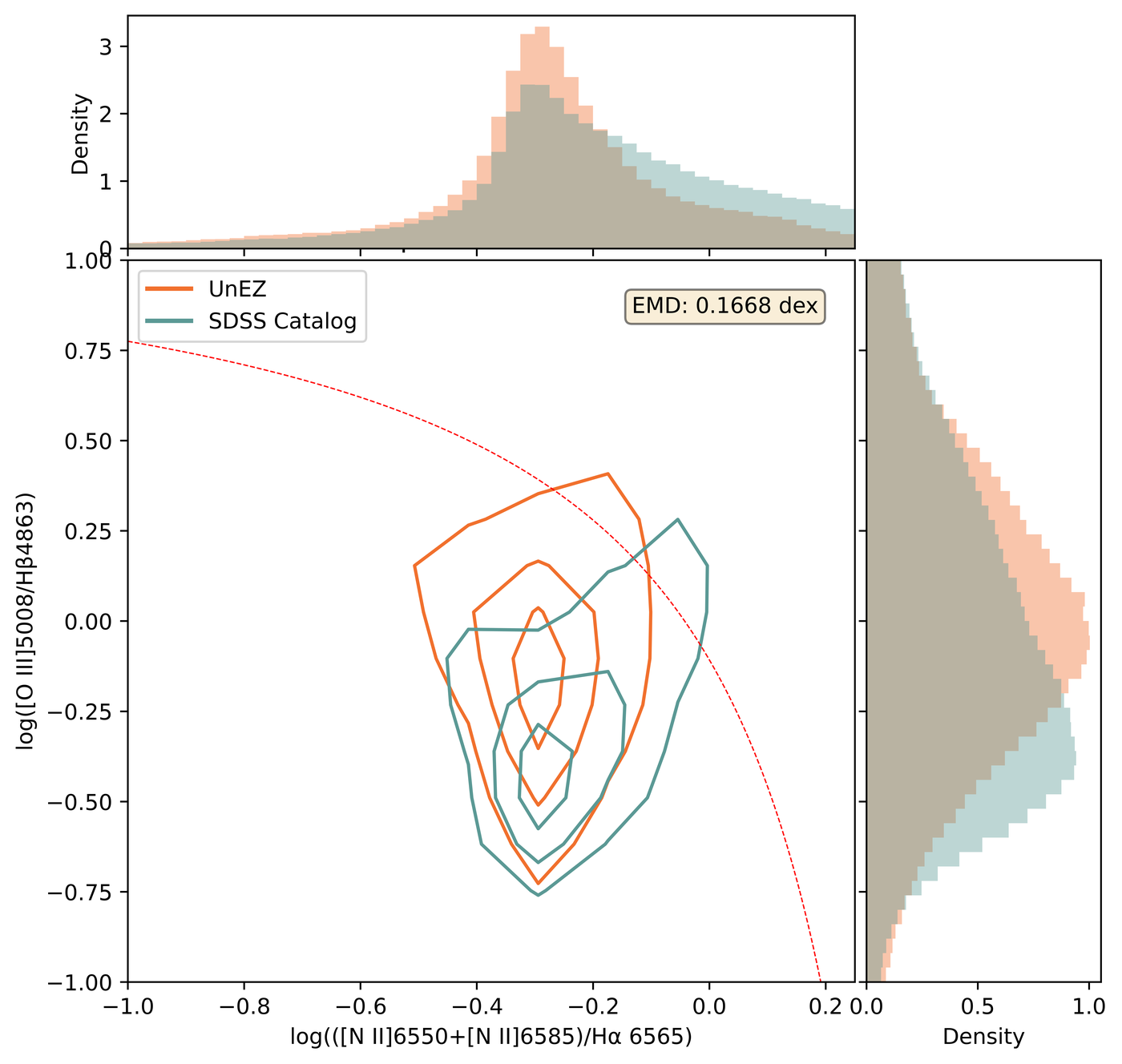}
    \caption{BPT Diagram composed from UnEZ's predicted line strength ratios with SDSS catalog line strength ratios overlaid. The red dashed line is an AGN diagnostic line separating AGN from starburst galaxies \citep{2001ApJ...556..121K}.} 
    \label{fig:bpt diagram}
\end{figure}

One method of comparing line ratio predictions is through a Baldwin, Phillips, and Terlevich (BPT) diagram \citep{1981PASP...93....5B}. BPT diagrams utilize ratios of various emission line strengths to classify spectra, making them an ideal candidate for understanding how UnEZ's systematic under-prediction of line fluxes translates to line ratios. The BPT diagram in Figure \ref{fig:bpt diagram} presents one such set of line ratios commonly used to distinguish AGN from star-forming galaxies \citep{10.1111/j.1365-2966.2006.10859.x}. 

The Earth Mover's Distance (EMD) is a measure of how much work is required to convert one distribution into another \citep{EMD_paper}. The EMD, unlike median offset, is sensitive to differences in shape and spread as well as location. It is also defined for non-overlapping distributions and is expressed in physical units (dex), unlike the KL divergence. Comparing UnEZ and SDSS catalog line strength ratios in Figure \ref{fig:bpt diagram} shows an EMD of approximately 0.1668 dex. The EMD is in units of the x and y axes of Figure \ref{fig:bpt diagram}. It can therefore be converted from log-space to linear-space to find UnEZ's line ratio predictions are, on average, off by a factor of 1.468. The SDSS pipeline fits template continua for continuum subtraction during its line fitting process \citep{Bolton_2012} as opposed to UnEZ's sixth-order polynomial fit. Choice of continuum removal has been shown to have an effect on spectral modeling \citep{2011ApJS..193...29A}, which can lead to the discrepancy in line-strength fits. Despite individual fluxes being systematically under-predicted, the line ratios differ by an EMD of 0.1668 dex, showing that the under-prediction largely cancels.

\subsubsection{Self Organizing Map Clustering}
\label{sec:self organizing map}
Another method to test the effectiveness of UnEZ's line strength predictions is to cluster similar spectra together by their predicted line strengths and see if any expected patterns are apparent. One expected pattern in clustering line strengths is a separation between galaxies and quasars. Galaxies are much dimmer, have much narrower lines, can contain predominantly or exclusively absorption lines, and are at generally lower redshifts. On the other hand, quasars have very broad lines, are extremely bright, have mostly emission lines, and are at generally higher redshifts. Considering UnEZ's predictions include line widths and strengths, quasars and galaxies should have substantially different predicted line strengths from UnEZ's pipeline.

\begin{figure}
	\includegraphics[width=\columnwidth]{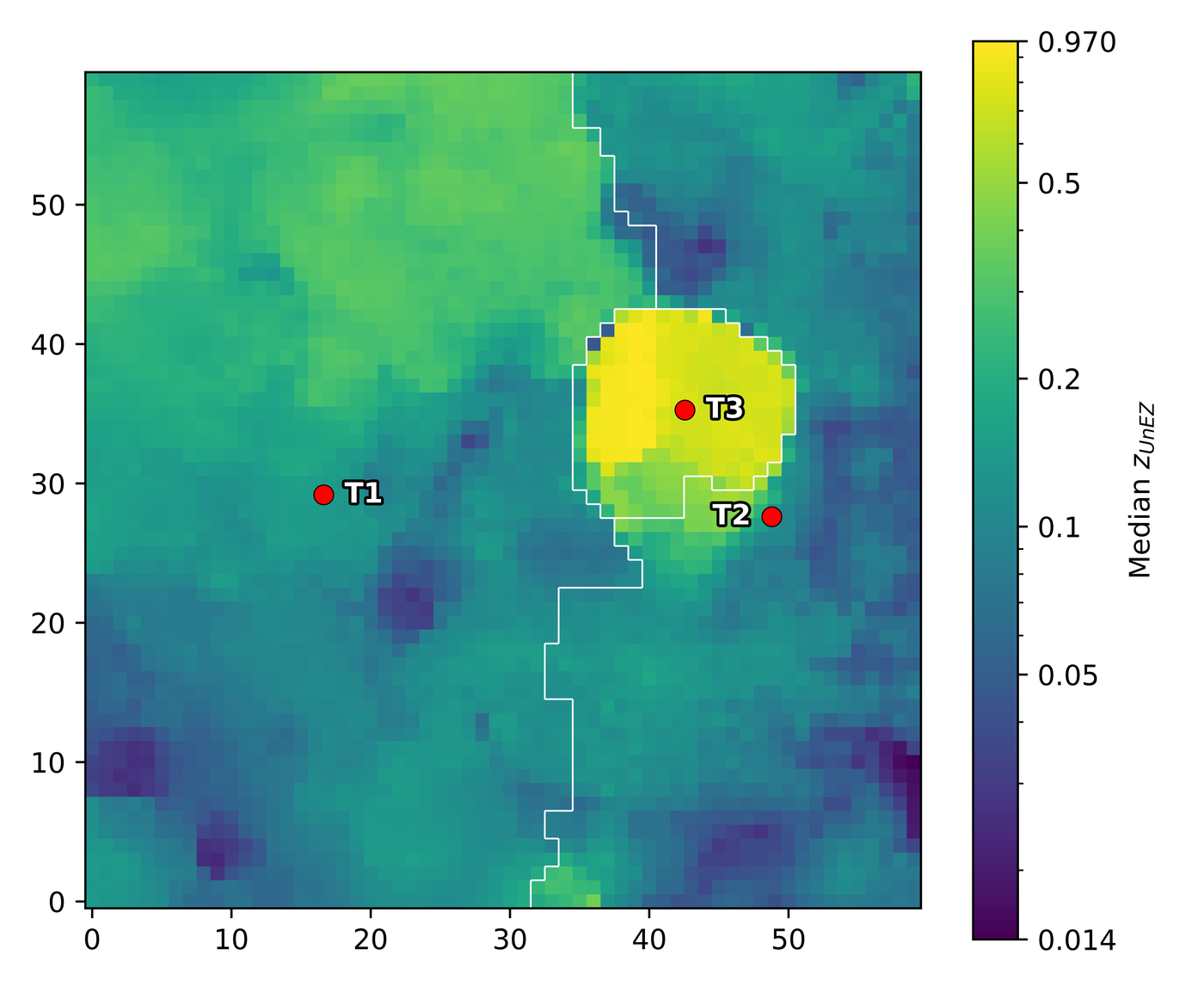}
    \caption{SOM map showing redshift topology across the map indices. The color represents the median $z_{\text{UnEZ}}$ of the spectra belonging to that node. The red scatter points, labeled by their corresponding template spectrum, are the cluster centroids of the map's node weights, while the white outlines show the cluster boundaries.} 
    \label{fig:z map som}
\end{figure}

\begin{figure}
	\includegraphics[width=\columnwidth]{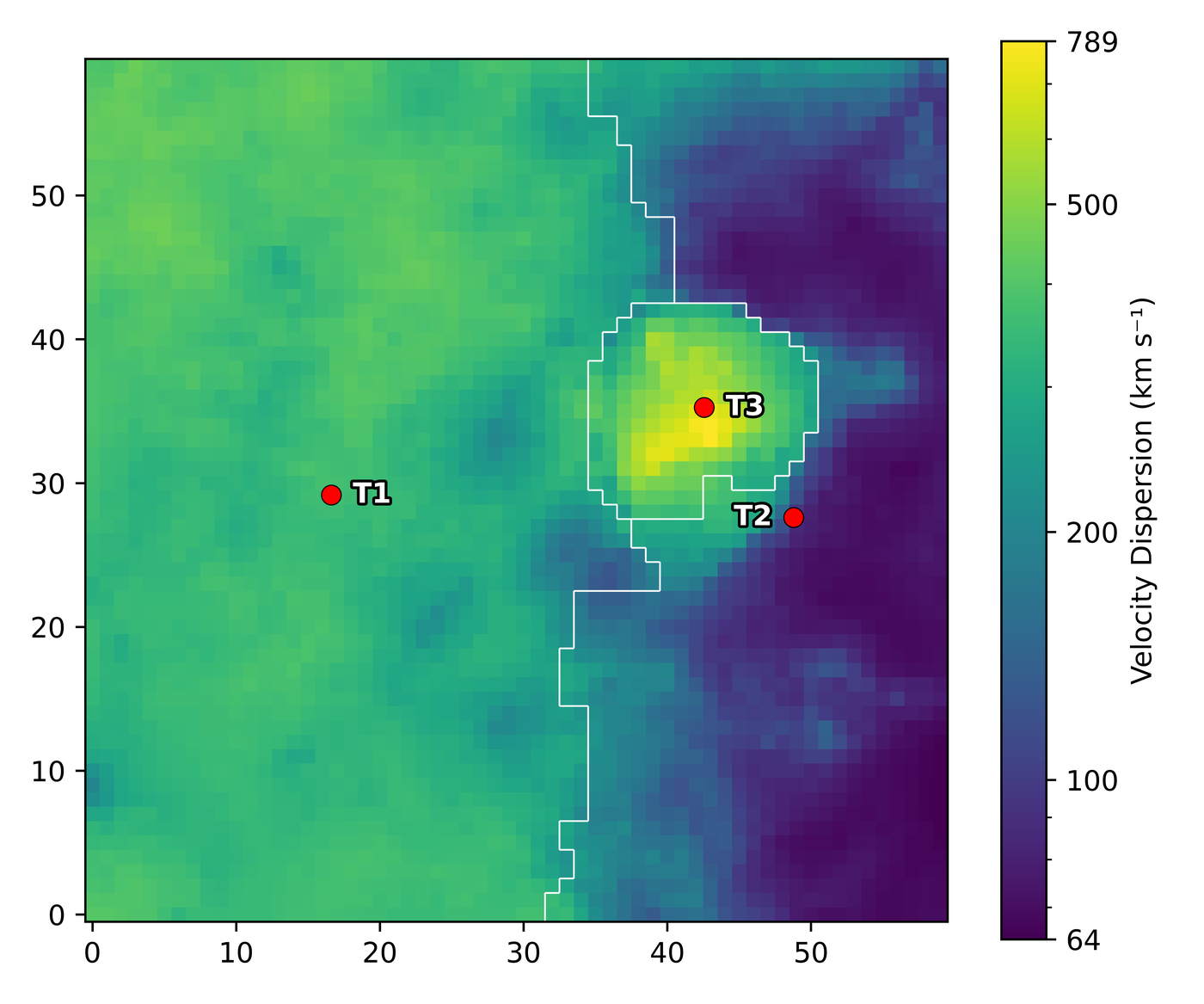}
    \caption{SOM map showing velocity dispersion topology across the map indices. The color represents the median velocity dispersion, predicted by UnEZ, of the spectra belonging to that node. The red scatter points, labeled by their corresponding template spectrum, are the cluster centroids, while the white outlines show the cluster boundaries.} 
    \label{fig:veldisp map som}
\end{figure}

A $60 \times 60$ SOM was trained for 200,000 training epochs on 491,323 UnEZ-predicted line strengths and widths to produce the clustering maps shown in Figure \ref{fig:z map som} and Figure \ref{fig:veldisp map som}. A clear cluster emerges in the center of the SOM that has a peculiarly high redshift and velocity dispersion when compared to the rest of the map. This cluster likely contains a majority of the quasars in the dataset and very few galaxies. 

To confirm this, a clustering algorithm was deployed on the SOM's weights. A connectivity-constrained agglomerative clustering algorithm \citep{MURTAGH1995399} was used to find 3 clusters denoted by the red points on Figure \ref{fig:z map som} and Figure \ref{fig:veldisp map som}. This clustering algorithm ensures that clusters do not overlap or cross gaps without connectivity. K-means clustering, for example, creates clusters that might be split on opposite sides of the map. The SOM clusters similar objects close together, therefore spatially disconnected clusters would group objects from different regions of the map, introducing severe cross-class contamination. One of these cluster centers falls onto the suspected quasar cluster in the center of the map. The median redshift and velocity dispersion of each cluster is then computed. By using a velocity dispersion cut of $\sigma > 700$ $\text{km}\cdot \text{s}^{-1}$, the suspected quasar cluster is separated and classified by UnEZ as quasars. Any cluster falling below the $\sigma$ cut is classified as a galaxy.

Table \ref{tab:som results} shows the results of the SOM clustering and classification technique. Of the 476,616 spectra classified as galaxies by SDSS, 1083 were classified as quasars by UnEZ. Of the 14,707 spectra classified as quasars by SDSS, 929 were classified as galaxies by UnEZ.

\begin{figure*}
    \centering
    \includegraphics[width=\textwidth]{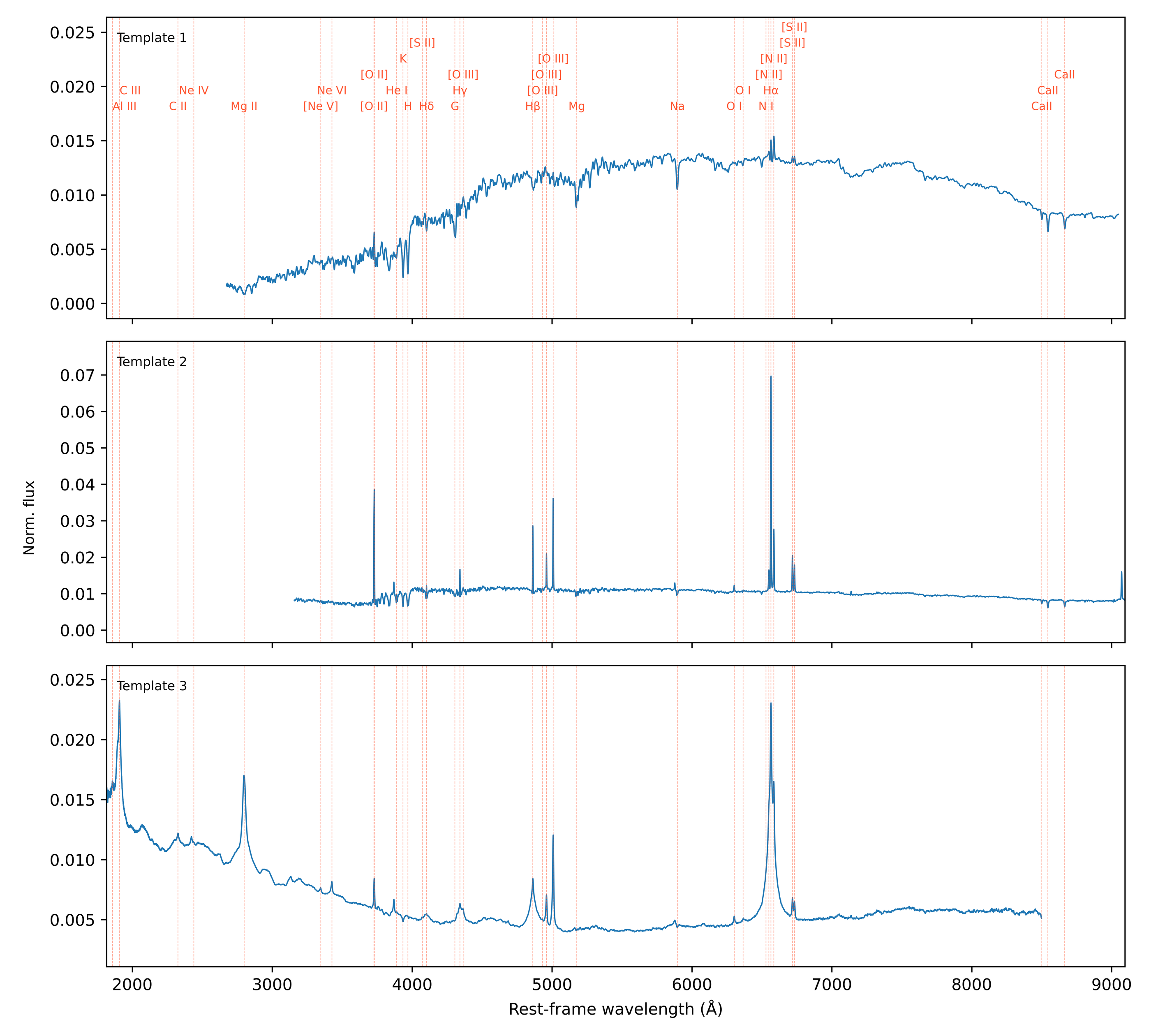}
    \caption{Three templates constructed from median spectrum of each cluster predicted by UnEZ line strengths.} 
    \label{fig:templates}
\end{figure*}

All of the spectra were then shifted to the rest-frame using UnEZ's predicted redshifts, and the median spectrum of each cluster was taken. Ideally, these median spectra would be the final template, however, the clusters do not perfectly separate quasars and galaxies. Some high redshift quasars can contaminate the galaxy templates leading to an unphysical transition between quasar and galaxy-like features. To mitigate this effect the number of spectra contributing to each pixel's median is counted. Any pixels with less than 5\% of the total count are omitted from the template. This creates different wavelength ranges in some templates shown in Figure \ref{fig:templates}. Template 1 exhibits a strong continuum with absorption lines as expected from high metallicity elliptical galaxies, template 2 shows strong and narrow emission lines from HII regions, and template 3 shows strong and wide emission lines expected from AGN.

\begin{table}
	\centering
	\caption{Table of SDSS classifications compared to UnEZ classifications.}
	\label{tab:som results}
	\begin{tabular}{lccr} 
		\hline
		QSO $\rightarrow$ QSO & Galaxy $\rightarrow$ Galaxy & QSO $\rightarrow$ Galaxy & Galaxy $\rightarrow$ QSO\\
		\hline
		13,778 & 475,533 & 929 & 1083\\
		\hline
	\end{tabular}
\end{table}

As no redshift, velocity dispersion, or classification labels were used during the training of the SOM, this extension of the UnEZ pipeline exhibits the ability to classify spectra and produce template spectra from large datasets without any prior labels.

\section{Discussion}
\label{sec:discussion} 

UnEZ serves as a compelling proof of concept that a neural network can learn the concept of redshift from first principles. An architecture designed to compress spectra into only scientifically significant values makes this possible. In contrast to conventional redshift estimation pipelines, the training of this model requires no redshift or line strength labels as it is primarily driven by its physics-informed constraints. This shows how competitive redshift performance can emerge from the structure of spectral data and minimal prior physics information. UnEZ creates a pathway toward scalable spectral analysis in new redshift regimes and where data labels are scarce.

\subsection{Output Quality}
\label{subsec: output quality}
UnEZ maintains high redshift accuracy even at relatively low signal-to-noise ratios, achieving over 95\% inlier performance down to SNR > 3 and over 99.5\% inlier performance at SNR > 10. It is important to note, the SDSS redshift estimation pipeline is not perfect and its reliability suffers particularly at lower SNR. Approximately 0.2\% of galaxy redshifts and 0.5\% of quasar redshifts have been found to be incorrect through visual inspection \citep{Bolton_2012}. This further highlights that the mentioned performance metrics, such as the outlier fraction, represent the lower limit of UnEZ's overall performance as mentioned in Section \ref{subsec: redshift}. Crucially, UnEZ does not rely on template spectra at any stage. Rather, the only priors required are the physics information that is mostly specific to the instrument being used and the spectral lines expected from different galaxies.

Despite its strength, there are some caveats to UnEZ's performance. UnEZ's performance is heavily linked to the quality of the data. Figure \ref{fig:snr vs nmad and outlier fraction} shows a stark increase in NMAD at SNR < 3, and a gradual increase in outlier fraction at SNR < 10. The increase in NMAD is likely attributed to how well individual lines are being fit, as noisier lines leave more room for error in the line fit. The increase in outlier fraction at lower SNR can be attributed to line misidentification. As the noise across the spectrum decreases, UnEZ is less likely to fit to the noise or the incorrect lines. Counterintuitively, there is also a slight increase in outlier fraction at very high SNR in Figure \ref{fig:snr vs nmad and outlier fraction}, likely due to selection effects or a lack of representation of such data in the training set. There are very few data points available at such high SNR which can prohibit UnEZ's ability to learn from such data properly. The plateau in both NMAD and outlier fraction at SNR > 10 represents UnEZ's best possible performance. Such catastrophic outliers at high SNR can either be attributed to bad data, or systematic shortcomings of UnEZ's ability to identify and fit lines.

Such systematic catastrophic errors in redshift prediction, as shown in Figure \ref{fig:z performance}, can be categorized into three main classes. The first class of catastrophic errors are the sky lines that appear as horizontal bands in Figure \ref{fig:z performance}. Because sky lines are redshift-invariant and occur at fixed observed wavelengths, if UnEZ mistakenly fits an emission line to a sky line feature, it will consistently assign the same incorrect redshift. The model can be modified in the future to predict a rest frame template of skylines in addition to the template of the target. Additionally, a precursor model could be trained to de-noise spectra before training and inference in the future \citep{Scourfield_2023}. 

Another systematic error is caused by a line confusion degeneracy that manifests as faint parallel lines (due to redshifts being plotted as $\text{log}_{10}(1+z)$) to the 1:1 line in Figure \ref{fig:z performance low snr}. Galaxies primarily suffer from a H$\alpha \rightarrow$[O III] 5008 confusion while quasars show primarily Mg II 2799$\rightarrow$C II 2326 and Mg II 2799$\rightarrow$[O III] 5008 line confusions. These degeneracy examples can typically be mitigated by reducing sky line contamination, which in turn allows the model to be more confident in its prediction of which lines are and are not real. Although this degeneracy is not yet corrected, future analysis may reveal a more effective method of addressing the degeneracy problem.

Accurate line strength prediction is important for UnEZ because they act as a prior in the redshift estimation process. Although these predictions are sufficiently accurate to enable precise and reliable redshift estimates, its systematic under-prediction of individual line fluxes in Figure \ref{fig:flux comparison} makes it unreliable for single-line measurements. However, as shown in Figure \ref{fig:bpt diagram}, the SDSS and UnEZ line-ratio distributions differ significantly less than individual line flux measurements. Future improvements to the normalization procedure used throughout the pipeline may help reduce this discrepancy in line flux estimations. UnEZ currently applies an L2 normalization to all spectra before training and inference. Other normalization techniques, such as min-max normalization, may be more suitable for line-strength fitting in the future. It is important to note, however, that UnEZ and SDSS use different continuum fitting methods during the preprocessing stage. This is known to cause discrepancies in line strength measurements \citep{2011ApJS..193...29A} and can be a potential cause of the discrepancy seen between UnEZ and the SDSS catalog.

\subsection{Benefits and Drawbacks}
\label{sec:benefits and drawbacks} 

UnEZ is a highly capable neural network with diverse applications. Although UnEZ's primary goal is to predict redshifts of astronomical spectra, it shows promise of also being used for line strength prediction and outlier detection. The model is successful in reliably and accurately predicting redshifts in a fully unsupervised manner. We re-emphasize that at no point did we include redshifts in the training data. The model learns how to precisely predict redshifts independently by trying to match the provided rest frame lines from the provided line list to features in spectra. 

Although unsupervised machine learning methods have been historically used for clustering, the results are not physically interpretable. Algorithms such as SOMz require further supervised steps to map a physical meaning (e.g. redshift) to an abstract latent space \citep{Carrasco_Kind_2014}. UnEZ, however, produces physically meaningful representations of data that require no supervision. Each latent space node is directly associated with a specific spectral line and wavelength, meaning the latent space can be directly interpreted and used for further scientific analysis. 

The unsupervised nature of the model's training and inference means that UnEZ is effectively learning what redshift is and how to predict it with no reference to any ground truths. This unsupervised theme is what makes UnEZ far less labor intensive to use and less prone to biases introduced by templates. However, like all data-driven models, unsupervised neural networks are not immune to all biases. For example, the choice of lines to fit can influence the model's performance. Without enough lines to fit, the model cannot predict an accurate redshift. Also, unsupervised neural networks are prone to the biases of the training data itself, such as selection effects. Fortunately, large-scale surveys help mitigate this issue by offering diverse and representative samples of astronomical objects for training.

In terms of performance, UnEZ outperforms conventional redshift estimation methods. Its redshift prediction precision and accuracy are slightly better than tools such as MARZ which cited a 3.5\% outlier fraction on high SNR spectra at z < 1. However, the model's lack of reliance on template spectra or redshift labels makes it a more appealing solution to the redshift estimation problem.

Some physics priors described in Section \ref{sec:physics priors} are specific to the spectroscopic instrument. For example, the maximum expected redshift, redshift sampling resolution, and maximum observed wavelength of the spectrograph are all specific to the instrument. The expected line list can also change depending on the redshift range being observed and wavelength range of the instrument. All of these parameters can be tuned in a configuration file, which should make adapting UnEZ to a new instrument straightforward and potentially allow it to be used in more diverse cases than traditional redshift estimation methods.

Like most unsupervised models, UnEZ requires a large amount of data for training. Unsupervised neural networks learn directly from the distribution of the dataset. This means they require large amounts of data to properly probe the parameter-space without overtraining. Luckily, larger surveys such as DESI \citep{levi2019darkenergyspectroscopicinstrument} and SDSS \citep{2023ApJS..267...44A} are perfect candidates for training UnEZ.

\subsection{Future Endeavors}
\label{sec:future endeavors} 

The fully unsupervised aspect of the model gives it the flexibility needed for new and unseen redshift regimes. If a new telescope, such as JWST, allows us to peer into the previously unseen universe, new templates are required for conventional redshift estimation algorithms. However, UnEZ's lack of templates is expected to allow quick and easy deployment on new surveys conducted with instruments such as JWST. This not only saves time but also reduces human-labor and biases inherent to template spectra.

UnEZ demonstrates that templates and vetting spectral fits by hand are not required for effective redshift prediction. There is plenty of room for improvement however. The biggest drawback of UnEZ that can be remedied is the line strength prediction. Currently the model uses L2 normalization throughout its pipeline, but there are other normalization techniques that could potentially be more effective for line strength prediction. The relationship between line strength prediction and redshift estimation accuracy is not yet fully understood. Future work will benefit greatly from understanding how line strength predictions directly impact redshift performance.

Another area for improvement is reducing reliance on manual hyperparameter tuning such as the top-k redshifts used for learning during training as mentioned in Section \ref{subsec:training loss}. This requires more tuning which counteracts the unsupervised theme of the model. Some of the hyperparameters can potentially be set to automatically scale with other parameters such as maximum expected redshift and number of redshift bins in each spectrum. This can allow the model to operate with even less tuning than is currently required. 

Beyond architectural changes, auxiliary pipelines can also be set up to separate interesting objects during inference. This can be done by either clustering the latent space or referencing another redshift catalog for disagreements. Some objects have already been found manually such as the superimposed galaxies and FeLoBAL quasar in Figure \ref{fig:outlier examples}. 

Finding outliers in large datasets utilizing UnEZ's predictions is the ideal next step. UnEZ outputs a wide range of data that should organically reveal interesting outliers in the SDSS dataset. Expanding the search for outliers beyond the current SDSS dataset will verify UnEZ's ability to directly encode physical properties of spectra spanning across different telescopes, datasets, and redshift regimes.

\section*{Acknowledgements}
This work was performed on the OzSTAR national facility at Swinburne University of Technology. The OzSTAR program receives funding in part from the Astronomy National Collaborative Research Infrastructure Strategy (NCRIS) allocation provided by the Australian Government, and from the Victorian Higher Education State Investment Fund (VHESIF) provided by the Victorian Government.

Funding for the Sloan Digital Sky Survey V has been provided by the Alfred P. Sloan Foundation, the Heising-Simons Foundation, the National Science Foundation, and the Participating Institutions. SDSS acknowledges support and resources from the Center for High-Performance Computing at the University of Utah. SDSS telescopes are located at Apache Point Observatory, funded by the Astrophysical Research Consortium and operated by New Mexico State University, and at Las Campanas Observatory, operated by the Carnegie Institution for Science. The SDSS web site is \url{www.sdss.org}.

SDSS is managed by the Astrophysical Research Consortium for the Participating Institutions of the SDSS Collaboration, including the Carnegie Institution for Science, Chilean National Time Allocation Committee (CNTAC) ratified researchers, Caltech, the Gotham Participation Group, Harvard University, Heidelberg University, The Flatiron Institute, The Johns Hopkins University, L'Ecole polytechnique f\'{e}d\'{e}rale de Lausanne (EPFL), Leibniz-Institut f\"{u}r Astrophysik Potsdam (AIP), Max-Planck-Institut f\"{u}r Astronomie (MPIA Heidelberg), Max-Planck-Institut f\"{u}r Extraterrestrische Physik (MPE), Nanjing University, National Astronomical Observatories of China (NAOC), New Mexico State University, The Ohio State University, Pennsylvania State University, Smithsonian Astrophysical Observatory, Space Telescope Science Institute (STScI), the Stellar Astrophysics Participation Group, Universidad Nacional Aut\'{o}noma de M\'{e}xico, University of Arizona, University of Colorado Boulder, University of Illinois at Urbana-Champaign, University of Toronto, University of Utah, University of Virginia, Yale University, and Yunnan University.  

\section*{Data Availability}

This work uses publicly available spectra from the Sloan Digital Sky Survey Data Release 18 \citep{2023ApJS..267...44A}, accessible through the \href{https://www.sdss.org/dr18/}{SDSS Science Archive Server}. The UnEZ predictions and code will be shared on reasonable request.



\bibliographystyle{mnras}
\bibliography{example} 

@ARTICLE{bezanson2016,
       author = {{Bezanson}, Rachel and {Wake}, David A. and {Brammer}, Gabriel B. and {van Dokkum}, Pieter G. and {Franx}, Marijn and {Labb{\'e}}, Ivo and {Leja}, Joel and {Momcheva}, Ivelina G. and {Nelson}, Erica J. and {Quadri}, Ryan F. and {Skelton}, Rosalind E. and {Weiner}, Benjamin J. and {Whitaker}, Katherine E.},
        title = "{Leveraging 3D-HST Grism Redshifts to Quantify Photometric Redshift Performance}",
      journal = {\apj},
         year = 2016,
        month = may,
       volume = {822},
       number = {1},
          eid = {30},
        pages = {30},
          doi = {10.3847/0004-637X/822/1/30},
archivePrefix = {arXiv},
       eprint = {1510.07049},
 primaryClass = {astro-ph.GA},
       adsurl = {https://ui.adsabs.harvard.edu/abs/2016ApJ...822...30B}
}

@misc{vaswani2023attentionneed,
      title={Attention Is All You Need}, 
      author={Ashish Vaswani and Noam Shazeer and Niki Parmar and Jakob Uszkoreit and Llion Jones and Aidan N. Gomez and Lukasz Kaiser and Illia Polosukhin},
      year={2023},
      eprint={1706.03762},
      archivePrefix={arXiv},
      primaryClass={cs.CL},
      url={https://arxiv.org/abs/1706.03762}, 
}

@ARTICLE{kollatschny2006,
       author = {{Kollatschny}, W. and {Zetzl}, M. and {Dietrich}, M.},
        title = "{Spectral line variability amplitudes in active galactic nuclei}",
      journal = {\aap},
         year = 2006,
        month = aug,
       volume = {454},
       number = {2},
        pages = {459-472},
          doi = {10.1051/0004-6361:20054357},
       adsurl = {https://ui.adsabs.harvard.edu/abs/2006A&A...454..459K}
}

@misc{damadi2023backpropagationalgorithmmathstudent,
      title={The Backpropagation algorithm for a math student}, 
      author={Saeed Damadi and Golnaz Moharrer and Mostafa Cham},
      year={2023},
      eprint={2301.09977},
      archivePrefix={arXiv},
      primaryClass={cs.LG},
      url={https://arxiv.org/abs/2301.09977}, 
}

@article{Hinton_2016,
   title={Marz: Manual and automatic redshifting software},
   volume={15},
   ISSN={2213-1337},
   url={http://dx.doi.org/10.1016/j.ascom.2016.03.001},
   DOI={10.1016/j.ascom.2016.03.001},
   journal={Astronomy and Computing},
   publisher={Elsevier BV},
   author={Hinton, S.R. and Davis, Tamara M. and Lidman, C. and Glazebrook, K. and Lewis, G.F.},
   year={2016},
   month=apr, pages={61--71} }

@article{Bolton_2012,
   title={SPECTRAL CLASSIFICATION AND REDSHIFT MEASUREMENT FOR THE SDSS-III BARYON OSCILLATION SPECTROSCOPIC SURVEY},
   volume={144},
   ISSN={1538-3881},
   url={http://dx.doi.org/10.1088/0004-6256/144/5/144},
   DOI={10.1088/0004-6256/144/5/144},
   number={5},
   journal={The Astronomical Journal},
   publisher={American Astronomical Society},
   author={Bolton, Adam S. and Schlegel, David J. and Aubourg, Éric and Bailey, Stephen and Bhardwaj, Vaishali and Brownstein, Joel R. and Burles, Scott and Chen, Yan-Mei and Dawson, Kyle and Eisenstein, Daniel J. and Gunn, James E. and Knapp, G. R. and Loomis, Craig P. and Lupton, Robert H. and Maraston, Claudia and Muna, Demitri and Myers, Adam D. and Olmstead, Matthew D. and Padmanabhan, Nikhil and Pâris, Isabelle and Percival, Will J. and Petitjean, Patrick and Rockosi, Constance M. and Ross, Nicholas P. and Schneider, Donald P. and Shu, Yiping and Strauss, Michael A. and Thomas, Daniel and Tremonti, Christy A. and Wake, David A. and Weaver, Benjamin A. and Wood-Vasey, W. Michael},
   year={2012},
   month=oct, pages={144} }

@article{Collister_2004,
   title={ANNz: Estimating Photometric Redshifts Using Artificial Neural Networks},
   volume={116},
   ISSN={1538-3873},
   url={http://dx.doi.org/10.1086/383254},
   DOI={10.1086/383254},
   number={818},
   journal={Publications of the Astronomical Society of the Pacific},
   publisher={IOP Publishing},
   author={Collister, Adrian A. and Lahav, Ofer},
   year={2004},
   month=apr, pages={345--351} }

@article{Carrasco_Kind_2014,
   title={SOMz: photometric redshift PDFs with self-organizing maps and random atlas},
   volume={438},
   ISSN={0035-8711},
   url={http://dx.doi.org/10.1093/mnras/stt2456},
   DOI={10.1093/mnras/stt2456},
   number={4},
   journal={Monthly Notices of the Royal Astronomical Society},
   publisher={Oxford University Press (OUP)},
   author={Carrasco Kind, Matias and Brunner, Robert J.},
   year={2014},
   month=jan, pages={3409--3421} }

@misc{veilleux2002spectroscopicdiagnosticsagns,
      title={Spectroscopic Diagnostics for AGNs}, 
      author={Sylvain Veilleux},
      year={2002},
      eprint={astro-ph/0201118},
      archivePrefix={arXiv},
      primaryClass={astro-ph},
      url={https://arxiv.org/abs/astro-ph/0201118}, 
}

@article{Hewett_2010,
   title={Improved redshifts for SDSS quasar spectra: Improved SDSS redshifts},
   ISSN={1365-2966},
   url={http://dx.doi.org/10.1111/j.1365-2966.2010.16648.x},
   DOI={10.1111/j.1365-2966.2010.16648.x},
   journal={Monthly Notices of the Royal Astronomical Society},
   publisher={Oxford University Press (OUP)},
   author={Hewett, Paul C. and Wild, Vivienne},
   year={2010},
   volume={405},
   month=jul, pages={2302--2316} }

@article{Pattnaik_2025,
   title={SpecPT (Spectroscopy Pre-trained Transformer) Model for Extragalactic Spectroscopy. I. Architecture and Automated Redshift Measurement},
   volume={988},
   ISSN={1538-4357},
   url={http://dx.doi.org/10.3847/1538-4357/ade053},
   DOI={10.3847/1538-4357/ade053},
   number={1},
   journal={The Astrophysical Journal},
   publisher={American Astronomical Society},
   author={Pattnaik, Rohan and Kartaltepe, Jeyhan S. and Binu, Clive},
   year={2025},
   month=jul, pages={139} }

@article{Molino_2014,
   title={The ALHAMBRA Survey: Bayesian photometric redshifts with 23 bands for 3 deg2},
   volume={441},
   ISSN={0035-8711},
   url={http://dx.doi.org/10.1093/mnras/stu387},
   DOI={10.1093/mnras/stu387},
   number={4},
   journal={Monthly Notices of the Royal Astronomical Society},
   publisher={Oxford University Press (OUP)},
   author={Molino, A. and Benítez, N. and Moles, M. and Fernández-Soto, A. and Cristóbal-Hornillos, D. and Ascaso, B. and Jiménez-Teja, Y. and Schoenell, W. and Arnalte-Mur, P. and Pović, M. and Coe, D. and López-Sanjuan, C. and Díaz-García, L. A. and Varela, J. and Stefanon, M. and Cenarro, J. and Matute, I. and Masegosa, J. and Márquez, I. and Perea, J. and Del Olmo, A. and Husillos, C. and Alfaro, E. and Aparicio-Villegas, T. and Cerviño, M. and Huertas-Company, M. and Aguerri, J. A. L. and Broadhurst, T. and Cabrera-Caño, J. and Cepa, J. and González, R. M. and Infante, L. and Martínez, V. J. and Prada, F. and Quintana, J. M.},
   year={2014},
   month=may, pages={2891--2922} }

@article{Wu_2022,
   title={A Catalog of Quasar Properties from Sloan Digital Sky Survey Data Release 16},
   volume={263},
   ISSN={1538-4365},
   url={http://dx.doi.org/10.3847/1538-4365/ac9ead},
   DOI={10.3847/1538-4365/ac9ead},
   number={2},
   journal={The Astrophysical Journal Supplement Series},
   publisher={American Astronomical Society},
   author={Wu, Qiaoya and Shen, Yue},
   year={2022},
   month=dec, pages={42} }

@ARTICLE{2023ApJS..267...44A,
       author = {{Almeida}, Andr{\'e}s and {Anderson}, Scott F. and {Argudo-Fern{\'a}ndez}, Maria and {Badenes}, Carles and {Barger}, Kat and {Barrera-Ballesteros}, Jorge K. and {Bender}, Chad F. and {Benitez}, Erika and {Besser}, Felipe and {Bird}, Jonathan C. and {Bizyaev}, Dmitry and {Blanton}, Michael R. and {Bochanski}, John and {Bovy}, Jo and {Brandt}, William Nielsen and {Brownstein}, Joel R. and {Buchner}, Johannes and {Bulbul}, Esra and {Burchett}, Joseph N. and {Cano D{\'\i}az}, Mariana and {Carlberg}, Joleen K. and {Casey}, Andrew R. and {Chandra}, Vedant and {Cherinka}, Brian and {Chiappini}, Cristina and {Coker}, Abigail A. and {Comparat}, Johan and {Conroy}, Charlie and {Contardo}, Gabriella and {Cortes}, Arlin and {Covey}, Kevin and {Crane}, Jeffrey D. and {Cunha}, Katia and {Dabbieri}, Collin and {Davidson}, James W. and {Davis}, Megan C. and {de Andrade Queiroz}, Anna Barbara and {De Lee}, Nathan and {M{\'e}ndez Delgado}, Jos{\'e} Eduardo and {Demasi}, Sebastian and {Di Mille}, Francesco and {Donor}, John and {Dow}, Peter and {Dwelly}, Tom and {Eracleous}, Mike and {Eriksen}, Jamey and {Fan}, Xiaohui and {Farr}, Emily and {Frederick}, Sara and {Fries}, Logan and {Frinchaboy}, Peter and {G{\"a}nsicke}, Boris T. and {Ge}, Junqiang and {Gonz{\'a}lez {\'A}vila}, Consuelo and {Grabowski}, Katie and {Grier}, Catherine and {Guiglion}, Guillaume and {Gupta}, Pramod and {Hall}, Patrick and {Hawkins}, Keith and {Hayes}, Christian R. and {Hermes}, J.~J. and {Hern{\'a}ndez-Garc{\'\i}a}, Lorena and {Hogg}, David W. and {Holtzman}, Jon A. and {Ibarra-Medel}, Hector Javier and {Ji}, Alexander and {Jofre}, Paula and {Johnson}, Jennifer A. and {Jones}, Amy M. and {Kinemuchi}, Karen and {Kluge}, Matthias and {Koekemoer}, Anton and {Kollmeier}, Juna A. and {Kounkel}, Marina and {Krishnarao}, Dhanesh and {Krumpe}, Mirko and {Lacerna}, Ivan and {Lago}, Paulo Jakson Assuncao and {Laporte}, Chervin and {Liu}, Chao and {Liu}, Ang and {Liu}, Xin and {Lopes}, Alexandre Roman and {Macktoobian}, Matin and {Majewski}, Steven R. and {Malanushenko}, Viktor and {Maoz}, Dan and {Masseron}, Thomas and {Masters}, Karen L. and {Matijevic}, Gal and {McBride}, Aidan and {Medan}, Ilija and {Merloni}, Andrea and {Morrison}, Sean and {Myers}, Natalie and {M{\'e}sz{\'a}ros}, Szabolcs and {Negrete}, C. Alenka and {Nidever}, David L. and {Nitschelm}, Christian and {Oravetz}, Daniel and {Oravetz}, Audrey and {Pan}, Kaike and {Peng}, Yingjie and {Pinsonneault}, Marc H. and {Pogge}, Rick and {Qiu}, Dan and {Ramirez}, Solange V. and {Rix}, Hans-Walter and {Fern{\'a}ndez Rosso}, Daniela and {Runnoe}, Jessie and {Salvato}, Mara and {Sanchez}, Sebastian F. and {Santana}, Felipe A. and {Saydjari}, Andrew and {Sayres}, Conor and {Schlaufman}, Kevin C. and {Schneider}, Donald P. and {Schwope}, Axel and {Serna}, Javier and {Shen}, Yue and {Sobeck}, Jennifer and {Song}, Ying-Yi and {Souto}, Diogo and {Spoo}, Taylor and {Stassun}, Keivan G. and {Steinmetz}, Matthias and {Straumit}, Ilya and {Stringfellow}, Guy and {S{\'a}nchez-Gallego}, Jos{\'e} and {Taghizadeh-Popp}, Manuchehr and {Tayar}, Jamie and {Thakar}, Ani and {Tissera}, Patricia B. and {Tkachenko}, Andrew and {Hernandez Toledo}, Hector and {Trakhtenbrot}, Benny and {Fern{\'a}ndez-Trincado}, Jos{\'e} G. and {Troup}, Nicholas and {Trump}, Jonathan R. and {Tuttle}, Sarah and {Ulloa}, Natalie and {Vazquez-Mata}, Jose Antonio and {Vera Alfaro}, Pablo and {Villanova}, Sandro and {Wachter}, Stefanie and {Weijmans}, Anne-Marie and {Wheeler}, Adam and {Wilson}, John and {Wojno}, Leigh and {Wolf}, Julien and {Xue}, Xiang-Xiang and {Ybarra}, Jason E. and {Zari}, Eleonora and {Zasowski}, Gail},
        title = "{The Eighteenth Data Release of the Sloan Digital Sky Surveys: Targeting and First Spectra from SDSS-V}",
      journal = {\apjs},
         year = 2023,
        month = aug,
       volume = {267},
       number = {2},
          eid = {44},
        pages = {44},
          doi = {10.3847/1538-4365/acda98},
archivePrefix = {arXiv},
       eprint = {2301.07688},
 primaryClass = {astro-ph.GA},
       adsurl = {https://ui.adsabs.harvard.edu/abs/2023ApJS..267...44A}
}

@misc{sdss_dr6_linelist,
  author       = {{Sloan Digital Sky Survey}},
  title        = {Table of Spectral Lines Used in SDSS spectro1d},
  year         = {2008},
  url          = {https://classic.sdss.org/dr6/algorithms/linestable.php}
}

@misc{brown2020languagemodelsfewshotlearners,
      title={Language Models are Few-Shot Learners}, 
      author={Tom B. Brown and Benjamin Mann and Nick Ryder and Melanie Subbiah and Jared Kaplan and Prafulla Dhariwal and Arvind Neelakantan and Pranav Shyam and Girish Sastry and Amanda Askell and Sandhini Agarwal and Ariel Herbert-Voss and Gretchen Krueger and Tom Henighan and Rewon Child and Aditya Ramesh and Daniel M. Ziegler and Jeffrey Wu and Clemens Winter and Christopher Hesse and Mark Chen and Eric Sigler and Mateusz Litwin and Scott Gray and Benjamin Chess and Jack Clark and Christopher Berner and Sam McCandlish and Alec Radford and Ilya Sutskever and Dario Amodei},
      year={2020},
      eprint={2005.14165},
      archivePrefix={arXiv},
      primaryClass={cs.CL},
      url={https://arxiv.org/abs/2005.14165}, 
}

@misc{baldry2018reinventingslideruleredshifts,
      title={Reinventing the slide rule for redshifts: the case for logarithmic wavelength shift}, 
      author={I. K. Baldry},
      year={2018},
      eprint={1812.05135},
      archivePrefix={arXiv},
      primaryClass={astro-ph.CO},
      url={https://arxiv.org/abs/1812.05135}, 
}

@article{Scourfield_2023,
   title={De-noising of galaxy optical spectra with autoencoders},
   volume={526},
   ISSN={1365-2966},
   url={http://dx.doi.org/10.1093/mnras/stad2709},
   DOI={10.1093/mnras/stad2709},
   number={2},
   journal={Monthly Notices of the Royal Astronomical Society},
   publisher={Oxford University Press (OUP)},
   author={Scourfield, M and Saintonge, A and de Mijolla, D and Viti, S},
   year={2023},
   month=sep, pages={3037--3050} }

@misc{levi2019darkenergyspectroscopicinstrument,
      title={The Dark Energy Spectroscopic Instrument (DESI)}, 
      author={Michael E. Levi and Lori E. Allen and Anand Raichoor and Charles Baltay and Segev BenZvi and Florian Beutler and Adam Bolton and Francisco J. Castander and Chia-Hsun Chuang and Andrew Cooper and Jean-Gabriel Cuby and Arjun Dey and Daniel Eisenstein and Xiaohui Fan and Brenna Flaugher and Carlos Frenk and Alma X. Gonzalez-Morales and Or Graur and Julien Guy and Salman Habib and Klaus Honscheid and Stephanie Juneau and Jean-Paul Kneib and Ofer Lahav and Dustin Lang and Alexie Leauthaud and Betta Lusso and Axel de la Macorra and Marc Manera and Paul Martini and Shude Mao and Jeffrey A. Newman and Nathalie Palanque-Delabrouille and Will J. Percival and Carlos Allende Prieto and Constance M. Rockosi and Vanina Ruhlmann-Kleider and David Schlegel and Hee-Jong Seo and Yong-Seon Song and Greg Tarle and Risa Wechsler and David Weinberg and Christophe Yeche and Ying Zu},
      year={2019},
      eprint={1907.10688},
      archivePrefix={arXiv},
      primaryClass={astro-ph.IM},
      url={https://arxiv.org/abs/1907.10688}, 
}

@ARTICLE{1981PASP...93....5B,
       author = {{Baldwin}, J.~A. and {Phillips}, M.~M. and {Terlevich}, R.},
        title = "{Classification parameters for the emission-line spectra of extragalactic objects.}",
      journal = {\pasp},
         year = 1981,
        month = feb,
       volume = {93},
        pages = {5-19},
          doi = {10.1086/130766},
       adsurl = {https://ui.adsabs.harvard.edu/abs/1981PASP...93....5B}
}

@ARTICLE{2001ApJ...556..121K,
       author = {{Kewley}, L.~J. and {Dopita}, M.~A. and {Sutherland}, R.~S. and {Heisler}, C.~A. and {Trevena}, J.},
        title = "{Theoretical Modeling of Starburst Galaxies}",
      journal = {\apj},
         year = 2001,
        month = jul,
       volume = {556},
       number = {1},
        pages = {121-140},
          doi = {10.1086/321545},
archivePrefix = {arXiv},
       eprint = {astro-ph/0106324},
 primaryClass = {astro-ph},
       adsurl = {https://ui.adsabs.harvard.edu/abs/2001ApJ...556..121K}
}

@article{10.1111/j.1365-2966.2006.10859.x,
    author = {Kewley, Lisa J. and Groves, Brent and Kauffmann, Guinevere and Heckman, Tim},
    title = {The host galaxies and classification of active galactic nuclei},
    journal = {Monthly Notices of the Royal Astronomical Society},
    volume = {372},
    number = {3},
    pages = {961-976},
    year = {2006},
    month = {09},
    issn = {0035-8711},
    doi = {10.1111/j.1365-2966.2006.10859.x},
    url = {https://doi.org/10.1111/j.1365-2966.2006.10859.x},
    eprint = {https://academic.oup.com/mnras/article-pdf/372/3/961/2943380/mnras0372-0961.pdf},
}

@article{EMD_paper,
author = {Rubner, Yossi and Tomasi, Carlo and Guibas, Leonidas},
year = {2000},
month = {01},
pages = {99-121},
title = {The Earth Mover’s Distance as a metric for image retrieval},
volume = {40},
journal = {International Journal of Computer Vision}
}

@article{Smee_2013,
   title={THE MULTI-OBJECT, FIBER-FED SPECTROGRAPHS FOR THE SLOAN DIGITAL SKY SURVEY AND THE BARYON OSCILLATION SPECTROSCOPIC SURVEY},
   volume={146},
   ISSN={1538-3881},
   url={http://dx.doi.org/10.1088/0004-6256/146/2/32},
   DOI={10.1088/0004-6256/146/2/32},
   number={2},
   journal={The Astronomical Journal},
   publisher={American Astronomical Society},
   author={Smee, Stephen A. and Gunn, James E. and Uomoto, Alan and Roe, Natalie and Schlegel, David and Rockosi, Constance M. and Carr, Michael A. and Leger, French and Dawson, Kyle S. and Olmstead, Matthew D. and Brinkmann, Jon and Owen, Russell and Barkhouser, Robert H. and Honscheid, Klaus and Harding, Paul and Long, Dan and Lupton, Robert H. and Loomis, Craig and Anderson, Lauren and Annis, James and Bernardi, Mariangela and Bhardwaj, Vaishali and Bizyaev, Dmitry and Bolton, Adam S. and Brewington, Howard and Briggs, John W. and Burles, Scott and Burns, James G. and Castander, Francisco Javier and Connolly, Andrew and Davenport, James R. A. and Ebelke, Garrett and Epps, Harland and Feldman, Paul D. and Friedman, Scott D. and Frieman, Joshua and Heckman, Timothy and Hull, Charles L. and Knapp, Gillian R. and Lawrence, David M. and Loveday, Jon and Mannery, Edward J. and Malanushenko, Elena and Malanushenko, Viktor and Merrelli, Aronne James and Muna, Demitri and Newman, Peter R. and Nichol, Robert C. and Oravetz, Daniel and Pan, Kaike and Pope, Adrian C. and Ricketts, Paul G. and Shelden, Alaina and Sandford, Dale and Siegmund, Walter and Simmons, Audrey and Smith, D. Shane and Snedden, Stephanie and Schneider, Donald P. and SubbaRao, Mark and Tremonti, Christy and Waddell, Patrick and York, Donald G.},
   year={2013},
   month=jul, pages={32} }

@article{Fluke_2019,
   title={Surveying the reach and maturity of machine learning and artificial intelligence in astronomy},
   volume={10},
   ISSN={1942-4795},
   url={http://dx.doi.org/10.1002/widm.1349},
   DOI={10.1002/widm.1349},
   number={2},
   journal={WIREs Data Mining and Knowledge Discovery},
   publisher={Wiley},
   author={Fluke, Christopher J. and Jacobs, Colin},
   year={2019},
   month=Dec }

@article{MURTAGH1995399,
title = {Interpreting the Kohonen self-organizing feature map using contiguity-constrained clustering},
journal = {Pattern Recognition Letters},
volume = {16},
number = {4},
pages = {399-408},
year = {1995},
issn = {0167-8655},
doi = {https://doi.org/10.1016/0167-8655(94)00113-H},
url = {https://www.sciencedirect.com/science/article/pii/016786559400113H},
author = {F. Murtagh}
}

@ARTICLE{2002AJ....124.1810S,
       author = {{Strauss}, Michael A. and {Weinberg}, David H. and {Lupton}, Robert H. and {Narayanan}, Vijay K. and {Annis}, James and {Bernardi}, Mariangela and {Blanton}, Michael and {Burles}, Scott and {Connolly}, A.~J. and {Dalcanton}, Julianne and {Doi}, Mamoru and {Eisenstein}, Daniel and {Frieman}, Joshua A. and {Fukugita}, Masataka and {Gunn}, James E. and {Ivezi{\'c}}, {\v{Z}}eljko and {Kent}, Stephen and {Kim}, Rita S.~J. and {Knapp}, G.~R. and {Kron}, Richard G. and {Munn}, Jeffrey A. and {Newberg}, Heidi Jo and {Nichol}, R.~C. and {Okamura}, Sadanori and {Quinn}, Thomas R. and {Richmond}, Michael W. and {Schlegel}, David J. and {Shimasaku}, Kazuhiro and {SubbaRao}, Mark and {Szalay}, Alexander S. and {Vanden Berk}, Dan and {Vogeley}, Michael S. and {Yanny}, Brian and {Yasuda}, Naoki and {York}, Donald G. and {Zehavi}, Idit},
        title = "{Spectroscopic Target Selection in the Sloan Digital Sky Survey: The Main Galaxy Sample}",
      journal = {\aj},
         year = 2002,
        month = sep,
       volume = {124},
       number = {3},
        pages = {1810-1824},
          doi = {10.1086/342343},
archivePrefix = {arXiv},
       eprint = {astro-ph/0206225},
 primaryClass = {astro-ph},
       adsurl = {https://ui.adsabs.harvard.edu/abs/2002AJ....124.1810S}
}

@ARTICLE{2001AJ....122.2267E,
       author = {{Eisenstein}, Daniel J. and {Annis}, James and {Gunn}, James E. and {Szalay}, Alexander S. and {Connolly}, Andrew J. and {Nichol}, R.~C. and {Bahcall}, Neta A. and {Bernardi}, Mariangela and {Burles}, Scott and {Castander}, Francisco J. and {Fukugita}, Masataka and {Hogg}, David W. and {Ivezi{\'c}}, {\v{Z}}eljko and {Knapp}, G.~R. and {Lupton}, Robert H. and {Narayanan}, Vijay and {Postman}, Marc and {Reichart}, Daniel E. and {Richmond}, Michael and {Schneider}, Donald P. and {Schlegel}, David J. and {Strauss}, Michael A. and {SubbaRao}, Mark and {Tucker}, Douglas L. and {Vanden Berk}, Daniel and {Vogeley}, Michael S. and {Weinberg}, David H. and {Yanny}, Brian},
        title = "{Spectroscopic Target Selection for the Sloan Digital Sky Survey: The Luminous Red Galaxy Sample}",
      journal = {\aj},
         year = 2001,
        month = nov,
       volume = {122},
       number = {5},
        pages = {2267-2280},
          doi = {10.1086/323717},
archivePrefix = {arXiv},
       eprint = {astro-ph/0108153},
 primaryClass = {astro-ph},
       adsurl = {https://ui.adsabs.harvard.edu/abs/2001AJ....122.2267E}
}

@ARTICLE{2002AJ....123.2945R,
       author = {{Richards}, Gordon T. and {Fan}, Xiaohui and {Newberg}, Heidi Jo and {Strauss}, Michael A. and {Vanden Berk}, Daniel E. and {Schneider}, Donald P. and {Yanny}, Brian and {Boucher}, Adam and {Burles}, Scott and {Frieman}, Joshua A. and {Gunn}, James E. and {Hall}, Patrick B. and {Ivezi{\'c}}, {\v{Z}}eljko and {Kent}, Stephen and {Loveday}, Jon and {Lupton}, Robert H. and {Rockosi}, Constance M. and {Schlegel}, David J. and {Stoughton}, Chris and {SubbaRao}, Mark and {York}, Donald G.},
        title = "{Spectroscopic Target Selection in the Sloan Digital Sky Survey: The Quasar Sample}",
      journal = {\aj},
         year = 2002,
        month = jun,
       volume = {123},
       number = {6},
        pages = {2945-2975},
          doi = {10.1086/340187},
archivePrefix = {arXiv},
       eprint = {astro-ph/0202251},
 primaryClass = {astro-ph},
       adsurl = {https://ui.adsabs.harvard.edu/abs/2002AJ....123.2945R}
}

@ARTICLE{2012ApJS..199....3R,
       author = {{Ross}, Nicholas P. and {Myers}, Adam D. and {Sheldon}, Erin S. and {Y{\`e}che}, Christophe and {Strauss}, Michael A. and {Bovy}, Jo and {Kirkpatrick}, Jessica A. and {Richards}, Gordon T. and {Aubourg}, {\'E}ric and {Blanton}, Michael R. and {Brandt}, W.~N. and {Carithers}, William C. and {Croft}, Rupert A.~C. and {da Silva}, Robert and {Dawson}, Kyle and {Eisenstein}, Daniel J. and {Hennawi}, Joseph F. and {Ho}, Shirley and {Hogg}, David W. and {Lee}, Khee-Gan and {Lundgren}, Britt and {McMahon}, Richard G. and {Miralda-Escud{\'e}}, Jordi and {Palanque-Delabrouille}, Nathalie and {P{\^a}ris}, Isabelle and {Petitjean}, Patrick and {Pieri}, Matthew M. and {Rich}, James and {Roe}, Natalie A. and {Schiminovich}, David and {Schlegel}, David J. and {Schneider}, Donald P. and {Slosar}, An{\v{z}}e and {Suzuki}, Nao and {Tinker}, Jeremy L. and {Weinberg}, David H. and {Weyant}, Anya and {White}, Martin and {Wood-Vasey}, W. Michael},
        title = "{The SDSS-III Baryon Oscillation Spectroscopic Survey: Quasar Target Selection for Data Release Nine}",
      journal = {\apjs},
         year = 2012,
        month = mar,
       volume = {199},
       number = {1},
          eid = {3},
        pages = {3},
          doi = {10.1088/0067-0049/199/1/3},
archivePrefix = {arXiv},
       eprint = {1105.0606},
 primaryClass = {astro-ph.CO},
       adsurl = {https://ui.adsabs.harvard.edu/abs/2012ApJS..199....3R}
}

@misc{leighly2025balmerabsorptionironlowionization,
      title={Balmer Absorption in Iron Low-Ionization Broad Absorption Line Quasars}, 
      author={Karen M. Leighly and Sarah C. Gallagher and Hyunseop Choi and Donald M. Terndrup and Julianna R. Voelker and Gordon T. Richards and Leah K. Morabito},
      year={2025},
      eprint={2509.07611},
      archivePrefix={arXiv},
      primaryClass={astro-ph.GA},
      url={https://arxiv.org/abs/2509.07611}, 
}

@article{Choi__2022,
   title={The Physical Properties of Low-redshift FeLoBAL Quasars. I. Spectral-synthesis Analysis of the Broad Absorption-line (BAL) Outflows Using SimBAL},
   volume={937},
   ISSN={1538-4357},
   url={http://dx.doi.org/10.3847/1538-4357/ac61d9},
   DOI={10.3847/1538-4357/ac61d9},
   number={2},
   journal={The Astrophysical Journal},
   publisher={American Astronomical Society},
   author={Choi , Hyunseop  and Leighly, Karen M. and Terndrup, Donald M. and Dabbieri, Collin and Gallagher, Sarah C. and Richards, Gordon T.},
   year={2022},
   month=Sept, pages={74} }

@misc{hogg2000distancemeasurescosmology,
      title={Distance measures in cosmology}, 
      author={David W. Hogg},
      year={2000},
      eprint={astro-ph/9905116},
      archivePrefix={arXiv},
      primaryClass={astro-ph},
      url={https://arxiv.org/abs/astro-ph/9905116}, 
}

@ARTICLE{2001ARA&A..39..137S,
       author = {{Sofue}, Yoshiaki and {Rubin}, Vera},
        title = "{Rotation Curves of Spiral Galaxies}",
      journal = {\araa},
         year = 2001,
        month = jan,
       volume = {39},
        pages = {137-174},
          doi = {10.1146/annurev.astro.39.1.137},
archivePrefix = {arXiv},
       eprint = {astro-ph/0010594},
 primaryClass = {astro-ph},
       adsurl = {https://ui.adsabs.harvard.edu/abs/2001ARA&A..39..137S}
}

@article{Abdul_Karim_2025,
   title={DESI DR2 results. II. Measurements of baryon acoustic oscillations and cosmological constraints},
   volume={112},
   ISSN={2470-0029},
   url={http://dx.doi.org/10.1103/tr6y-kpc6},
   DOI={10.1103/tr6y-kpc6},
   number={8},
   journal={Physical Review D},
   publisher={American Physical Society (APS)},
   author={Abdul Karim, M. and Aguilar, J. and Ahlen, S. and Alam, S. and Allen, L. and Prieto, C. Allende and Alves, O. and Anand, A. and Andrade, U. and Armengaud, E. and Aviles, A. and Bailey, S. and Baltay, C. and Bansal, P. and Bault, A. and Behera, J. and BenZvi, S. and Bianchi, D. and Blake, C. and Brieden, S. and Brodzeller, A. and Brooks, D. and Buckley-Geer, E. and Burtin, E. and Calderon, R. and Canning, R. and Rosell, A. Carnero and Carrilho, P. and Casas, L. and Castander, F. J. and Charles, M. and Chaussidon, E. and Chaves-Montero, J. and Chebat, D. and Chen, X. and Claybaugh, T. and Cole, S. and Cooper, A. P. and Cuceu, A. and Dawson, K. S. and de la Macorra, A. and de Mattia, A. and Deiosso, N. and Della Costa, J. and Demina, R. and Dey, A. and Dey, B. and Ding, Z. and Doel, P. and Edelstein, J. and Eisenstein, D. J. and Elbers, W. and Fagrelius, P. and Fanning, K. and Fernández-García, E. and Ferraro, S. and Font-Ribera, A. and Forero-Romero, J. E. and Frenk, C. S. and Garcia-Quintero, C. and Garrison, L. H. and Gaztañaga, E. and Gil-Marín, H. and Gontcho, S. Gontcho A. and Gonzalez, D. and Gonzalez-Morales, A. X. and Gordon, C. and Green, D. and Gutierrez, G. and Guy, J. and Hadzhiyska, B. and Hahn, C. and He, S. and Herbold, M. and Herrera-Alcantar, H. K. and Ho, M.-F. and Honscheid, K. and Howlett, C. and Huterer, D. and Ishak, M. and Juneau, S. and Kamble, N. V. and Karaçayl��, N. G. and Kehoe, R. and Kent, S. and Kim, A. G. and Kirkby, D. and Kisner, T. and Koposov, S. E. and Kremin, A. and Krolewski, A. and Lahav, O. and Lamman, C. and Landriau, M. and Lang, D. and Lasker, J. and Le Goff, J. M. and Le Guillou, L. and Leauthaud, A. and Levi, M. E. and Li, Q. and Li, T. S. and Lodha, K. and Lokken, M. and Lozano-Rodríguez, F. and Magneville, C. and Manera, M. and Martini, P. and Matthewson, W. L. and Meisner, A. and Mena-Fernández, J. and Menegas, A. and Mergulhão, T. and Miquel, R. and Moustakas, J. and Muñoz-Gutiérrez, A. and Muñoz-Santos, D. and Myers, A. D. and Nadathur, S. and Naidoo, K. and Napolitano, L. and Newman, J. A. and Niz, G. and Noriega, H. E. and Paillas, E. and Palanque-Delabrouille, N. and Pan, J. and Peacock, J. A. and Ibanez, M. P. and Percival, W. J. and Pérez-Fernández, A. and Pérez-Ràfols, I. and Pieri, M. M. and Poppett, C. and Prada, F. and Rabinowitz, D. and Raichoor, A. and Ramírez-Pérez, C. and Rashkovetskyi, M. and Ravoux, C. and Rich, J. and Rocher, A. and Rockosi, C. and Rohlf, J. and Román-Herrera, J. O. and Ross, A. J. and Rossi, G. and Ruggeri, R. and Ruhlmann-Kleider, V. and Samushia, L. and Sanchez, E. and Sanders, N. and Schlegel, D. and Schubnell, M. and Seo, H. and Shafieloo, A. and Sharples, R. and Silber, J. and Sinigaglia, F. and Sprayberry, D. and Tan, T. and Tarlé, G. and Taylor, P. and Turner, W. and Ureña-López, L. A. and Vaisakh, R. and Valdes, F. and Valogiannis, G. and Vargas-Magaña, M. and Verde, L. and Walther, M. and Weaver, B. A. and Weinberg, D. H. and White, M. and Wolfson, M. and Yèche, C. and Yu, J. and Zaborowski, E. A. and Zarrouk, P. and Zhai, Z. and Zhang, H. and Zhao, C. and Zhao, G. B. and Zhou, R. and Zou, H. and },
   year={2025},
   month=Oct }

@ARTICLE{2000ApJ...536..571B,
       author = {{Ben{\'\i}tez}, Narciso},
        title = "{Bayesian Photometric Redshift Estimation}",
      journal = {\apj},
         year = 2000,
        month = jun,
       volume = {536},
       number = {2},
        pages = {571-583},
          doi = {10.1086/308947},
archivePrefix = {arXiv},
       eprint = {astro-ph/9811189},
 primaryClass = {astro-ph},
       adsurl = {https://ui.adsabs.harvard.edu/abs/2000ApJ...536..571B}
}

@article{Massarotti_2001,
   title={A critical appraisal of the SED fitting method to estimate photometric redshifts},
   volume={368},
   ISSN={1432-0746},
   url={http://dx.doi.org/10.1051/0004-6361:20000553},
   DOI={10.1051/0004-6361:20000553},
   number={1},
   journal={Astronomy \& Astrophysics},
   publisher={EDP Sciences},
   author={Massarotti, M. and Iovino, A. and Buzzoni, A.},
   year={2001},
   month=Mar, pages={74--85} }

@article{Newman_2022,
   title={Photometric Redshifts for Next-Generation Surveys},
   volume={60},
   ISSN={1545-4282},
   url={http://dx.doi.org/10.1146/annurev-astro-032122-014611},
   DOI={10.1146/annurev-astro-032122-014611},
   number={1},
   journal={Annual Review of Astronomy and Astrophysics},
   publisher={Annual Reviews},
   author={Newman, Jeffrey A. and Gruen, Daniel},
   year={2022},
   month=Aug, pages={363--414} }

@INPROCEEDINGS{1988IAUS..130..105H,
       author = {{Huchra}, J.~P. and {Geller}, M.~J. and {de Lapparent}, V. and {Burg}, R.},
        title = "{The CFA Redshift Survey}",
    booktitle = {Large Scale Structures of the Universe},
         year = 1988,
       editor = {{Audouze}, J. and {Pelletan}, M. -C. and {Szalay}, A. and {Zel'dovich}, Ya. B. and {Peebles}, P.~J.~E.},
       series = {IAU Symposium},
       volume = {130},
        month = jan,
        pages = {105},
       adsurl = {https://ui.adsabs.harvard.edu/abs/1988IAUS..130..105H}
}

@ARTICLE{2003astro.ph..6581C,
       author = {{Colless}, Matthew and {Peterson}, Bruce A. and {Jackson}, Carole and {Peacock}, John A. and {Cole}, Shaun and {Norberg}, Peder and {Baldry}, Ivan K. and {Baugh}, Carlton M. and {Bland-Hawthorn}, Joss and {Bridges}, Terry and {Cannon}, Russell and {Collins}, Chris and {Couch}, Warrick and {Cross}, Nicholas and {Dalton}, Gavin and {De Propris}, Roberto and {Driver}, Simon P. and {Efstathiou}, George and {Ellis}, Richard S. and {Frenk}, Carlos S. and {Glazebrook}, Karl and {Lahav}, Ofer and {Lewis}, Ian and {Lumsden}, Stuart and {Maddox}, Steve and {Madgwick}, Darren and {Sutherland}, Will and {Taylor}, Keith},
        title = "{The 2dF Galaxy Redshift Survey: Final Data Release}",
      journal = {arXiv e-prints},
         year = 2003,
        month = jun,
          eid = {astro-ph/0306581},
        pages = {astro-ph/0306581},
          doi = {10.48550/arXiv.astro-ph/0306581},
archivePrefix = {arXiv},
       eprint = {astro-ph/0306581},
 primaryClass = {astro-ph},
       adsurl = {https://ui.adsabs.harvard.edu/abs/2003astro.ph..6581C}
}

@misc{desicollaboration2026datarelease1dark,
      title={Data Release 1 of the Dark Energy Spectroscopic Instrument}, 
      author={{DESI Collaboration} and M. Abdul Karim and A. G. Adame and D. Aguado and J. Aguilar and S. Ahlen and S. Alam and G. Aldering and D. M. Alexander and R. Alfarsy and L. Allen and C. Allende Prieto and O. Alves and A. Anand and U. Andrade and E. Armengaud and S. Avila and A. Aviles and H. Awan and S. Bailey and A. Baleato Lizancos and O. Ballester and A. Bault and J. Bautista and R. Bean and J. Behera and S. BenZvi and L. Beraldo e Silva and J. R. Bermejo-Climent and F. Beutler and D. Bianchi and C. Blake and R. Blum and A. S. Bolton and M. Bonici and S. Brieden and A. Brodzeller and D. Brooks and E. Buckley-Geer and E. Burtin and A. Byström and R. Canning and A. Carnero Rosell and A. Carr and P. Carrilho and L. Casas and F. J. Castander and R. Cereskaite and J. L. Cervantes-Cota and E. Chaussidon and J. Chaves-Montero and S. Chen and X. Chen and C. Circosta and T. Claybaugh and S. Cole and A. P. Cooper and M. -C. Cousinou and A. Cuceu and T. M. Davis and K. S. Dawson and R. de Belsunce and R. de la Cruz and A. de la Macorra and A. de Mattia and N. Deiosso and J. Della Costa and R. Demina and U. Demirbozan and J. DeRose and A. Dey and B. Dey and J. Ding and Z. Ding and P. Doel and K. Douglass and M. Dowicz and H. Ebina and J. Edelstein and D. J. Eisenstein and W. Elbers and N. Emas and S. Escoffier and P. Fagrelius and X. Fan and K. Fanning and G. Favole and V. A. Fawcett and E. Fernández-García and S. Ferraro and N. Findlay and A. Font-Ribera and J. E. Forero-Romero and D. Forero-Sánchez and C. S. Frenk and B. T. Gänsicke and L. Galbany and J. García-Bellido and C. Garcia-Quintero and L. H. Garrison and E. Gaztañaga and H. Gil-Marín and A. Gloudemans and O. Y. Gnedin and S. Gontcho A Gontcho and D. Gonzalez and A. X. Gonzalez-Morales and V. Gonzalez-Perez and C. Gordon and O. Graur and D. Green and D. Gruen and R. Gsponer and C. Guandalin and G. Gutierrez and J. Guy and C. Hahn and J. J. Han and J. Han and S. He and H. K. Herrera-Alcantar and S. Heydenreich and K. Honscheid and J. Hou and C. Howlett and D. Huterer and V. Iršič and M. Ishak and A. Jacques and L. Jiang and J. Jimenez and Y. P. Jing and B. Joachimi and S. Joudaki and R. Joyce and E. Jullo and S. Juneau and N. G. Karaçaylı and T. Karim and R. Kehoe and S. Kent and A. Khederlarian and D. Kirkby and T. Kisner and F. -S. Kitaura and N. Kizhuprakkat and H. Kong and S. E. Koposov and A. Kremin and A. Krolewski and O. Lahav and Y. Lai and C. Lamman and T. -W. Lan and M. Landriau and D. Lang and J. U. Lange and J. Lasker and J. M. Le Goff and L. Le Guillou and A. Leauthaud and M. E. Levi and S. Li and T. S. Li and W. Liu and K. Lodha and M. Lokken and Y. Luo and Y. Luo and C. Magneville and M. Manera and C. J. Manser and D. Margala and P. Martini and M. Maus and J. McCullough and P. McDonald and G. E. Medina and L. Medina-Varela and A. Meisner and J. Mena-Fernández and A. Menegas and J. Meneses-Rizo and M. Mezcua and R. Miquel and P. Montero-Camacho and J. Moon and J. Moustakas and A. Muñoz-Gutiérrez and D. Muñoz-Santos and A. D. Myers and J. Myles and S. Nadathur and J. Najita and L. Napolitano and J. A. Newman and F. Nikakhtar and R. Nikutta and G. Niz and H. E. Noriega and P. Nugent and N. Padmanabhan and E. Paillas and N. Palanque-Delabrouille and A. Palmese and J. Pan and Z. Pan and D. Parkinson and J. A. Peacock and M. P. Ibanez and W. J. Percival and A. Pérez-Fernández and I. Pérez-Ràfols and P. Peterson and J. Piat and M. M. Pieri and M. Pinon and C. Poppett and A. Porredon and F. Prada and R. Pucha and F. Qin and D. Rabinowitz and A. Raichoor and C. Ramírez-Pérez and S. Ramirez-Solano and M. Rashkovetskyi and C. Ravoux and B. Ried Guachalla and A. H. Riley and A. Rocher and C. Rockosi and J. Rohlf and A. J. Rosado-Marín and A. J. Ross and C. Ross and G. Rossi and R. Ruggeri and V. Ruhlmann-Kleider and C. G. Sabiu and K. Said and N. Sailer and A. Saintonge and Y. Salcedo Hernandez and L. Samushia and E. Sanchez and N. Sanders and N. Sandford and S. Satyavolu and C. Saulder and A. K. Saydjari and E. F. Schlafly and D. Schlegel and D. Scholte and M. Schubnell and A. Semenaite and H. Seo and A. Shafieloo and R. Sharples and J. Silber and F. Sinigaglia and M. Siudek and Z. Slepian and A. Smith and M. Soumagnac and D. Sprayberry and J. Suárez-Pérez and J. Swanson and T. Tan and G. Tarlé and P. Taylor and G. Thomas and R. Tojeiro and R. J. Turner and W. Turner and L. A. Ureña-López and R. Vaisakh and M. Valluri and G. Valogiannis and M. Vargas-Magaña and L. Verde and P. Vielzeuf and M. Walther and B. Wang and M. S. Wang and W. Wang and B. A. Weaver and N. Weaverdyck and R. H. Wechsler and D. H. Weinberg and M. White and A. Whitford and M. Wolfson and J. Yang and C. Yèche and S. Youles and J. Yu and S. Yuan and E. A. Zaborowski and P. Zarrouk and H. Zhang and C. Zhao and R. Zhao and Z. Zheng and C. Zhou and R. Zhou and Y. Zhou and H. Zou and S. Zou and Y. Zu},
      year={2026},
      eprint={2503.14745},
      archivePrefix={arXiv},
      primaryClass={astro-ph.CO},
      url={https://arxiv.org/abs/2503.14745}, 
}

@ARTICLE{1956AJ.....61...97H,
       author = {{Humason}, M.~L. and {Mayall}, N.~U. and {Sandage}, A.~R.},
        title = "{Redshifts and magnitudes of extragalactic nebulae.}",
      journal = {\aj},
         year = 1956,
        month = jan,
       volume = {61},
        pages = {97-162},
          doi = {10.1086/107297},
       adsurl = {https://ui.adsabs.harvard.edu/abs/1956AJ.....61...97H}
}

@article{Baldry_2014,
   title={Galaxy And Mass Assembly (GAMA): AUTOZ spectral redshift measurements, confidence and errors},
   volume={441},
   ISSN={1365-2966},
   url={http://dx.doi.org/10.1093/mnras/stu727},
   DOI={10.1093/mnras/stu727},
   number={3},
   journal={Monthly Notices of the Royal Astronomical Society},
   publisher={Oxford University Press (OUP)},
   author={Baldry, I. K. and Alpaslan, M. and Bauer, A. E. and Bland-Hawthorn, J. and Brough, S. and Cluver, M. E. and Croom, S. M. and Davies, L. J. M. and Driver, S. P. and Gunawardhana, M. L. P. and Holwerda, B. W. and Hopkins, A. M. and Kelvin, L. S. and Liske, J. and Lopez-Sanchez, A. R. and Loveday, J. and Norberg, P. and Peacock, J. and Robotham, A. S. G. and Taylor, E. N.},
   year={2014},
   month=May, pages={2440--2451} }

@ARTICLE{1979AJ.....84.1511T,
       author = {{Tonry}, J. and {Davis}, M.},
        title = "{A survey of galaxy redshifts. I. Data reduction techniques.}",
      journal = {\aj},
         year = 1979,
        month = oct,
       volume = {84},
        pages = {1511-1525},
          doi = {10.1086/112569},
       adsurl = {https://ui.adsabs.harvard.edu/abs/1979AJ.....84.1511T}
}

@article{10.1093/mnras/sty3020,
    author = {Liu, W and Zhu, M and Dai, C and He, D Y and Yao, Jiawen and Tian, H F and Wang, B Y and Wu, K and Zhan, Y and Chen, B-Q and Luo, A-Li and Wang, R and Cao, Y and Yu, X C},
    title = {Classification of large-scale stellar spectra based on deep convolutional neural network},
    journal = {Monthly Notices of the Royal Astronomical Society},
    volume = {483},
    number = {4},
    pages = {4774-4783},
    year = {2019},
    month = {03},
    issn = {0035-8711},
    doi = {10.1093/mnras/sty3020},
    url = {https://doi.org/10.1093/mnras/sty3020},
    eprint = {https://academic.oup.com/mnras/article-pdf/483/4/4774/27441003/sty3020.pdf},
}

@article{10.1093/mnras/stae1461,
    author = {Zhong, Fucheng and Napolitano, Nicola R and Heneka, Caroline and Li, Rui and Bauer, Franz Erik and Bouche, Nicolas and Comparat, Johan and Kim, Young-Lo and Krogager, Jens-Kristian and Longhetti, Marcella and Loveday, Jonathan and Roukema, Boudewijn F and Rouse, Benedict L and Salvato, Mara and Tortora, Crescenzo and Assef, Roberto J and Cassarà, Letizia P and Costantin, Luca and Croom, Scott M and Davies, Luke J M and Fritz, Alexander and Guiglion, Guillaume and Humphrey, Andrew and Pompei, Emanuela and Ricci, Claudio and Sifón, Cristóbal and Tempel, Elmo and Zafar, Tayyaba},
    title = {Galaxy Spectra neural Network (GaSNet). II. Using deep learning for spectral classification and redshift predictions},
    journal = {Monthly Notices of the Royal Astronomical Society},
    volume = {532},
    number = {1},
    pages = {643-665},
    year = {2024},
    month = {07},
    issn = {0035-8711},
    doi = {10.1093/mnras/stae1461},
    url = {https://doi.org/10.1093/mnras/stae1461},
    eprint = {https://academic.oup.com/mnras/article-pdf/532/1/643/58366128/stae1461.pdf},
}

@ARTICLE{2023ApJ...943...68L,
       author = {{Lan}, Ting-Wen and {Tojeiro}, R. and {Armengaud}, E. and {Prochaska}, J. Xavier and {Davis}, T.~M. and {Alexander}, David M. and {Raichoor}, A. and {Zhou}, Rongpu and {Y{\`e}che}, Christophe and {Balland}, C. and {BenZvi}, S. and {Berti}, A. and {Canning}, R. and {Carr}, A. and {Chittenden}, H. and {Cole}, S. and {Cousinou}, M.-C. and {Dawson}, K. and {Dey}, Biprateep and {Douglass}, K. and {Edge}, A. and {Escoffier}, S. and {Glanville}, A. and {A Gontcho}, S. Gontcho and {Guy}, J. and {Hahn}, C. and {Howlett}, C. and {Hwang}, Ho Seong and {Jiang}, L. and {Kov{\'a}cs}, A. and {Mezcua}, M. and {Moore}, S. and {Nadathur}, S. and {Oh}, M. and {Parkinson}, D. and {Rocher}, A. and {Ross}, A.~J. and {Ruhlmann-Kleider}, V. and {Sabiu}, C.~G. and {Said}, K. and {Saulder}, C. and {Sierra-Porta}, D. and {Weiner}, B. and {Yu}, J. and {Zarrouk}, P. and {Zhang}, Y. and {Zou}, H. and {Ahlen}, S. and {Bailey}, S. and {Brooks}, D. and {Cooper}, A.~P. and {de la Macorra}, A. and {Dey}, A. and {Dhungana}, G. and {Doel}, P. and {Eftekharzadeh}, S. and {Fanning}, K. and {Font-Ribera}, A. and {Garrison}, L. and {Gazta{\~n}aga}, E. and {Kehoe}, R. and {Kisner}, T. and {Kremin}, A. and {Landriau}, M. and {Le Guillou}, L. and {Levi}, Michael E. and {Magneville}, C. and {Meisner}, Aaron M. and {Miquel}, R. and {Moustakas}, J. and {Myers}, Adam D. and {Newman}, Jeffrey A. and {Nie}, J.~D. and {Palanque-Delabrouille}, N. and {Percival}, W.~J. and {Poppett}, C. and {Prada}, F. and {Schubnell}, M. and {Tarl{\'e}}, Gregory and {Weaver}, B.~A. and {Zhang}, K. and {Zhou}, Zhimin},
        title = "{The DESI Survey Validation: Results from Visual Inspection of Bright Galaxies, Luminous Red Galaxies, and Emission-line Galaxies}",
      journal = {\apj},
         year = 2023,
        month = jan,
       volume = {943},
       number = {1},
          eid = {68},
        pages = {68},
          doi = {10.3847/1538-4357/aca5fa},
archivePrefix = {arXiv},
       eprint = {2208.08516},
 primaryClass = {astro-ph.CO},
       adsurl = {https://ui.adsabs.harvard.edu/abs/2023ApJ...943...68L}
}

@ARTICLE{2023AJ....165..124A,
       author = {{Alexander}, David M. and {Davis}, Tamara M. and {Chaussidon}, E. and {Fawcett}, V.~A. and {X. Gonzalez-Morales}, Alma and {Lan}, Ting-Wen and {Y{\`e}che}, Christophe and {Ahlen}, S. and {Aguilar}, J.~N. and {Armengaud}, E. and {Bailey}, S. and {Brooks}, D. and {Cai}, Z. and {Canning}, R. and {Carr}, A. and {Chabanier}, S. and {Cousinou}, Marie-Claude and {Dawson}, K. and {de la Macorra}, A. and {Dey}, A. and {Dey}, Biprateep and {Dhungana}, G. and {Edge}, A.~C. and {Eftekharzadeh}, S. and {Fanning}, K. and {Farr}, James and {Font-Ribera}, A. and {Garcia-Bellido}, J. and {Garrison}, Lehman and {Gazta{\~n}aga}, E. and {A Gontcho}, Satya Gontcho and {Gordon}, C. and {Medellin Gonzalez}, Stefany Guadalupe and {Guy}, J. and {Herrera-Alcantar}, Hiram K. and {Jiang}, L. and {Juneau}, S. and {Kara{\c{c}}ayl{\i}}, N.~G. and {Kehoe}, R. and {Kisner}, T. and {Kov{\'a}cs}, A. and {Landriau}, M. and {Levi}, Michael E. and {Magneville}, C. and {Martini}, P. and {Meisner}, Aaron M. and {Mezcua}, M. and {Miquel}, R. and {Camacho}, P. Montero and {Moustakas}, J. and {Mu{\~n}oz-Guti{\'e}rrez}, Andrea and {Myers}, Adam D. and {Nadathur}, S. and {Napolitano}, L. and {Nie}, J.~D. and {Palanque-Delabrouille}, N. and {Pan}, Z. and {Percival}, W.~J. and {P{\'e}rez-R{\`a}fols}, I. and {Poppett}, C. and {Prada}, F. and {Ram{\'\i}rez-P{\'e}rez}, C{\'e}sar and {Ravoux}, C. and {Rosario}, D.~J. and {Schubnell}, M. and {Tarl{\'e}}, Gregory and {Walther}, M. and {Weiner}, B. and {Youles}, S. and {Zhou}, Zhimin and {Zou}, H. and {Zou}, Siwei},
        title = "{The DESI Survey Validation: Results from Visual Inspection of the Quasar Survey Spectra}",
      journal = {\aj},
         year = 2023,
        month = mar,
       volume = {165},
       number = {3},
          eid = {124},
        pages = {124},
          doi = {10.3847/1538-3881/acacfc},
archivePrefix = {arXiv},
       eprint = {2208.08517},
 primaryClass = {astro-ph.GA},
       adsurl = {https://ui.adsabs.harvard.edu/abs/2023AJ....165..124A}
}

@article{Kohonen2004SelforganizedFO,
  title={Self-organized formation of topologically correct feature maps},
  author={Teuvo Kohonen},
  journal={Biological Cybernetics},
  year={1982},
  volume={43},
  pages={59--69},
  url={https://api.semanticscholar.org/CorpusID:206775459}
}

@ARTICLE{2024AJ....168..124A,
       author = {{Anand}, Abhijeet and {Guy}, Julien and {Bailey}, Stephen and {Moustakas}, John and {Aguilar}, J. and {Ahlen}, S. and {Bolton}, A.~S. and {Brodzeller}, A. and {Brooks}, D. and {Claybaugh}, T. and {Cole}, S. and {de la Macorra}, A. and {Dey}, Biprateep and {Fanning}, K. and {Forero-Romero}, J.~E. and {Gazta{\~n}aga}, E. and {Gontcho A Gontcho}, S. and {Gutierrez}, G. and {Honscheid}, K. and {Howlett}, C. and {Juneau}, S. and {Kirkby}, D. and {Kisner}, T. and {Kremin}, A. and {Lambert}, A. and {Landriau}, M. and {Le Guillou}, L. and {Manera}, M. and {Meisner}, A. and {Miquel}, R. and {Mueller}, E. and {Niz}, G. and {Palanque-Delabrouille}, N. and {Percival}, W.~J. and {Poppett}, C. and {Prada}, F. and {Raichoor}, A. and {Rezaie}, M. and {Rossi}, G. and {Sanchez}, E. and {Schlafly}, E.~F. and {Schlegel}, D. and {Schubnell}, M. and {Sprayberry}, D. and {Tarl{\'e}}, G. and {Warner}, C. and {Weaver}, B.~A. and {Zhou}, R. and {Zou}, H.},
        title = "{Archetype-based Redshift Estimation for the Dark Energy Spectroscopic Instrument Survey}",
      journal = {\aj},
         year = 2024,
        month = sep,
       volume = {168},
       number = {3},
          eid = {124},
        pages = {124},
          doi = {10.3847/1538-3881/ad60c2},
archivePrefix = {arXiv},
       eprint = {2405.19288},
 primaryClass = {astro-ph.CO},
       adsurl = {https://ui.adsabs.harvard.edu/abs/2024AJ....168..124A}
}

@ARTICLE{2023MNRAS.518..752M,
       author = {{Mong}, Y.-L. and {Ackley}, K. and {Killestein}, T.~L. and {Galloway}, D.~K. and {Vassallo}, C. and {Dyer}, M. and {Cutter}, R. and {Brown}, M.~J.~I. and {Lyman}, J. and {Ulaczyk}, K. and {Steeghs}, D. and {Dhillon}, V. and {O'Brien}, P. and {Ramsay}, G. and {Noysena}, K. and {Kotak}, R. and {Breton}, R. and {Nuttall}, L. and {Pall{\'e}}, E. and {Pollacco}, D. and {Thrane}, E. and {Awiphan}, S. and {Burhanudin}, U. and {Chote}, P. and {Chrimes}, A. and {Daw}, E. and {Duffy}, C. and {Eyles-Ferris}, R. and {Gompertz}, B.~P. and {Heikkil{\"a}}, T. and {Irawati}, P. and {Kennedy}, M. and {Levan}, A. and {Littlefair}, S. and {Makrygianni}, L. and {Marsh}, T. and {Mata S{\'a}nchez}, D. and {Mattila}, S. and {Maund}, J.~R. and {McCormac}, J. and {Mkrtichian}, D. and {Mullaney}, J. and {Rol}, E. and {Sawangwit}, U. and {Stanway}, E. and {Starling}, R. and {Str{\o}m}, P. and {Tooke}, S. and {Wiersema}, K.},
        title = "{Self-supervised clustering on image-subtracted data with deep-embedded self-organizing map}",
      journal = {\mnras},
         year = 2023,
        month = jan,
       volume = {518},
       number = {1},
        pages = {752-762},
          doi = {10.1093/mnras/stac3103},
archivePrefix = {arXiv},
       eprint = {2209.06375},
 primaryClass = {cs.CV},
       adsurl = {https://ui.adsabs.harvard.edu/abs/2023MNRAS.518..752M}
}

@article{autoencoder,
author = {Bourlard, Herve and Kamp, Y},
year = {1988},
month = {02},
pages = {291-4},
title = {Auto-Association by Multilayer Perceptrons and Singular Value Decomposition},
volume = {59},
journal = {Biological cybernetics},
doi = {10.1007/BF00332918}
}

@ARTICLE{1993VA.....36..141M,
       author = {{Miller}, A.~S.},
        title = "{A review of neural network applications in Astronomy}",
      journal = {Vistas in Astronomy},
         year = 1993,
        month = jan,
       volume = {36},
       number = {2},
        pages = {141-161},
          doi = {10.1016/0083-6656(93)90118-4},
       adsurl = {https://ui.adsabs.harvard.edu/abs/1993VA.....36..141M}
}

@article{Dieleman_2015,
   title={Rotation-invariant convolutional neural networks for galaxy morphology prediction},
   volume={450},
   ISSN={0035-8711},
   url={http://dx.doi.org/10.1093/mnras/stv632},
   DOI={10.1093/mnras/stv632},
   number={2},
   journal={Monthly Notices of the Royal Astronomical Society},
   publisher={Oxford University Press (OUP)},
   author={Dieleman, Sander and Willett, Kyle W. and Dambre, Joni},
   year={2015},
   month=apr, pages={1441--1459} }

@article{Teimoorinia_2022,
   title={Mapping the Diversity of Galaxy Spectra with Deep Unsupervised Machine Learning},
   volume={163},
   ISSN={1538-3881},
   url={http://dx.doi.org/10.3847/1538-3881/ac4039},
   DOI={10.3847/1538-3881/ac4039},
   number={2},
   journal={The Astronomical Journal},
   publisher={American Astronomical Society},
   author={Teimoorinia, Hossen and Archinuk, Finn and Woo, Joanna and Shishehchi, Sara and Bluck, Asa F. L.},
   year={2022},
   month=Jan, pages={71} }

@article{Tegmark_2004,
   title={The Three‐Dimensional Power Spectrum of Galaxies from the Sloan Digital Sky Survey},
   volume={606},
   ISSN={1538-4357},
   url={http://dx.doi.org/10.1086/382125},
   DOI={10.1086/382125},
   number={2},
   journal={The Astrophysical Journal},
   publisher={American Astronomical Society},
   author={Tegmark, Max and Blanton, Michael R. and Strauss, Michael A. and Hoyle, Fiona and Schlegel, David and Scoccimarro, Roman and Vogeley, Michael S. and Weinberg, David H. and Zehavi, Idit and Berlind, Andreas and Budavari, Tamas and Connolly, Andrew and Eisenstein, Daniel J. and Finkbeiner, Douglas and Frieman, Joshua A. and Gunn, James E. and Hamilton, Andrew J. S. and Hui, Lam and Jain, Bhuvnesh and Johnston, David and Kent, Stephen and Lin, Huan and Nakajima, Reiko and Nichol, Robert C. and Ostriker, Jeremiah P. and Pope, Adrian and Scranton, Ryan and Seljak, Uroš and Sheth, Ravi K. and Stebbins, Albert and Szalay, Alexander S. and Szapudi, Istvan and Verde, Licia and Xu, Yongzhong and Annis, James and Bahcall, Neta A. and Brinkmann, J. and Burles, Scott and Castander, Francisco J. and Csabai, Istvan and Loveday, Jon and Doi, Mamoru and Fukugita, Masataka and Gott III, J. Richard and Hennessy, Greg and Hogg, David W. and Ivezić, Željko and Knapp, Gillian R. and Lamb, Don Q. and Lee, Brian C. and Lupton, Robert H. and McKay, Timothy A. and Kunszt, Peter and Munn, Jeffrey A. and O’Connell, Liam and Peoples, John and Pier, Jeffrey R. and Richmond, Michael and Rockosi, Constance and Schneider, Donald P. and Stoughton, Christopher and Tucker, Douglas L. and Vanden Berk, Daniel E. and Yanny, Brian and York, Donald G. and },
   year={2004},
   month=May, pages={702--740} }

@ARTICLE{1986Natur.323..533R,
       author = {{Rumelhart}, David E. and {Hinton}, Geoffrey E. and {Williams}, Ronald J.},
        title = "{Learning representations by back-propagating errors}",
      journal = {\nat},
         year = 1986,
        month = oct,
       volume = {323},
       number = {6088},
        pages = {533-536},
          doi = {10.1038/323533a0},
       adsurl = {https://ui.adsabs.harvard.edu/abs/1986Natur.323..533R}
}

@ARTICLE{2000AJ....120.1579Y,
       author = {{York}, Donald G. and {Adelman}, J. and {Anderson}, Jr., John E. and {Anderson}, Scott F. and {Annis}, James and {Bahcall}, Neta A. and {Bakken}, J.~A. and {Barkhouser}, Robert and {Bastian}, Steven and {Berman}, Eileen and {Boroski}, William N. and {Bracker}, Steve and {Briegel}, Charlie and {Briggs}, John W. and {Brinkmann}, J. and {Brunner}, Robert and {Burles}, Scott and {Carey}, Larry and {Carr}, Michael A. and {Castander}, Francisco J. and {Chen}, Bing and {Colestock}, Patrick L. and {Connolly}, A.~J. and {Crocker}, J.~H. and {Csabai}, Istv{\'a}n and {Czarapata}, Paul C. and {Davis}, John Eric and {Doi}, Mamoru and {Dombeck}, Tom and {Eisenstein}, Daniel and {Ellman}, Nancy and {Elms}, Brian R. and {Evans}, Michael L. and {Fan}, Xiaohui and {Federwitz}, Glenn R. and {Fiscelli}, Larry and {Friedman}, Scott and {Frieman}, Joshua A. and {Fukugita}, Masataka and {Gillespie}, Bruce and {Gunn}, James E. and {Gurbani}, Vijay K. and {de Haas}, Ernst and {Haldeman}, Merle and {Harris}, Frederick H. and {Hayes}, J. and {Heckman}, Timothy M. and {Hennessy}, G.~S. and {Hindsley}, Robert B. and {Holm}, Scott and {Holmgren}, Donald J. and {Huang}, Chi-hao and {Hull}, Charles and {Husby}, Don and {Ichikawa}, Shin-Ichi and {Ichikawa}, Takashi and {Ivezi{\'c}}, {\v{Z}}eljko and {Kent}, Stephen and {Kim}, Rita S.~J. and {Kinney}, E. and {Klaene}, Mark and {Kleinman}, A.~N. and {Kleinman}, S. and {Knapp}, G.~R. and {Korienek}, John and {Kron}, Richard G. and {Kunszt}, Peter Z. and {Lamb}, D.~Q. and {Lee}, B. and {Leger}, R. French and {Limmongkol}, Siriluk and {Lindenmeyer}, Carl and {Long}, Daniel C. and {Loomis}, Craig and {Loveday}, Jon and {Lucinio}, Rich and {Lupton}, Robert H. and {MacKinnon}, Bryan and {Mannery}, Edward J. and {Mantsch}, P.~M. and {Margon}, Bruce and {McGehee}, Peregrine and {McKay}, Timothy A. and {Meiksin}, Avery and {Merelli}, Aronne and {Monet}, David G. and {Munn}, Jeffrey A. and {Narayanan}, Vijay K. and {Nash}, Thomas and {Neilsen}, Eric and {Neswold}, Rich and {Newberg}, Heidi Jo and {Nichol}, R.~C. and {Nicinski}, Tom and {Nonino}, Mario and {Okada}, Norio and {Okamura}, Sadanori and {Ostriker}, Jeremiah P. and {Owen}, Russell and {Pauls}, A. George and {Peoples}, John and {Peterson}, R.~L. and {Petravick}, Donald and {Pier}, Jeffrey R. and {Pope}, Adrian and {Pordes}, Ruth and {Prosapio}, Angela and {Rechenmacher}, Ron and {Quinn}, Thomas R. and {Richards}, Gordon T. and {Richmond}, Michael W. and {Rivetta}, Claudio H. and {Rockosi}, Constance M. and {Ruthmansdorfer}, Kurt and {Sandford}, Dale and {Schlegel}, David J. and {Schneider}, Donald P. and {Sekiguchi}, Maki and {Sergey}, Gary and {Shimasaku}, Kazuhiro and {Siegmund}, Walter A. and {Smee}, Stephen and {Smith}, J. Allyn and {Snedden}, S. and {Stone}, R. and {Stoughton}, Chris and {Strauss}, Michael A. and {Stubbs}, Christopher and {SubbaRao}, Mark and {Szalay}, Alexander S. and {Szapudi}, Istvan and {Szokoly}, Gyula P. and {Thakar}, Anirudda R. and {Tremonti}, Christy and {Tucker}, Douglas L. and {Uomoto}, Alan and {Vanden Berk}, Dan and {Vogeley}, Michael S. and {Waddell}, Patrick and {Wang}, Shu-i. and {Watanabe}, Masaru and {Weinberg}, David H. and {Yanny}, Brian and {Yasuda}, Naoki and {SDSS Collaboration}},
        title = "{The Sloan Digital Sky Survey: Technical Summary}",
      journal = {\aj},
         year = 2000,
        month = sep,
       volume = {120},
       number = {3},
        pages = {1579-1587},
          doi = {10.1086/301513},
archivePrefix = {arXiv},
       eprint = {astro-ph/0006396},
 primaryClass = {astro-ph},
       adsurl = {https://ui.adsabs.harvard.edu/abs/2000AJ....120.1579Y}
}

@ARTICLE{2011ApJS..193...29A,
       author = {{Aihara}, Hiroaki and {Allende Prieto}, Carlos and {An}, Deokkeun and {Anderson}, Scott F. and {Aubourg}, {\'E}ric and {Balbinot}, Eduardo and {Beers}, Timothy C. and {Berlind}, Andreas A. and {Bickerton}, Steven J. and {Bizyaev}, Dmitry and {Blanton}, Michael R. and {Bochanski}, John J. and {Bolton}, Adam S. and {Bovy}, Jo and {Brandt}, W.~N. and {Brinkmann}, J. and {Brown}, Peter J. and {Brownstein}, Joel R. and {Busca}, Nicolas G. and {Campbell}, Heather and {Carr}, Michael A. and {Chen}, Yanmei and {Chiappini}, Cristina and {Comparat}, Johan and {Connolly}, Natalia and {Cortes}, Marina and {Croft}, Rupert A.~C. and {Cuesta}, Antonio J. and {da Costa}, Luiz N. and {Davenport}, James R.~A. and {Dawson}, Kyle and {Dhital}, Saurav and {Ealet}, Anne and {Ebelke}, Garrett L. and {Edmondson}, Edward M. and {Eisenstein}, Daniel J. and {Escoffier}, Stephanie and {Esposito}, Massimiliano and {Evans}, Michael L. and {Fan}, Xiaohui and {Femen{\'\i}a Castell{\'a}}, Bruno and {Font-Ribera}, Andreu and {Frinchaboy}, Peter M. and {Ge}, Jian and {Gillespie}, Bruce A. and {Gilmore}, G. and {Gonz{\'a}lez Hern{\'a}ndez}, Jonay I. and {Gott}, J. Richard and {Gould}, Andrew and {Grebel}, Eva K. and {Gunn}, James E. and {Hamilton}, Jean-Christophe and {Harding}, Paul and {Harris}, David W. and {Hawley}, Suzanne L. and {Hearty}, Frederick R. and {Ho}, Shirley and {Hogg}, David W. and {Holtzman}, Jon A. and {Honscheid}, Klaus and {Inada}, Naohisa and {Ivans}, Inese I. and {Jiang}, Linhua and {Johnson}, Jennifer A. and {Jordan}, Cathy and {Jordan}, Wendell P. and {Kazin}, Eyal A. and {Kirkby}, David and {Klaene}, Mark A. and {Knapp}, G.~R. and {Kneib}, Jean-Paul and {Kochanek}, C.~S. and {Koesterke}, Lars and {Kollmeier}, Juna A. and {Kron}, Richard G. and {Lampeitl}, Hubert and {Lang}, Dustin and {Le Goff}, Jean-Marc and {Lee}, Young Sun and {Lin}, Yen-Ting and {Long}, Daniel C. and {Loomis}, Craig P. and {Lucatello}, Sara and {Lundgren}, Britt and {Lupton}, Robert H. and {Ma}, Zhibo and {MacDonald}, Nicholas and {Mahadevan}, Suvrath and {Maia}, Marcio A.~G. and {Makler}, Martin and {Malanushenko}, Elena and {Malanushenko}, Viktor and {Mandelbaum}, Rachel and {Maraston}, Claudia and {Margala}, Daniel and {Masters}, Karen L. and {McBride}, Cameron K. and {McGehee}, Peregrine M. and {McGreer}, Ian D. and {M{\'e}nard}, Brice and {Miralda-Escud{\'e}}, Jordi and {Morrison}, Heather L. and {Mullally}, F. and {Muna}, Demitri and {Munn}, Jeffrey A. and {Murayama}, Hitoshi and {Myers}, Adam D. and {Naugle}, Tracy and {Neto}, Angelo Fausti and {Nguyen}, Duy Cuong and {Nichol}, Robert C. and {O'Connell}, Robert W. and {Ogando}, Ricardo L.~C. and {Olmstead}, Matthew D. and {Oravetz}, Daniel J. and {Padmanabhan}, Nikhil and {Palanque-Delabrouille}, Nathalie and {Pan}, Kaike and {Pandey}, Parul and {P{\^a}ris}, Isabelle and {Percival}, Will J. and {Petitjean}, Patrick and {Pfaffenberger}, Robert and {Pforr}, Janine and {Phleps}, Stefanie and {Pichon}, Christophe and {Pieri}, Matthew M. and {Prada}, Francisco and {Price-Whelan}, Adrian M. and {Raddick}, M. Jordan and {Ramos}, Beatriz H.~F. and {Reyl{\'e}}, C{\'e}line and {Rich}, James and {Richards}, Gordon T. and {Rix}, Hans-Walter and {Robin}, Annie C. and {Rocha-Pinto}, Helio J. and {Rockosi}, Constance M. and {Roe}, Natalie A. and {Rollinde}, Emmanuel and {Ross}, Ashley J. and {Ross}, Nicholas P. and {Rossetto}, Bruno M. and {S{\'a}nchez}, Ariel G. and {Sayres}, Conor and {Schlegel}, David J. and {Schlesinger}, Katharine J. and {Schmidt}, Sarah J. and {Schneider}, Donald P. and {Sheldon}, Erin and {Shu}, Yiping and {Simmerer}, Jennifer and {Simmons}, Audrey E. and {Sivarani}, Thirupathi and {Snedden}, Stephanie A. and {Sobeck}, Jennifer S. and {Steinmetz}, Matthias and {Strauss}, Michael A. and {Szalay}, Alexander S. and {Tanaka}, Masayuki and {Thakar}, Aniruddha R. and {Thomas}, Daniel and {Tinker}, Jeremy L. and {Tofflemire}, Benjamin M. and {Tojeiro}, Rita and {Tremonti}, Christy A. and {Vandenberg}, Jan and {Vargas Maga{\~n}a}, M. and {Verde}, Licia and {Vogt}, Nicole P. and {Wake}, David A. and {Wang}, Ji and {Weaver}, Benjamin A. and {Weinberg}, David H. and {White}, Martin and {White}, Simon D.~M. and {Yanny}, Brian and {Yasuda}, Naoki and {Yeche}, Christophe and {Zehavi}, Idit},
        title = "{The Eighth Data Release of the Sloan Digital Sky Survey: First Data from SDSS-III}",
      journal = {\apjs},
         year = 2011,
        month = apr,
       volume = {193},
       number = {2},
          eid = {29},
        pages = {29},
          doi = {10.1088/0067-0049/193/2/29},
archivePrefix = {arXiv},
       eprint = {1101.1559},
 primaryClass = {astro-ph.IM},
       adsurl = {https://ui.adsabs.harvard.edu/abs/2011ApJS..193...29A}
}

@article{Glazebrook_1998,
doi = {10.1086/305039},
url = {https://doi.org/10.1086/305039},
year = {1998},
month = {jan},
publisher = {},
volume = {492},
number = {1},
pages = {98},
author = {Glazebrook, Karl and Offer, Alison R. and Deeley, Kathryn},
title = {Automatic Redshift Determination by Use of Principal Component Analysis. I. Fundamentals},
journal = {The Astrophysical Journal}
}




\appendix

\section{Spectral Line List}
\label{app:linelist}

Table~\ref{tab:linelist} lists the spectral lines supplied to UnEZ as
physics priors. Rest-frame vacuum wavelengths are taken from the SDSS
\textsc{spectro1d} line table \citep{sdss_dr6_linelist}. The index column
gives the position of each line in the configuration file, which is also
the index of the corresponding node in UnEZ's latent space.

Each line is assigned a classification that determines the activation
function applied to its predicted strength
(Section~\ref{sec:encoder}): emission-only lines (E) are passed through a
ReLU, absorption-only lines (A) through a negative ReLU, and lines that
may appear in either emission or absorption (B) through a linear
activation. Classifications were assigned based on the physical origin of
each transition rather than being taken from any catalogue: nebular
forbidden lines are restricted to emission, stellar photospheric features
to absorption, and resonance lines that may appear in absorption in broad
absorption line quasars are left unconstrained. The four sky lines listed
by \citet{sdss_dr6_linelist} are not fitted; they are masked during
preprocessing (Section~\ref{sec:preprocessing}).

\begin{table}
    \centering
    \caption{Spectral lines used as physics priors in UnEZ. Wavelengths
    are rest-frame vacuum values from the SDSS \textsc{spectro1d} line
    table \citep{sdss_dr6_linelist}. Classifications: E = emission-only,
    A = absorption-only, B = emission or absorption.}
    \label{tab:linelist}
    \begin{tabular}{rlcc @{\hskip 1.5em} rlcc}
        \toprule
        \# & Species & $\lambda_{\rm vac}$ (\AA) & Cl. &
        \# & Species & $\lambda_{\rm vac}$ (\AA) & Cl. \\
        \midrule
         1 & O\,\textsc{vi}                  & 1033.82 & B &
        23 & H$\gamma$                       & 4341.68 & B \\
         2 & Ly$\alpha$                      & 1215.24 & B &
        24 & [O\,\textsc{iii}]               & 4364.44 & E \\
         3 & N\,\textsc{v}                   & 1240.81 & B &
        25 & H$\beta$                        & 4862.68 & B \\
         4 & O\,\textsc{i}                   & 1305.53 & E &
        26 & [O\,\textsc{iii}]               & 4932.60 & E \\
         5 & C\,\textsc{ii}                  & 1335.31 & E &
        27 & [O\,\textsc{iii}]               & 4960.30 & E \\
         6 & Si\,\textsc{iv}                 & 1397.61 & E &
        28 & [O\,\textsc{iii}]               & 5008.24 & E \\
         7 & Si\,\textsc{iv}\,+\,O\,\textsc{iv}] & 1399.80 & E &
        29 & [O\,\textsc{i}]                 & 6302.05 & E \\
         8 & C\,\textsc{iv}                  & 1549.48 & B &
        30 & [O\,\textsc{i}]                 & 6365.54 & E \\
         9 & He\,\textsc{ii}                 & 1640.40 & B &
        31 & N\,\textsc{i}                 & 6529.03 & E \\
        10 & O\,\textsc{iii}]                & 1665.85 & E &
        32 & [N\,\textsc{ii}]                & 6549.86 & E \\
        11 & Al\,\textsc{iii}                & 1857.40 & B &
        33 & H$\alpha$                       & 6564.61 & B \\
        12 & C\,\textsc{iii}]                & 1908.73 & E &
        34 & [N\,\textsc{ii}]                & 6585.27 & E \\
        13 & C\,\textsc{ii}]                 & 2326.00 & E &
        35 & [S\,\textsc{ii}]                & 6718.29 & E \\
        14 & Ne\,\textsc{iv}               & 2439.50 & B &
        36 & [S\,\textsc{ii}]                & 6732.67 & E \\
        15 & Mg\,\textsc{ii}                 & 2799.12 & B &
        37 & Ca\,\textsc{ii}\,K              & 3934.78 & A \\
        16 & [Ne\,\textsc{v}]                & 3346.79 & B &
        38 & Ca\,\textsc{ii}\,H              & 3969.59 & A \\
        17 & Ne\,\textsc{vi}               & 3426.85 & B &
        39 & G band                          & 4305.61 & A \\
        18 & [O\,\textsc{ii}]                & 3727.09 & E &
        40 & Mg\,b                           & 5176.70 & A \\
        19 & [O\,\textsc{ii}]                & 3729.88 & E &
        41 & Na\,D                           & 5895.60 & A \\
        20 & He\,\textsc{i}                  & 3889.00 & B &
        42 & Ca\,\textsc{ii}                 & 8500.36 & A \\
        21 & [S\,\textsc{ii}]                & 4072.30 & E &
        43 & Ca\,\textsc{ii}                 & 8544.44 & A \\
        22 & H$\delta$                       & 4102.89 & B &
        44 & Ca\,\textsc{ii}                 & 8664.52 & A \\
        \bottomrule
    \end{tabular}
\end{table}

\section{Brute Force Experiment}
\label{app:brute force}
UnEZ deterministically estimates redshift from a neural network line strength prediction. As posed in Section \ref{sec:decoder}, we test whether the line strengths can be estimated deterministically. Here we show why a neural network that has learned the distribution of spectra performs better than brute force fitting every line at every redshift.

The brute force method we use measures $\chi^{2}$, indicated in equation \ref{eq: brute force chi squared}
\begin{equation}
    \chi^{2} = \sum_{i} \left( \sum_{n=0}^{n_{e}-1} S_{n} \, e^{-\frac{(\lambda_{i} - \mu_{n})^{2}}{2\sigma^{2}}} - F_{i} \right)^{2}
    \label{eq: brute force chi squared}
\end{equation}
at trial redshifts and line widths $\sigma$. At a fixed $z$ and $\sigma$ the model in Equation~\ref{eq: brute force chi squared} is linear in the line strengths. Each Gaussian is a fixed basis vector, and the spectrum is modeled as a linear combination of them. Therefore, the line strengths that minimize $\chi^{2}$ are given by a single linear least-squares projection of the spectrum onto this basis, with no iteration required. This makes the computation fast and exact at each trial. It recovers the optimal amplitudes at every trial $(z, \sigma)$, making it the strongest non-learned counterpart to UnEZ's decoder rather than an approximation of it.

This is the same linear-in-amplitude method used by traditional pipelines. The SDSS redshift pipeline \citep{Bolton_2012} and PCAz \citep{Glazebrook_1998} project each spectrum onto a fixed basis of PCA eigenspectra at each trial redshift. UnEZ, the brute force method, and these traditional algorithms differ only in basis and in how the amplitudes are obtained. The brute force projects onto a fixed line basis, SDSS onto fixed eigenspectra, and UnEZ predicts per-object amplitudes over the physics-constrained line set with an encoder trained on the distribution of real spectra. Comparing UnEZ against this baseline tests the contribution of that learned prediction.

The same preprocessing pipeline was used for both UnEZ and the brute force method. We also sampled the same redshifts as UnEZ for the brute force fit so the two methods can easily be compared. The brute force method samples 32 trial line widths across the same line width range as UnEZ.

\begin{figure*}
    \centering
    \includegraphics[width=\textwidth]{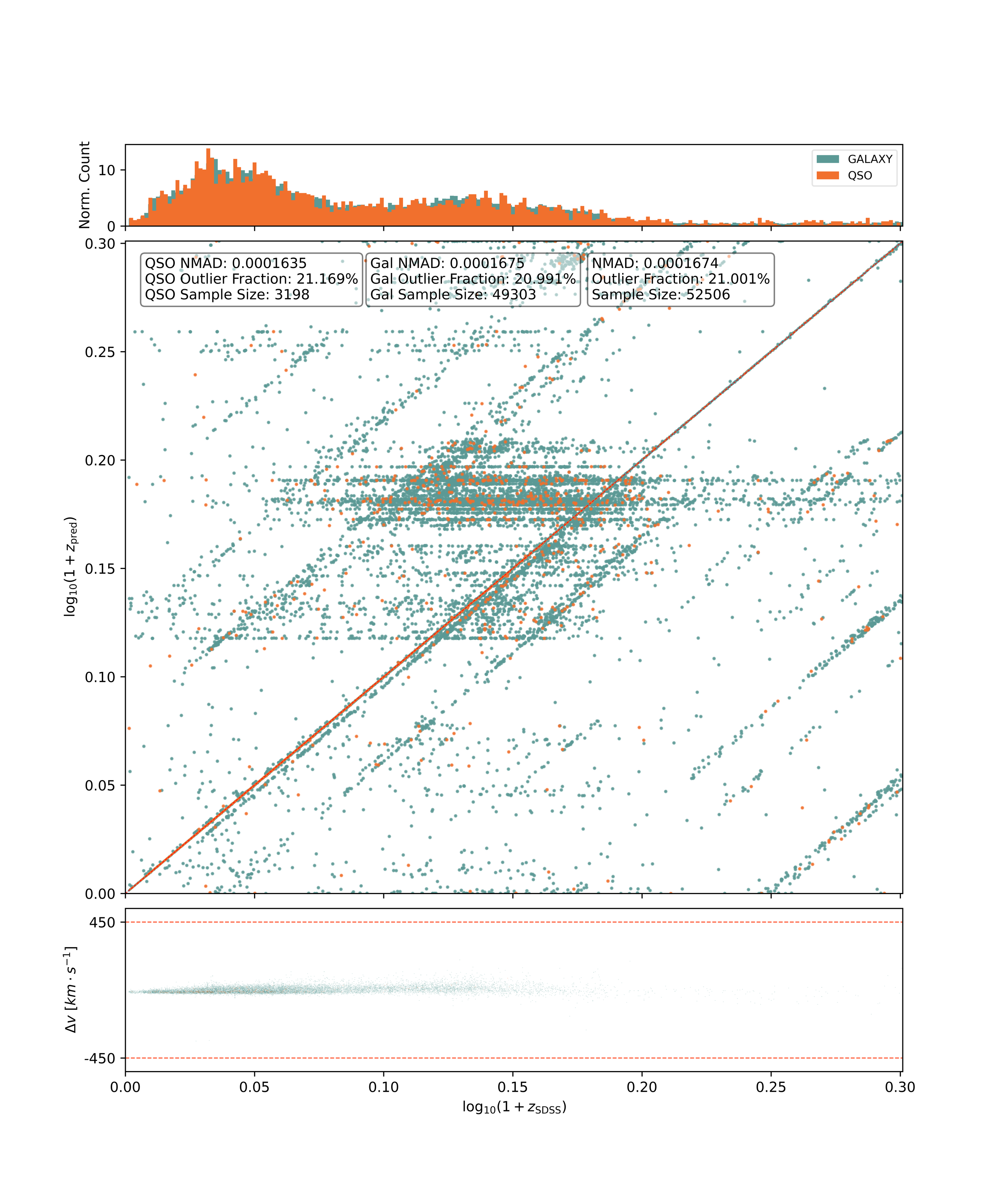}
    \caption{As Figure \ref{fig:z performance low snr} but for the brute force method. Line confusion degeneracies and skyline contamination become much more prevalent using the brute force method.}
    \label{fig:brute force z performance}    

\end{figure*}

\begin{figure}
    \includegraphics[width=\columnwidth]{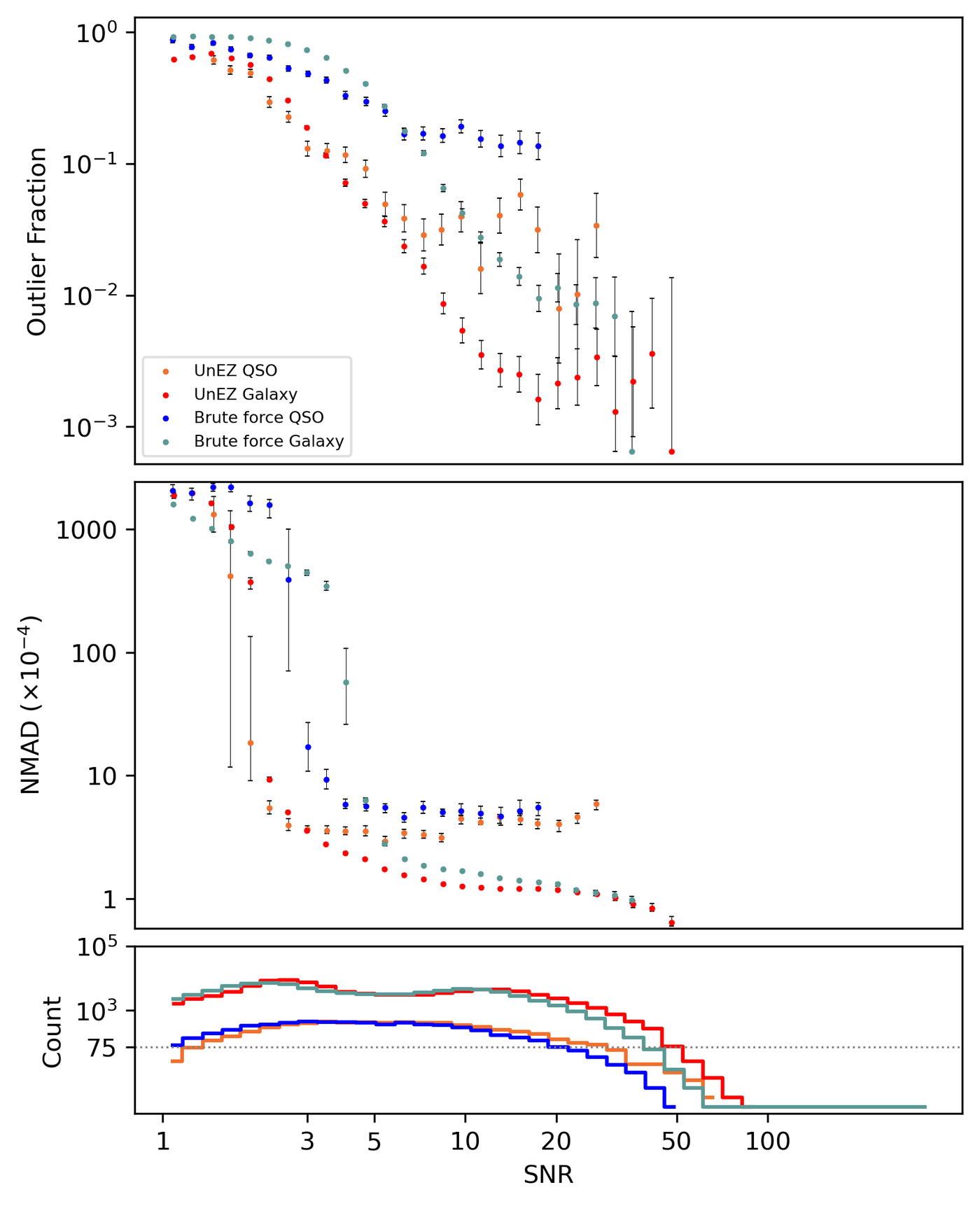}
    \caption{As Figure \ref{fig:snr vs nmad and outlier fraction} with the brute force method plotted as well. Only 100,000 random spectra were selected for each method due to the long compute time required for the brute force method. UnEZ consistently performs better across the entire SNR range and converges toward peak performance at a lower SNR.}
    \label{fig:brute force snr vs nmad vs outlier fraction}

\end{figure}

100,000 randomly selected spectra were chosen across the entire SNR range for fitting. Of those, approximately 52,500 fall above the SNR>3 threshold as seen in Figure \ref{fig:brute force z performance}. We found that the brute force method is much more prone to fitting skylines and confusing lines than UnEZ, achieving an outlier fraction of approximately 21\% as opposed to UnEZ's 3.9\% at the same SNR range. Additionally, Figure \ref{fig:brute force snr vs nmad vs outlier fraction} shows that UnEZ performs better than the brute force method in all performance metrics across the entire SNR range. Not only does UnEZ consistently perform better, but the performance tends towards peak performance at a lower SNR than the brute force. Although this method is very efficient at individual trial redshifts and line widths, due to the hundreds of thousands of trials it ends up far less efficient than UnEZ. UnEZ's inference takes approximately 25 milliseconds per spectrum, while the brute force method takes approximately 250 milliseconds. This order of magnitude increase in inference speed and redshift prediction performance makes UnEZ a much more viable candidate for modern large-scale surveys.

This experiment does systematically calculate the redshift with the lowest $\chi^{2}$, however, that does not always lead to the correct redshift. If skyline or noise contamination is too strong, $\chi^{2}$ can be lower when fitting skylines as they drown out spectral features native to the observed object. UnEZ on the other hand has learned from the distribution of spectra, and can recognize and encode spectral features native to the object being observed, rather than just searching for the strongest features and fitting to those.

We conclude that the neural network in UnEZ's architecture is critical to ensuring the pipeline can effectively predict redshift as it reduces the likelihood of fitting noise or confusing different lines while also reducing the amount of compute time needed. This experiment serves to uphold the need for UnEZ's machine learning architecture.

\bsp	
\label{lastpage}
\end{document}